\documentclass[10pt,pre,twocolumn,tightenlines,longbibliography,notitlepage,superscriptaddress,nofootinbib]{revtex4-2}
\pdfoutput=1

\usepackage{xcolor}
\usepackage{verbatim}
\usepackage{graphicx}
\usepackage{natbib}
\usepackage{siunitx}
\usepackage{float}
\usepackage{xr}
\usepackage{pifont}
\usepackage{amssymb}
\usepackage[textsize=tiny]{todonotes}

\makeatletter
\newcommand*{\addFileDependency}[1]{
  \typeout{(#1)}
  \@addtofilelist{#1}
  \IfFileExists{#1}{}{\typeout{No file #1.}}
}
\makeatother

\newcommand*{\myexternaldocument}[1]{%
    \externaldocument{#1}%
    \addFileDependency{#1.tex}%
    \addFileDependency{#1.aux}%
}
\myexternaldocument{main-shortSI}

\newcommand*{\rom}[1]{\uppercase\expandafter{\romannumeral #1\relax}}

\definecolor{cryomarker}{RGB}{0,129,167} 

\begin{document}
\title{Modular programming of interaction and geometric specificity enables assembly of complex DNA origami nanostructures}

\author{Rupam Saha*}
\affiliation{Martin A. Fisher School of Physics, Brandeis University, Waltham, Massachusetts 02453, USA}
\author{Daichi Hayakawa*}
\affiliation{Martin A. Fisher School of Physics, Brandeis University, Waltham, Massachusetts 02453, USA}
\author{Thomas E. Videb\ae k}
\affiliation{Martin A. Fisher School of Physics, Brandeis University, Waltham, Massachusetts 02453, USA}
\author{Mason Price}
\affiliation{Martin A. Fisher School of Physics, Brandeis University, Waltham, Massachusetts 02453, USA}
\author{Wei-Shao Wei}
\affiliation{Martin A. Fisher School of Physics, Brandeis University, Waltham, Massachusetts 02453, USA}
\author{Juanita Pombo}
\affiliation{Thomas Lord Department of Mechanical Engineering and Materials Science, Duke University, Durham, North Carolina 27705, USA}
\author{Daniel Duke}
\affiliation{Thomas Lord Department of Mechanical Engineering and Materials Science, Duke University, Durham, North Carolina 27705, USA}
\author{Gaurav Arya}
\affiliation{Thomas Lord Department of Mechanical Engineering and Materials Science, Duke University, Durham, North Carolina 27705, USA}
\author{Gregory M. Grason}
\affiliation{Department of Polymer Science and Engineering, University of Massachusetts, Amherst, Massachusetts 01003, USA}
\author{W. Benjamin Rogers}
\email{wrogers@brandeis.edu}
\affiliation{Martin A. Fisher School of Physics, Brandeis University, Waltham, Massachusetts 02453, USA}
\author{Seth Fraden}
\email{fraden@brandeis.edu}
\affiliation{Martin A. Fisher School of Physics, Brandeis University, Waltham, Massachusetts 02453, USA}

\begin{abstract}
We present a modular DNA origami design approach to address the challenges of assembling geometrically complex nanoscale structures, including those with nonuniform Gaussian curvature. This approach features a core structure that completely conserves the scaffold routing across different designs and preserves more than 70\% of the DNA staples between designs, dramatically reducing both cost and effort, while enabling precise and independent programming of subunit interactions and binding angles through adjustable overhang lengths and sequences. Using cryogenic electron microscopy, gel electrophoresis, and coarse-grained molecular dynamics simulations, we validate a set of robust design rules. We demonstrate the method's utility by assembling a variety of self-limiting structures, including anisotropic shells with controlled inter-subunit interactions and curvature, and a toroid with globally varying curvature.  Our strategy is both cost-effective and versatile, providing a promising and efficient solution for the synthetic fabrication of complex nanostructures.

\end{abstract}

\maketitle
\def\thefootnote{*}\footnotetext{These authors contributed equally to this work}\def\thefootnote{\arabic{footnote}}

Nature is replete with functional biomolecular structures, such as  viral capsids and microtubules, which are self-assembled with remarkable precision, relying on specific geometries and interactions encoded in molecular subunits \cite{ caspar_physical_1962, Perlmutter2015Apr, Wade2009Oct}. Inspired by Nature, scientists have developed synthetic building blocks with programmable geometries and interactions---such as DNA tiles~\cite{wei_complex_2012}, DNA-coated colloids~\cite{Mirkin1996Aug, Alivisatos1996Aug,rogers2016using,Jacobs2024arXiv}, DNA origami~\cite{rothemund_folding_2006}, DNA bricks~\cite{ke_three-dimensional_2012, ong2017programmable}, and \textit{de novo}-designed proteins~\cite{Huang2016Sep, Jumper2021Aug}---to assemble functional nanostructures. Among the available platforms, DNA origami stands alone in its ability to simultaneously encode the block geometry, binding angles, and interaction specificity with nanometer- and $k_\mathrm{B}T$-scale precision~\cite{dietz2009folding, douglas_rapid_2009, Ke2009Nov, gerling_dynamic_2015, tikhomirov_programmable_2017, Wintersinger2023Mar, wagenbauer_gigadalton-scale_2017, sigl2021programmable, Hayakawa2022Oct, videbaek2023economical, Hayakawa2024Mar, wei2024hierarchical, Luu2024Nov, Lee2025Jan}, making it a promising tool in nanotechnological applications, such as photonics~\cite{posnjak2024diamond, liu2024inverse}, plasmonics~\cite{wang2016programming, Kahn2022Sep}, and anti-viral therapeutics~\cite{sigl2021programmable}.

However, such applications are yet to be realized in complex 2D surfaces with spatially varying Gaussian curvature, such as helical tubules or triply periodic minimal surfaces. Among programmable 2D surface-like structures, existing DNA-origami approaches have, so far, been limited to relatively simple geometries, specifically those with spatially uniform Gaussian curvature, like sheets~\cite{Wintersinger2023Mar,Hayakawa2024Mar}, shells~\cite{sigl2021programmable}, and cylindrical tubules~\cite{wagenbauer_gigadalton-scale_2017, Hayakawa2022Oct, videbaek2023economical}. Programming the assembly of 3D architectures with nonuniform Gaussian curvature remains a significant challenge. Targeting a unique surface with a spatial curvature variation requires precise control over both {\it type specificity} (the local rules for which particle types bind to one another) and {\it geometric specificity} (the local geometry of the bonds between neighboring subunits) simultaneously in a complex mixture of many distinct building block species. The type specificity determines which particles are neighbors in the ``graph'' of a given target structure, while the geometric specificity determines the distance and angles of the bonds between bound particles~\cite{duque2024limits}.  Without geometrically specific interactions (e.g. isotropic DNA-functionalized spheres~\cite{Jacobs2024arXiv}), the graph of type-specific interactions, in general, has multiple degenerate embeddings~\cite{zeravcic2017colloquium}, leading to the absence of 3D-shape control.  

While DNA origami has been used to implement both type- and geometry-specific interactions to program multicomponent structures, in nearly all cases, the structures are built from many copies of one or a few block geometries, limiting the accessible structures to those with uniform curvatures~\cite{rothemund_folding_2006, gerling_dynamic_2015, wagenbauer_gigadalton-scale_2017, sigl2021programmable, Hayakawa2022Oct, Wintersinger2023Mar, videbaek2023economical}. The limited number of block geometries results from the fact that, historically, creating multiple block shapes required designing multiple distinct scaffold routings---one per block shape---which is laborious and costly. 

Recent efforts by us~\cite{wei2024economical} and Karfuscher \textit{et al.}~\cite{karfusehr2024self} demonstrate the feasibility of designing modular DNA-origami building blocks with programmable geometric and type specificity using modular single-stranded spacers. However, these approaches have so far targeted structures composed of a small number of building block types. Furthermore, because they use single-stranded DNA spacers, it could be challenging to use these methods to precisely control the intersubunit angles in a complex structure that requires accurately prescribed variations in the spatial curvature. Therefore, going beyond existing classes of assemblies to make arbitrarily complex 2D manifolds requires developing additional approaches that enable the efficient design of building blocks whose type specificity and geometric specificity can be programmed on demand while preserving the scaffold routing and making minimal changes to the staple sequences.

\begin{figure*}[th!]
 \centering
 \includegraphics[width=\textwidth]{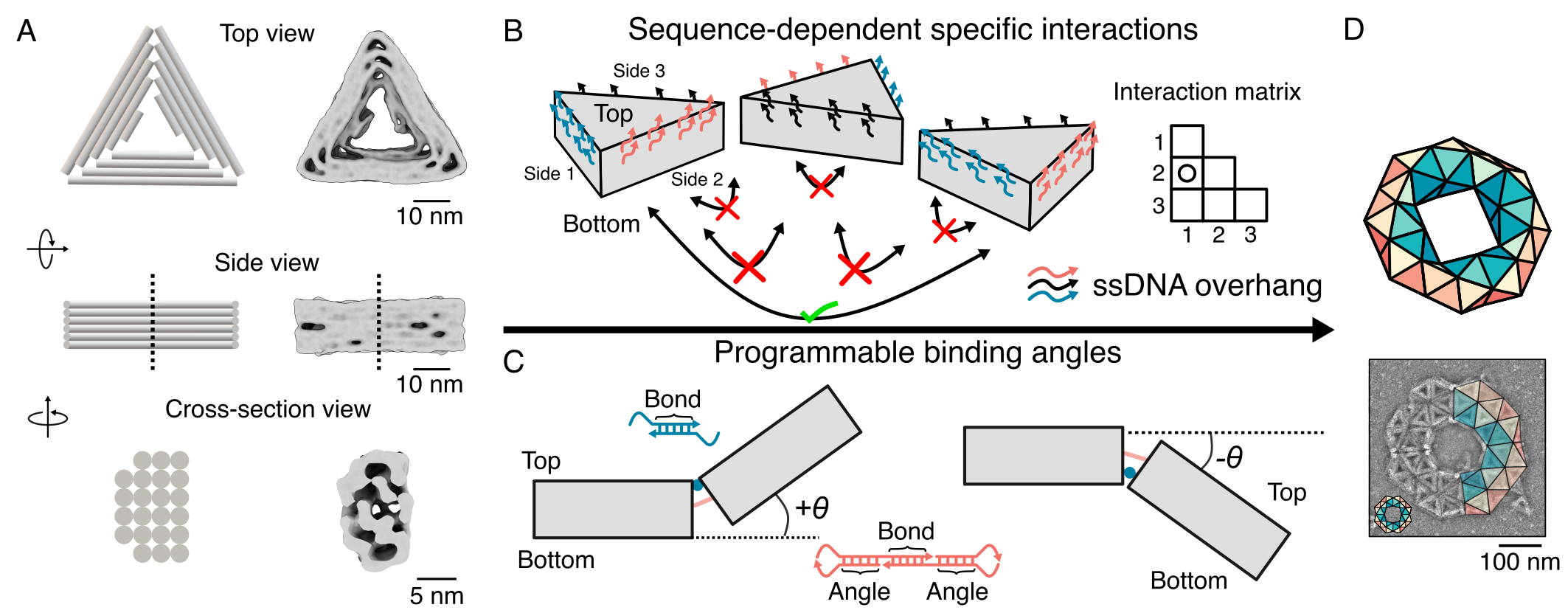}
 \caption{\textbf{Modular DNA origami building block with programmable interactions and binding angles.}
(A)  Schematic and cryo-EM reconstruction of the modular DNA origami building block, a hollow right equilateral triangular prism, but lacking up-down symmetry in the bonds. Cylinders represent dsDNA helices and are packed on a square lattice~\cite{Ke2009Nov, douglas_rapid_2009}. Top, side and section views. (B)  Single-stranded DNA overhangs protrude from the sides of identical triangles. The symbol \ding{51} indicates a binding pair and \ding{53} a non-binding pair. Only 1 of 6 possible bonds form in this example. Colors denote members of a set of extruded overhangs on a single side. The interaction matrix contains the symbol $\circ$ when two sides bond and is blank otherwise. (C)   The binding angle between two bound triangles, $\theta$, is determined by the relative length of the two rows of overhangs on a single side. The binding angle can be made positive or negative by changing the location of the overhangs. (D) A schematic and EM image of a toroid. Different colors indicate different species of triangles encoded with a unique set of interactions and binding angles. 
}
 \label{fig:1}
\end{figure*}

\section*{Results and discussions}


\subsection{Modular design concept}

\begin{figure*}[th!]
 \centering
 \includegraphics[width=0.9\linewidth]{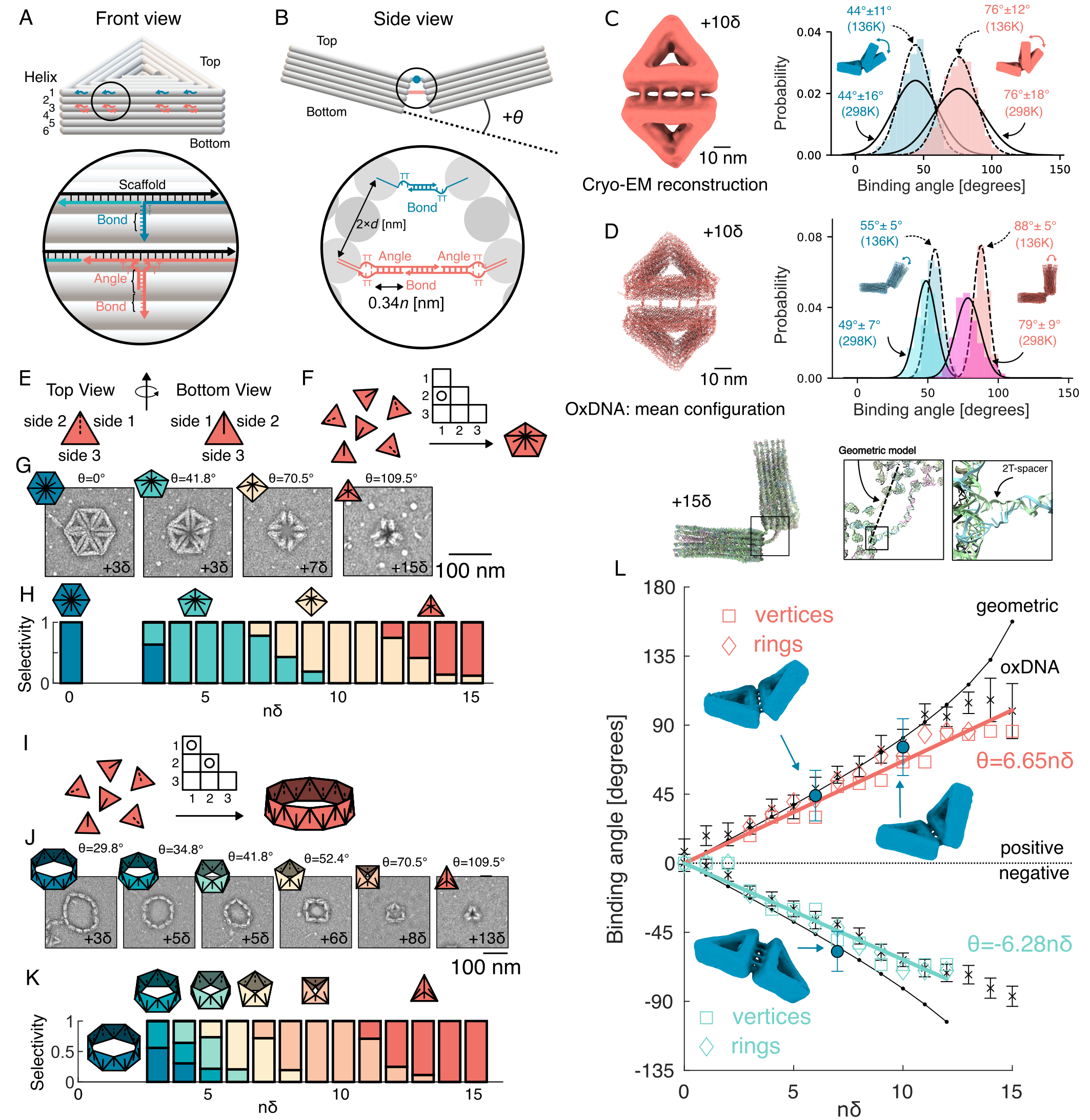}
 \caption{\textbf{Design and characterization of binding angles.}
 (A)  Front view of the DNA origami building block. (B)  Side view  of bonded pair at the binding angle $\theta$, with zoomed-in views of extruded overhangs. (C)  Left: cryo-EM reconstruction of $+10\delta$ dimers. Right: binding angle distributions for $+6\delta$ (blue) and $+10\delta$ (red) dimers. Dashed line are measurements assumed to be equilibrated at 136~K, the solid line is rescaled to 298~K.  Two extreme conformations are shown for each. (D)  Top row: (left) oxDNA generated mean configuration of $+10\delta$ dimers at 298~K,  (right) binding angle distributions for $+6\delta$ (blue) and $+10\delta$ (red). Dashed lines are for 136~K, solid lines are for 298~K. For clarity, histograms are displayed only for 136~K. Two extreme conformations at 136~K are shown for each. Bottom row: Mean configuration of $+15\delta$ dimers from oxDNA with zoomed-in views of overhangs and two-thymine spacers. 
 (E) Anisotropic triangles, labeled with a line to indicate the orientation: top view (dotted line) and bottom view (solid line). (F) Monomers with side 1 binding to side 2 assemble into vertices. (G) TEM images of vertices. (H) Selectivity of vertices of different sizes as angle-domain length varies, obtained from gel electrophoresis. Each color signifies an unique vertex. (I) Monomers with sides 1 and 2 both binding to themselves assemble into strips or rings. (J) TEM images of rings. (K) Selectivity of rings of different sizes as angle-domain length varies, derived from gel electrophoresis. (L) Summary of binding angles from geometrical prediction ($\bullet$), oxDNA dimer simulation ($\times$), cryo-EM multibody analysis (${\textcolor{cryomarker}{\ensuremath\bullet}}$), and vertex ($\square$) and ring ($\lozenge$) gel electrophoresis assembly analysis (red for positive and green for negative angles). Solid line indicates the linear fit to the vertex and ring assembly results. Error bars indicate fluctuations. All values are either obtained or scaled to 298~K.}
 \label{fig:2}
\end{figure*}

 Here, we introduce a predictable and efficient design strategy for creating modular building blocks whose interaction specificity and local curvature can be programmed both independently and precisely to assemble geometrically complex 2D manifolds from many distinct block species. We design a modular DNA-origami building block in the form of a hollow right equilateral triangular prism whose interactions and binding angles can be programmed independently while completely conserving the scaffold routing, as well as 70\% of the staple sequences (Figs.~\ref{fig:1}A, \ref{fig:2}A and S25).  By varying the sequences of the 30\% of the staples that are not attached to the scaffold, denoted as ``overhangs'', we encode both the specific interactions among the building blocks (Fig.~\ref{fig:1}B), as well as the binding angles between neighboring triangles (Fig.~\ref{fig:1}C). This modular approach offers an enormous technical benefit as changing the scaffold routing almost always requires many new staples, which raises the cost and prolongs the time needed to design and optimize folding conditions of the structure.  We demonstrate the potential of our economical design approach by assembling various complex 2D manifolds, including Platonic and non-Platonic shells, and a toroid, which has non-uniform Gaussian curvature (Fig.~\ref{fig:1}D), using different mixtures of a variety of block species made from the same modular template. 



The hollow right equilateral triangular prism was chosen as the building block for three key reasons. First, mathematical theorems establish that any 2D manifold can be triangulated, making the method universally applicable. Triangulation algorithms are  practical and widely used in applied mathematics for creating efficient meshes, particularly in computer graphics and finite-element analysis. Second, triangles are the only structurally rigid polygon, a property exploited in engineering to construct lightweight and stable structures such as trusses in bridges and towers, exemplified by the Eiffel Tower. Third, accurate rendering of right triangular prisms in the form of DNA origami is straightforward, as all DNA double helices lie in parallel planes.

In previous work, triangular prisms were used to assemble capsids and cylinders by designing prisms with tilted sides. While these designs successfully self-assembled, each tilt angle required rerouting the scaffold and replacing all staples. Furthermore, the use of shape complementarity to bind triangles introduced two significant limitations; it restricted the number of distinct binding sites and prevented independent control of interaction strength, binding angle, and bond flexibility, often resulting in off-target assemblies~\cite{Hayakawa2022Oct}. Although replacing shape-complementary bonds with ssDNA expanded the number of distinct binding sites, each new tilt angle still required a redesign of the triangle~\cite{videbaek2023economical}. Our new modular approach overcomes these constraints.

In our design, an 8064-nucleotide scaffold strand is folded by two types of staple strands: universal `core staples' in the interior of the triangular prism and `interface staples' that have portions that protrude out of the triangle sides at assigned locations. The 148 unique core staples remain constant across different triangle designs; the 60 unique interface staples, 20 per side, vary from design to design and program the interaction specificity and binding angles. Each unique design utilizes two rows of interface staples per side (Fig.~\ref{fig:2}A).

The interface staples are conceptually divided into two parts; one part binds to the scaffold and remains invariant from design to design, while the second binds to other interface staples and can vary between designs. This variable part contains two domains, a `bond' domain that programs the interaction specificity and an `angle' domain that programs the binding angles (Fig.~\ref{fig:2}A, Fig.~\ref{fig:2}B). 

Although a triangular prism has top-bottom symmetry (Fig.~\ref{fig:1}A,B), we include bonding sites that make the top distinguishable from the bottom, a necessary condition to form structures with a well defined inside and outside, such as capsids. The bond-domain of a single staple is a five-nucleotide-long sequence located at the end of an interface staple; two examples are shown in Fig.~\ref{fig:2}A.  Four pairs of bonds are located on each of the three sides of a triangle; one pair is illustrated in Fig.~\ref{fig:2}B.   The eight bonds that bind two triangles together are designed to have a high off-rate to promote rapid equilibration. We further design the interactions between two triangles to be highly specific, either binding or non-binding, as represented by an interaction matrix such as in Fig.~\ref{fig:1}B.  See SI Section~IX for more details.

 The angle domain, formed by hybridizing adjacent interface staples, is a double-stranded region that controls the binding angle $\theta$ between neighboring triangles and can be tuned from 0 to 15 base pairs (Figs.~\ref{fig:2}A, B). We program the binding angle by choosing the difference in lengths of the angle domains in two different rows of interface staples. To create positive angles, we set one angle domain to have zero length, in which case only one staple extrudes from the core, while the other angle domain requires two adjacent staples extruding from the core (Fig.~\ref{fig:2}B). The sign of the angle is controlled by the relative location of the shorter angle domains with respect to the longer angle domains. We classify angle domains as $+n\delta$ for positive and $-n\delta$ for negative angles, where $n\delta$ is the length difference of the angle domain in units of the number of base pairs. All interface staples have a two-thymine-long spacer where the overhang protrudes from the building block. 
The design details and sequences of the overhangs are organized in SI Section~IX.

\subsection{Designing and characterizing the binding angle}

We start by developing a simple geometrical model to build an intuition for the relationship between the difference in lengths of the angle domains and the resultant binding angle. By assuming that the bridges linking neighboring triangles are rigid rods, we derive an expression relating the binding angle $\theta$ to the  length difference of the angle domains $n\delta$: 
\begin{equation}
    \theta=2\arcsin{\frac{0.34~\textrm{nm} \times n\delta}{2 \times d}},
\end{equation}
where $d=2.6$~nm is the interhelical distance~\cite{Ke2009Nov, Bai2012Dec}, assuming the helices are close-packed (Fig.~\ref{fig:2}B). We also assume that the bridges approximate rigid rods because their lengths are much shorter than the persistence length of double-stranded DNA~\cite{wang1997stretching}. The factor of 0.34~nm/bp is the length of double-stranded DNA per base pair~\cite{hagerman1988flexibility}. For small angles, $\theta$ is approximately linear with $n\delta$  (Fig.~\ref{fig:2}L).


We test the geometric relationship between binding angle ($\theta$) and angle-domain length ($n\delta$) using three complementary approaches. First,  we experimentally assemble dimers and use cryogenic electron microscopy (cryo-EM) reconstruction to characterize the binding angle distributions as a function of $n\delta$ (Fig.~\ref{fig:2}C). Second, we experimentally assemble two classes of self-closing assemblies, vertices and rings, and compare the yields of different polymorphs, whose sizes are governed by the binding angles, bending rigidity, and effective binding free energy, to a thermodynamic model. Third, we use the oxDNA coarse-grained model to represent the dimers and carry out molecular dynamics simulations to predict their binding angle distributions (Fig.~\ref{fig:2}D)~\cite{Snodin2015Jun,shi2017conformational}.  

 We obtain 3D structures of dimers of particles using single-particle cryo-EM. The average reconstructions for three designs with angle-domain lengths of +6$\delta$, +10$\delta$ and -7$\delta$ are shown in Figs.~\ref{fig:2}C~\&~S9. From these average reconstructions, we fit the separate triangular bodies and find the opening angle between them. The measured average angle is in excellent agreement with the geometric model (Fig.~\ref{fig:2}L). To quantify the fluctuations in the system, we also perform a multibody analysis~\cite{nakane2018characterisation, videbaek_measuring_2025} on the dimer, assuming that each triangular monomer is a rigid body (Figs.~\ref{fig:2}C). In cryo-EM, samples are quenched to the vitrification temperature of water, 136~K, in about $10^{-4}$ seconds, which we assume is enough time for the bond angle distribution to equilibrate as the rotational diffusion time of a monomer is much less than the quench time. Furthermore, we assume that the physical properties of DNA are independent of the temperature to relate the fluctuations that we measure at the vitrification temperature to the fluctuations we would expect at room temperature. Thus, in Fig.~\ref{fig:2}C we assume the measured bond angle distribution is equilibrated at 136~K and then, to obtain the distribution at 298~K, we scale the standard deviation by temperature as ${\sigma_{298\mbox{\scriptsize{K}}}^{\mbox{\scriptsize{EM}}}}/{\sigma_{136\mbox{\scriptsize{K}}}^{\mbox{\scriptsize{EM}}}} = \sqrt{298~\mathrm{K}/136~\mathrm{K}}$. More details are in SI section~IV.


 
To test the assumption that the physical properties of dsDNA are independent of temperature, ranging from room temperature to the vitrification temperature of water, we design an experiment to infer the binding angles experimentally at 298~K from the assembly of self-closing structures whose polymorphs are predictable. For example, vertices ranging from hexamers to trimers self-assemble when side 1 binds to side 2 of the triangle (with side 3 inert), depending on the angle-domain lengths (Fig.~\ref{fig:2} E-H and Fig.~S22).  Rings form when side 1 binds to side 1 and side 2 binds to side 2 (with side 3 inert) (Fig.~\ref{fig:2} I-K and Fig.~S23). We conduct these experiments using both positive and negative binding angles. 

We measure the equilibrium room temperature distributions of the vertex and ring sizes using gel electrophoresis, and infer the average binding angle for each angle-domain length from a thermodynamic model. By analyzing the band intensities from gel electrophoresis, we obtain the relative yields of all of the self-closing structures that assemble (Fig.~\ref{fig:2}H, K and Fig.~S15). We find a monotonic decrease in the most probable structure size with increasing angle-domain lengths. This trend is expected because increasing $n\delta$ should increase $\theta$, resulting in closed structures with a smaller radius of curvature and fewer subunits. To infer the preferred binding angles from the experiments, we fit the selectivity of the various self-closing structures, defined as the relative yield of a given polymorph to that of all possible self-closing structures, to a theoretical selectivity estimated by considering the energies of all allowed polymorphs in a constant-temperature (canonical) ensemble (see SI section~VI)~\cite{wagenbauer_gigadalton-scale_2017}. The absolute values of the binding angles deduced from electrophoresis experiments are systematically smaller than the predictions of the geometrical model, with the deviation increasing for longer angle-domain lengths~(Fig.~\ref{fig:2}L). 

Next, we analyze the binding angles computed from oxDNA simulations (Fig.~\ref{fig:2}D). We first run simulations to determine the equilibrium distribution of binding angles at room temperature, 298~K, which is close to the temperature at which assembly experiments are performed, and then we quench the system to 136~K, which is the temperature at which we presume the dimers are equilibrated in the cryo-EM measurements.  The average angles obtained from oxDNA at 298~K are consistent with the vertex and ring measurements.  The average angles vary with temperature, with the bond angle being larger at low temperature. Notably, as shown in Figs.~\ref{fig:2}C~and~2D, the standard deviation of the distributions determined by cryo-EM  at 136~K are about two-times greater than predicted by oxDNA.   Consequently, oxDNA predicts a roughly 4-fold greater bending modulus than obtained by cryo-EM.    For discussion of the oxDNA results and possible explanations of the discrepancy with cryo-EM, see SI sections~II and~V.

Finally, we derive a design rule for encoding angles via a linear fit to our data points from the assembly-size experiments. For positive and negative angles, we obtain 6.65 and -6.28 degrees per base pair (Fig.~\ref{fig:2}L), which are roughly 15\% and 22\% smaller than one would predict from the geometrical model over its linear regime, respectively, but are in good agreement with cryo-EM measurements, vertex and ring assembly experiments, and oxDNA predictions. Going forward, we use this linear relationship to program the angles necessary to assemble geometrically complex structures.


\subsection{Self-assembly of deltahedral shells}

\begin{figure*}[th!]
 \centering
 \includegraphics[width=0.9\textwidth]{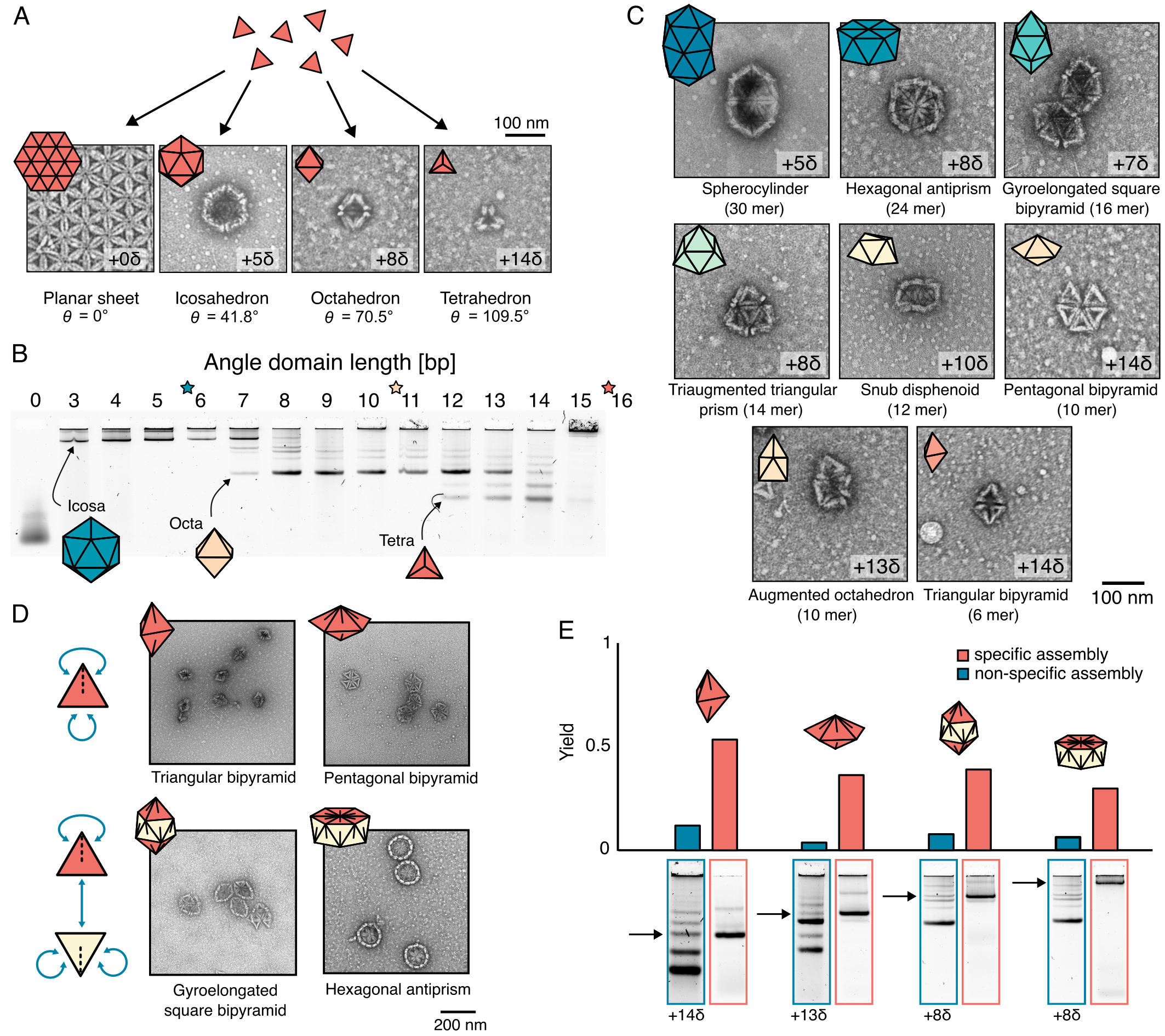}
 \caption{\textbf{Shell assembly by tuning the angle-domain length.}
 (A) Every side of the monomer is encoded with identical, self-complementary interaction and angle-domain length. By changing the angle-domain length, monomers assemble into various structures from 2D sheets to self-closed Platonic solids, shown in TEM micrographs. Note that the bottom face of the triangle always points outwards. (B) Laser-scanned image of gel electrophoresis where the monomers are assembled while varying the relative angle-domain length from 0~bp and 3~bp to 15~bp. Stars indicate angle-domain lengths whose corresponding binding angles are predicted to match closely with icosahedron, octahedron, and tetrahedron. (C) TEM micrographs of various non-Platonic shells that are observed in TEM at different angle-domain lengths. (D) Programmable self-assembly of non-Platonic shells by controlling the interaction specificity and binding angles, using up to two species of triangles. Blue arrows indicate interacting sides. For each assembly, we show a representative TEM micrograph. (E) The yield of specific and non-specific assemblies of non-Platonic shells is compared side-by-side, accompanied by the corresponding gel scan. The arrows indicate the location of the target assemblies. For non-specific assembly, we reference the angle-domain length with the highest yield for the target structure from (B). 
}
 \label{fig:3}
\end{figure*}

To test our design scheme, we start by assembling triangles with identical angle-domain lengths and self-complementary sequences on all three sides. Because all the sides bind to each other and the average binding angles of all three sides are the same, we expect to assemble symmetric structures with constant binding angles between neighboring subunits, e.g., 2D sheets, icosahedra, octahedra, and tetrahedra, with binding angles of $0^\circ$, $41.8^\circ$, $70.5^\circ$, and $109.5^\circ$, respectively (Fig.~\ref{fig:3}A). Given the empirical design relationship between binding angle, $\theta$, and the difference in the number of bases in the angle domain, $n\delta$ (Fig.~\ref{fig:2}L), we expect these structures to emerge for $+0\delta$, $+6\delta$, $+11\delta$, and $+16\delta$, respectively. However, as each bond has a distribution of binding angles, other structures could emerge as well. 
 
By varying the angle-domain length from 0 to 15 base pairs, we observe the formation of four distinct structures, which we characterize using gel electrophoresis and TEM micrographs (see SI section~III). Platonic solids, such as icosahedra, octahedra, and tetrahedra, consistently dominate across all angle-domain lengths, with yields of 30\% at $+4\delta$, 35\% at $+12\delta$, and 26\% at $+14\delta$ (Fig.~\ref{fig:3}B). Each solid forms over a range of angle-domain lengths, but the distributions are shifted toward shorter lengths than predicted by their designed $n\delta$ values. We hypothesize that this shift arises from the finite width of the dimer angle distribution, which allows closed shells to form at sizes larger or smaller than their target angles. Furthermore, the higher stability of closed structures compared to open ones may favor their formation. Smaller closed structures are also likely kinetically favored, as they require fewer subunits and form more readily. For the longest angle domain ($+15\delta$), monomers aggregate into large, irregular structures and rarely form stable, self-closed assemblies (Fig.~S21).

The finite bending rigidity of inter-particle bonds allows the system to assemble non-Platonic polymorphs, whose binding angles vary throughout the structure (Fig.~\ref{fig:3}C and Fig.~S7). We observe the formation of at least eight distinct non-Platonic shells, though these structures appear at lower yields compared to the Platonic forms. This behavior highlights the inherent flexibility of the system, which can accommodate a range of curvatures and structural arrangements. To address the observed polymorphism and further enhance the yield of targeted structures, we next incorporate both type specificity and geometric specificity into our design strategy.

The ability to program type-specific  interactions and inter-particle binding angles enables the targeted assembly of asymmetric, non-Platonic shells with increased yield. We select four such shells---triangular bipyramid, pentagonal bipyramid, gyroelongated square bipyramid, and hexagonal antiprism (Fig. \ref{fig:3}D)---classified into two groups. The first two are formed using a single triangle species with differing binding angles, while the latter two use two triangle species with four unique interactions and varying binding angles.  Details on the symmetry-guided inverse design and the required specific interaction are in the supplementary information (SI sections VIII and IX).

Optimizing type and geometric specificity enhances yields and minimizes polymorphism in non-Platonic shells. The triangular and pentagonal bipyramids achieve yields of 54\% and 36\%, up from 12\% and 6\%, respectively (Fig.~\ref{fig:3}E). Off-target octahedra account for 3\% during pentagonal bipyramid assembly. Similarly, the gyroelongated square bipyramids and hexagonal antiprisms reach 39\% and 30\% yields, compared to 8\% and 7\% previously. Further improvements may be possible by optimizing annealing rates, assembly temperature, and ionic strength.

\subsection{Programmable self-assembly of complex, self-limiting structures}

\begin{figure}[th!]
 \centering
 \includegraphics[width=0.95\linewidth]{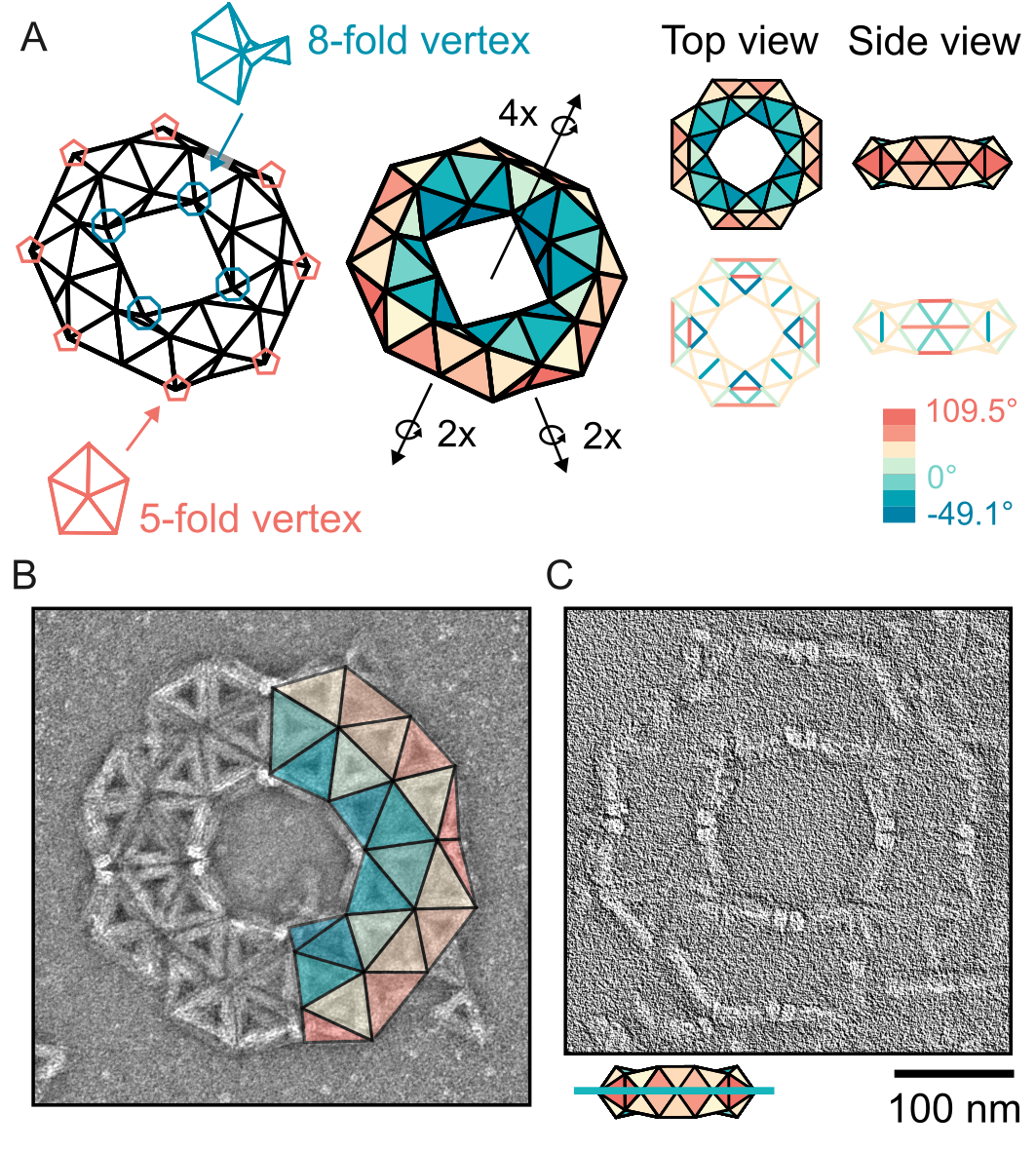}
 \caption{\textbf{Self-assembly of toroids using specific interaction and binding angle.}
(A) The design of toroids with 12 unique species of triangles and 7 unique angles. The geometry of the toroid with the locations of all disclination defects (left), unique rotational symmetries of the toroid (middle), and the species and binding angles between triangles (right). (B) TEM micrograph of the toroid. False color is overlaid to guide the eye. (C) Tomography reconstruction of a toroid scanning through the middle of the toroid.}
 \label{fig:4}
\end{figure}

Finally, we apply our design approach to assemble a toroid---a complex topology requiring both positive and negative binding angles and a correspondingly complex mixture of unique block species to encode those local angles. 
To this end, we choose to build a hollow toroid that contains both positive and negative Gaussian curvature~\cite{Lee2014Feb}. While this topology does not require defects (i.e. non 6-fold coordinated vertices), in their absence the variable Gaussian curvature of the toroid would require variable side-length triangles~\cite{bowick2009two}.  Instead, we show that self-closing toroids can be assembled from equilateral triangles by way of a specific high-symmetry pattern of disclinations: eight 5-fold vertices on the positively curved exterior and four 8-fold vertices within the negatively-curved inner region. As a result, the toroid consists of 96 equilateral triangles arranged in 422 symmetry (Fig.~\ref{fig:4}A). Using this symmetry, the number of unique species of triangles required to assemble the toroid can be reduced to 12, containing 36 unique interactions and 7 unique binding angles ranging from 109.5 to -49.1$^\circ$. The details of this design are described in SI section~VIII.



Next, we demonstrate the assembly of complete toroids under TEM, thereby validating our design approach for assembling complex 2D manifolds with spatially varying Gaussian curvature. Since multispecies assemblies generally have slower kinetics \cite{Wintersinger2023Mar, videbaek2023economical, Hayakawa2024Mar}, 
we increase the concentration of monomers by about four times and reduce the annealing rate by about half compared to assemblies previously demonstrated (SI section~I). In TEM micrographs, we observe fully assembled toroids (Fig.~\ref{fig:4}B). Figure~\ref{fig:4}C shows the equatorial cross-section from a TEM tomography reconstruction, which clearly indicates the hollow interior and four-fold symmetry of the toroid. Notably, we find that the number of fully-formed toroids is relatively small compared to previous non-Platonic shell assemblies (Fig.~S16). We speculate that two mechanisms could have reduced the yield: kinetic deadlock \cite{Deeds2012Feb} and competing polymorphic structures \cite{videbaek2023economical}. These two failure modes are often encountered in multispecies assembly and may be combated through adopting different strategies in the future, including seeded growth or hierarchical assembly \cite{Deeds2012Feb, Whitelam2015Apr, wei2024hierarchical, Tikhomirov2017Dec, Wintersinger2023Mar}. See SI section~VII for detailed results for toroid self-assembly.

The financial cost of ordering DNA staples increases with the complexity of interactions. For instance, assembling toroids with our modular approach requires 520 unique staples, costing approximately \$8,000. In comparison, ordering 12 unique triangle designs raises the cost substantially to around \$36,000. Furthermore, a key advantage of our approach is that the DNA strands from one structure can be repurposed to assemble a diverse range of shapes without incurring any additional cost.  This flexibility significantly enhances the cost-efficiency of our method, as an initial investment in a library of interaction motifs and binding angles grants access to a broad spectrum of possible nanostructures. By preserving a single core structure adaptable to multiple geometries, our modular DNA origami design strategy not only reduces costs and design effort but also expands the accessible structural landscape, streamlining the manufacturing process for a wide range of applications.\\


\section*{Conclusions}


We presented a cost-effective modular design strategy for assembling DNA origami into complex 2D manifolds with independently programmable interaction specificity, strength, and binding angles. We establish robust design rules through a combination of cryo-EM measurements and oxDNA simulations to determine the structures and binding-angle distributions, and gel electrophoresis to assess the assembly yields. This approach enables the efficient assembly of intricate self-closing structures, including non-Platonic shells and a hollow toroid, a rare and complex object with nonuniform curvature. Notably, the precise control over interparticle interactions and curvature achieved by this method facilitates the successful construction of such challenging geometries.

The modularity of this strategy extends beyond the triangular subunits described here. By tuning the sequence and length of double-stranded DNA at the interfaces, this approach can be generalized to a wide range of multilayer DNA origami designs. Looking ahead, this universal and adaptable design framework offers a pathway to accurately program the local curvature of geometrically complex architectures, reducing both synthesis costs and experimental time while broadening the possibilities for nanoscale engineering.

\begin{acknowledgments}
This work is supported by the Brandeis University Materials Research Science and Engineering Center, which is funded by the National Science Foundation under award number DMR-2011846. G.A. acknowledges support from the National Science Foundation (Grant no. CMMI-2323969). TEM images are prepared and imaged at the Brandeis Electron Microscopy facility. Computational resources were provided by the Duke Computing Cluster and the ACCESS program supported by the National Science Foundation (Grants nos. ACI-2138259, 2138286, 2138307, 2137603, and 2138296). We thank Berith Isaac and Amanda Tiano for their technical support at the Brandeis Electron Microscopy facility.  We thank Shibani Dalal and Ian Murphy for their help with the scaffold preparation. We acknowledge Madhurima Roy, Simon Liu, Michael Stehnach, Katsu Nishiyama, Quang Tran, and Pragya Arora for the helpful discussions. 
\end{acknowledgments}

\bibliography{main.bib}

\end{document}


\title{Supplemental Information for ``Modular programming of interaction and geometric specificity enables assembly of complex DNA origami nanostructures''}

\author{Rupam Saha\footnote{These authors contributed equally to this work}}
\affiliation{Martin A. Fisher School of Physics, Brandeis University, Waltham, Massachusetts 02453, USA}
\author{Daichi Hayakawa$^{\mathrm{a}}$}
\affiliation{Martin A. Fisher School of Physics, Brandeis University, Waltham, Massachusetts 02453, USA}
\author{Thomas E. Videb\ae k}
\affiliation{Martin A. Fisher School of Physics, Brandeis University, Waltham, Massachusetts 02453, USA}
\author{Mason Price}
\affiliation{Martin A. Fisher School of Physics, Brandeis University, Waltham, Massachusetts 02453, USA}
\author{Wei-Shao Wei}
\affiliation{Martin A. Fisher School of Physics, Brandeis University, Waltham, Massachusetts 02453, USA}
\author{Juanita Pombo}
\affiliation{Thomas Lord Department of Mechanical Engineering and Materials Science, Duke University, Durham, North Carolina 27705, USA}
\author{Daniel Duke}
\affiliation{Thomas Lord Department of Mechanical Engineering and Materials Science, Duke University, Durham, North Carolina 27705, USA}
\author{Gaurav Arya}
\affiliation{Thomas Lord Department of Mechanical Engineering and Materials Science, Duke University, Durham, North Carolina 27705, USA}
\author{Gregory M. Grason}
\affiliation{Department of Polymer Science and Engineering, University of Massachusetts, Amherst, Massachusetts 01003, USA}
\author{W. Benjamin Rogers}
\email{wrogers@brandeis.edu}
\affiliation{Martin A. Fisher School of Physics, Brandeis University, Waltham, Massachusetts 02453, USA}
\author{Seth Fraden}
\email{fraden@brandeis.edu}
\affiliation{Martin A. Fisher School of Physics, Brandeis University, Waltham, Massachusetts 02453, USA}

\maketitle

\section{Methods: Experiment and simulation}\label{sec:methods}

\subsection{Folding DNA origami}\label{subsec:folding}
Each DNA origami particle is folded by mixing 50 nM of p8064 scaffold DNA (Tilibit) and 200 nM each of staple strands with folding buffer and annealed through a temperature ramp starting at 65~$^{\circ}$C for 15 minutes, then 58 to 50~$^{\circ}$C, $-1~^{\circ}$C per hour. Our folding buffer, contains 5 mM Tris Base, 1 mM EDTA, 5 mM NaCl, and 15 mM MgCl$_2$. We use a Tetrad (Bio-Rad) thermocycler for annealing the solutions. 

\subsection{Agarose gel electrophoresis}\label{subsec:electrophoresis}
To assess the outcome of folding, we separate the folding mixture using agarose gel electrophoresis. Gel electrophoresis requires the preparation of the gel and the buffer. The gel is prepared by heating a solution of 1.5\% w/w agarose, 0.5x TBE to boiling in a microwave. The solution is cooled to 60 $^{\circ}$C. At this point, we add MgCl$_2$ solution and SYBR-safe (Invitrogen) to adjust the concentration of the gel to 5.5 mM MgCl$_2$ and 0.5x SYBR-safe. The solution is then quickly cast into an Owl B2 gel cast, and further cooled to room temperature. The buffer solution contains 0.5x TBE and 5.5 mM MgCl$_2$, and is heated to 45~$^{\circ}$C before use. Agarose gel electrophoresis is performed at 90 V for 1.5 hours in room temperature. The gel is then scanned with a Typhoon FLA 9500 laser scanner (GE Healthcare).

\subsection{Gel purification and resuspension}\label{subsec:purification}
After folding, DNA-origami particles are purified to remove all excess staples and misfolded aggregates using gel purification. The folded particles are run through an agarose gel (now at a 1xSYBR-safe concentration for visualization) using a custom gel comb, which can hold around 4 mL of solution per gel. We use a blue fluorescent table to identify the gel band containing the monomers. The monomer band is then extracted using a razor blade, which is further crushed into smaller pieces by passing through a syringe. We place the gel pieces into a Freeze 'N Squeeze spin column (Bio-Rad), freeze it in a -80$^\circ$C freezer for 30 minutes, thaw at room temperature, and then spin the solution down for 5 minutes at 13 krcf. 

Since the concentration of particles obtained after gel purification is typically not high enough for assembly, we concentrate the solution through ultrafiltration \cite{wagenbauer_how_2017}. First, a 0.5 mL Amicon 100kDA ultrafiltration spin column is equilibrated by centrifuging down 0.5 mL of the folding buffer at 5 krcf for 7 minutes. Then, the DNA origami solution is added up to 0.5 mL and centrifuged at 14 krcf for 15 minutes. Finally, we flip the filter upside down into a new Amicon tube and spin down the solution at 1 krcf for 2 minutes. The concentration of the DNA origami particles is measured using a Nanodrop (Thermofisher), assuming that the solution consists only of monomers, where each monomer has 8064 base pairs.

\subsection{Assembly experiment}\label{subsec:assembly}
All assembly experiments are conducted at a DNA-origami particle concentration of 10~nM.  For two species assembly, the total DNA origami concentration is 10~nM, whereas for toroids, the total concentration is 36~nM, and each triangular species is mixed in a stoichiometric ratio of the target structure. By mixing the concentrated DNA-origami solution after purification with buffer solution, we make 50 $\upmu$L of 10 nM DNA origami (e.g. for the deltahedral shell assembly experiment) at 20 to 30~mM MgCl$_2$, depending on the sample. The solution is carefully pipetted into 0.2 ml strip tubes (Thermo Scientific) and annealed from 40~$^\circ$C to 25~$^\circ$C at -0.1~$^\circ$C per 20 minutes, using a Tetrad (Bio-Rad) thermocycler. For assembly of toroids, the annealing rate is set to -0.1~$^\circ$C per 45 minutes.

\subsection{Negative stain TEM}\label{subsec:TEM}
We first prepare a solution of uranyl formate (UFo). ddH$_2$O is boiled to deoxygenate it and then mixed with uranyl formate powder to create a 2\% w/w UFo solution. The solution is covered with aluminum foil to avoid light exposure, then vortexed vigorously for 20 minutes. The solution is filtered using a 0.2 $\upmu$m filter. The solution is divided into 0.2 mL aliquots, which are stored in a -80~$^\circ$C freezer until further use.

Prior to each negative-stain TEM experiment, a 0.2 mL aliquot is taken out from the freezer to thaw at room temperature. We add 4 $\upmu$L of 1 M NaOH and vortex the solution vigorously for 15 seconds. The solution is centrifuged at 4~$^\circ$C and 16 krcf for 8 minutes. We extract 170 $\upmu$L of the supernatant for staining and discard the rest. 

The EM samples are prepared using FCF400-Cu grids (Electron Microscopy Sciences). We glow discharge the grid prior to use at -20 mA for 30 seconds at 0.1 mbar, using a Quorum Emitech K100X glow discharger. We place 4 $\upmu$L of the sample on the grid for 1 minute to allow adsorption of the sample to the grid. During this time 5 $\upmu$L and 18 $\upmu$L droplets of UFo solution are placed on a piece of parafilm. After the adsorption period, the remaining sample solution is blotted on a Whatman filter paper. We then touch the carbon side of the grid to the 5 $\upmu$L drop and blot it away immediately to wash away any buffer solution from the grid. This step is followed by picking up the 18 $\upmu$L UFo drop onto the carbon side of the grid and letting it rest for 30 seconds to deposit the stain. The UFo solution is then blotted to remove excess fluid. Grids are allowed to dry for a minimum of 15 minutes before insertion into the TEM.

We image the grids using an FEI Morgagni TEM operated at 80 kV with a Nanosprint5 CMOS camera (AMT). The microscope is operated at 80 kV and images are acquired between x8,000 to x28,000. The images are high-pass filtered and the contrast is adjusted using ImageJ.

\subsection{Cryo-electron microscopy}\label{subsec:cryo}
Higher concentrations of DNA origami are used for cryo-EM grids than for assembly experiments. To avoid assembly and aggregation of the subunits, we removed single-stranded DNA strands protruding from the faces of the DNA origami. To prepare samples we fold 2 mL of the folding mixture, gel purify it, and concentrate the sample by ultrafiltration, as described above, targeting a concentration of 300 nM of DNA origami. EM samples are prepared on glow-discharged C-flat 1.2/1.3 400 mesh grids (Protochip). Plunge-freezing of grids in liquid ethane is performed with an FEI Vitrobot IV with sample volumes of 3 $\upmu$L, blot times of 16 s, a blot force of -1, and a drain time of 0 s at 20C and 100\% humidity.


Cryo-EM images for the modular block DNA origami were acquired with a Tecnai F30 TEM with a field emission gun electron source operated at 300 kV and Compustage, equipped with an FEI Falcon II direct electron detector at a magnification of x39000. Particle acquisition is performed with SerialEM. The defocus is set to -2 $\upmu$m for all acquisitions with a pixel size of 2.87 Angstrom.

Cryo-EM images for DNA origami dimers were acquired with a Tecnai F20 TEM with a field emission gun electron source operated at 200 kV and Compustage, equipped with a Gatan Oneview CMOS camera at a magnification of x29000. Particle acquisition is performed with SerialEM. The defocus is set between -1.5 and -4 $\upmu$m for all acquisitions with a pixel size of 3.757 Angstrom.

\subsection{Single-particle reconstruction}\label{subsec:recontruction}
Image processing is performed using RELION-4~\cite{kimanius2021RELION4}. Contrast-transfer-function (CTF) estimation is performed using CTFFIND4.1~\cite{ctffind}. After picking single particles we performed a reference-free 2D classification from which the best 2D class averages are selected for processing, estimated by visual inspection. The particles in these 2D class averages are used to calculate an initial 3D model. A single round of 3D classification is used to remove heterogeneous monomers and the remaining particles are used for 3D auto-refinement and post-processing.  Our reconstructions of the dimers use 2838, 3465, and 1936 particles with resolutions of 35.9~\r{A}, 36.7~\r{A}, and 36.7~\r{A} for $+6\delta$, $+10\delta$ and $-7\delta$ respectively, Fig.~\ref{Sfig:cryo-dimer}. Our reconstruction of the monomer uses 2650 particles and has a resolution of 21.3~\r{A}, Fig.~\ref{Sfig:cryo-monomer}. The post-processed maps are deposited in the Electron Microscopy Data Bank with accession codes EMD-48566, EMD-48565, EMD-48567, and EMD-48569.

\subsection{Cryo-EM multibody analysis} \label{subsec:multibody}
Fluctuations of subunits were processed using RELION-4's~\cite{kimanius2021RELION4} multibody refinement~\cite{nakane2018characterisation, videbaek_measuring_2025}. After getting a postprocessed reconstruction of a dimer using single-particle reconstruction, we create masks around the two triangular bodies using the eraser tool in ChimeraX~\cite{goddard2018ucsf}. These were used in the ``3D multi-body" job in RELION 4 to get the set of fluctuations in translation and rotation of the two bodies in the dimer.

\subsection{oxDNA simulation}
\label{subsec:simulation}
TacoxDNA tools and an in-house script were used for generating the topology and configuration files for the DNA-origami equilateral triangle from the caDNAno design~\cite{Suma2019Nov}. Rigid-body dynamics in oxView were carried out to align the DNA bundle subunits into a conformation more representative of the correct global structure~\cite{Bohlin2022Aug, Poppleton2020Jul}. Overhangs matching the angle- and bond-domains were created protruding from a single outside face of the triangle. The structure was then duplicated in oxView to form a dimer~\cite{Bohlin2022Aug, Poppleton2020Jul}. This process of creating overhangs and dimers was performed for all explored angle-domain lengths to generate the initial configurations needed for carrying out subsequent molecular dynamics simulations using the oxDNA2 package~\cite{Snodin2015Jun}.

To prepare equilibrated structures for the production stage of the simulation, mutual traps between paired scaffold and staple bases were applied, and the structures were subjected to 10,000 steps of gradient descent minimization followed by dynamic relaxation. The initial stages of dynamic relaxation involved substituting the DNA backbone potential with linear springs while applying mutual traps. The maximum applied spring force was gradually increased over 1.52~ns to a force value of 57.09~nN/nm. Subsequently, the backbone spring potential was maintained at 57.09~nN/nm while the time step was gradually increased from $\Delta$t= 0.0303~fs to $\Delta$t= 9.09~fs over 321.18~ns while preserving mutual traps. For the final stage of dynamic relaxation, the full finitely extensible nonlinear elastic (FENE) potential was enforced by removing the spring force on the backbone while maintaining mutual traps on the base pairs for 90.9~ns at $\Delta$t = 9.09 fs.

Dynamic relaxation was followed by a production stage in which mutual traps were removed, and the simulation continued for 0.909~$\upmu$s at $\Delta$t = 9.09 fs at a monovalent salt concentration of 1 M. It is important to note that this simulation time does not directly correspond to physical time, due to the implicit treatment of the solvent and the coarsened resolution of the model, which effectively smoothens the energy landscape \cite{shi2017conformational}. By applying a previously derived scaling factor \cite{shi2017conformational} to estimate the physical time of oxDNA simulations with $\alpha \approx 330$, the simulations in this study correspond to approximately 300~$\upmu$s. The John thermostat with diffusion coefficient and Newtonian step settings of 2.5 and 103 was used to maintain a constant temperature of 27~$^\circ$C. Coordinates were stored every 0.909 ns of simulated time into a trajectory file containing  1000 frames. The generated trajectory files were used for subsequent analyses, which were conducted using a combination of oxDNA analysis tools and in-house scripts~\cite{Bohlin2022Aug}.

Additional simulations were performed for +6$\delta$ and +10$\delta$ at 136~K using the same thermostat settings applied for the simulation at 298~K temperature. Three independent dimers for +6$\delta$ and +10$\delta$ were subjected to dynamic relaxation and production at 298~Ktemperature following the protocol previously described. These dimers equilibrated at 298~K were used as the starting configurations for the simulations at 136~K. The dimers were then simulated under production conditions at 136~K for 9.09~$\upmu$s with a time step ($\Delta$t) of 9.09~fs. Coordinates were saved every 9.09~ns into trajectory files for further analysis of the developed binding angles.

The binding angle between the triangular origami subunits was computed using a custom script. This script takes as input the list of indices of the nucleotides on each face of the side from which the overhangs are extended, along with two additional nucleotides belonging to one of the triangle faces, selected to establish a consistent positive vector pointing from helix 3 to helix 1. The index selection was carried out in oxView~\cite{Bohlin2022Aug}. The first two principal components of each face were computed, with their corresponding eigenvectors defining the directions along the width and height of each face, respectively. From these directions, the normal vectors to each face were derived, and their directions were adjusted to ensure a consistent definition of the binding angle. The binding angle was then calculated using the dot product of the normal vectors, with the sign corrected based on the predefined positive vector pointing from helix 3 to helix 1. This process was repeated for every frame in the trajectory file to obtain the distribution of binding angles for each simulated system.

\clearpage


\section{Angle distribution analysis from \lowercase{ox}DNA simulation}
\label{sec:oxDNA}

By analyzing the angle distributions and inspecting the simulated trajectories, we  gain insights into how our DNA-origami particles can be redesigned to more precisely target specific binding angles. While this iterative design process for DNA origami could be led experimentally using cryo-EM reconstructions, in our design, the binding angle is encoded through a handle of isolated double-stranded DNA that is difficult to visualize at molecular resolutions. To overcome this we analyze the dimer binding angle distributions and visualize the representative configurations obtained from molecular dynamics simulations. To achieve the latter, we examine the mean configuration from each simulated trajectory. This approach allows us to explore some of the behaviors reflected in the binding angles seen in Fig.~2L, and sheds light on the factors contributing to the observed binding angle deviations from the geometric model. It is important to note that the mean configurations are derived \textit{via} the Singular Value Decomposition (SVD) of the corresponding trajectory. While these SVD-derived mean configurations represent the most significant conformations, they may differ from the corresponding mean configuration extracted directly from the binding angle distributions. 

\begin{figure}[h]
 \centering
 \includegraphics[width=0.9\textwidth]{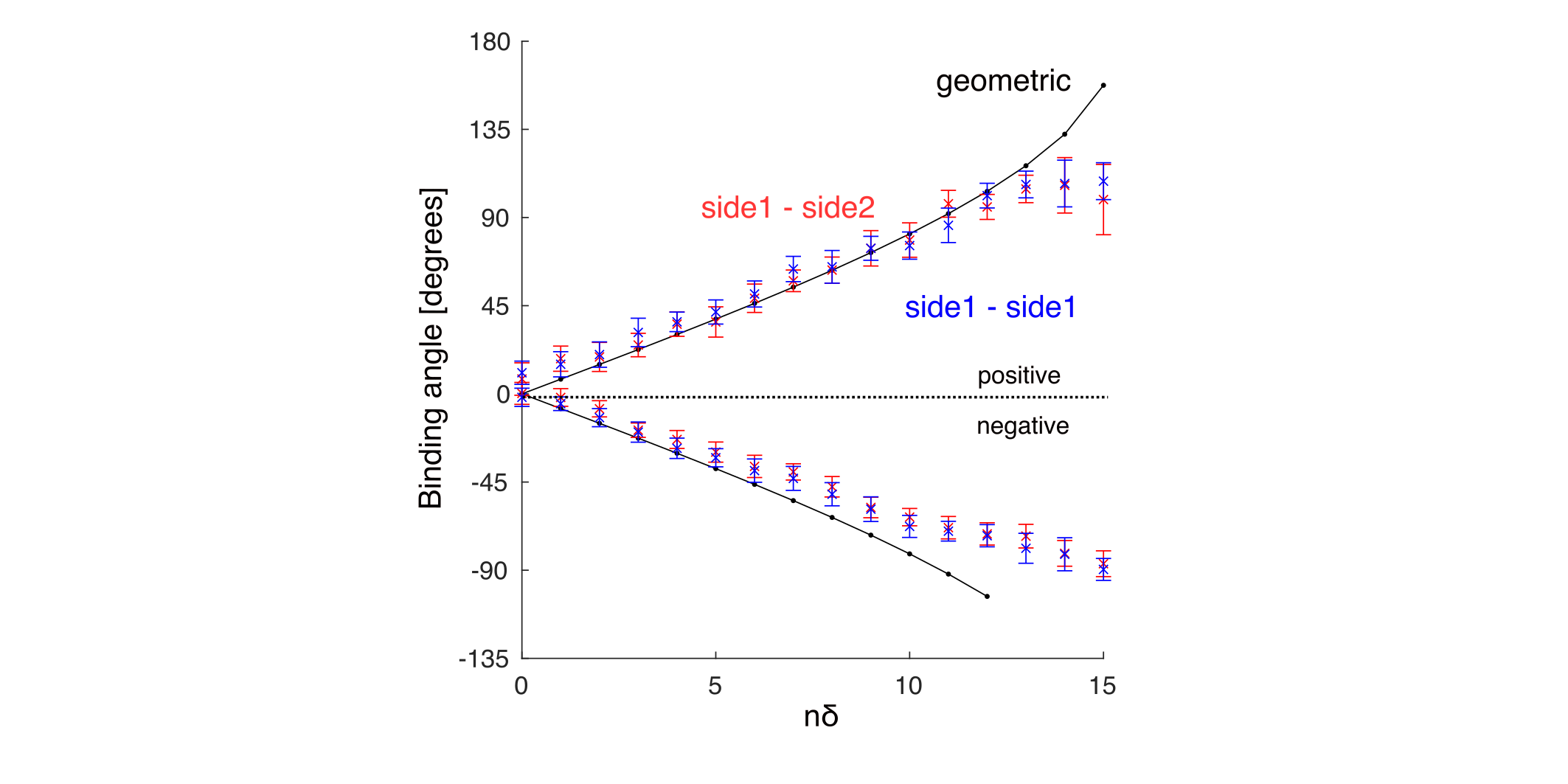}
 \caption{\textbf{Summary of binding angles derived from oxDNA simulation.}
The angles derived for side 1-side 2 binding, shown in red, correspond to the data shown in Fig.~2C. The blue data points indicate binding angles obtained from dimers bound side 1-side 1.
}
\label{sFIG:oxDNA_angle}
\end{figure}

The binding angle analysis of MD-simulated dimers reveals a slight bias towards the positive-angle for small angle-domain lengths compared to the geometric prediction. Specifically, for angle-domain lengths shorter than $5\delta$, we find that both positive and negative binding angle conforming dimers exhibit more positive binding angles than expected from the geometric prediction. This trend is unexpected, as the angle-domain lengths at which this bias occurs are significantly shorter than the persistence length of double-stranded DNA, supporting the assumption that DNA handles behave as rigid rods upon which the geometric model is based. We hypothesized that this bias might be an artifact of the sides of triangles not being precisely perpendicular with respect to the top or bottom surface but slightly tilted. However, the binding angles of dimers bound solely through side 1, as well as those in which side 1 binds to side 2 (Fig.~\ref{sFIG:oxDNA_angle}), follow remarkably similar distributions, suggesting the possibility of an alternative source for this bias.

\begin{figure}[h]
 \centering
 \includegraphics[width=\textwidth]{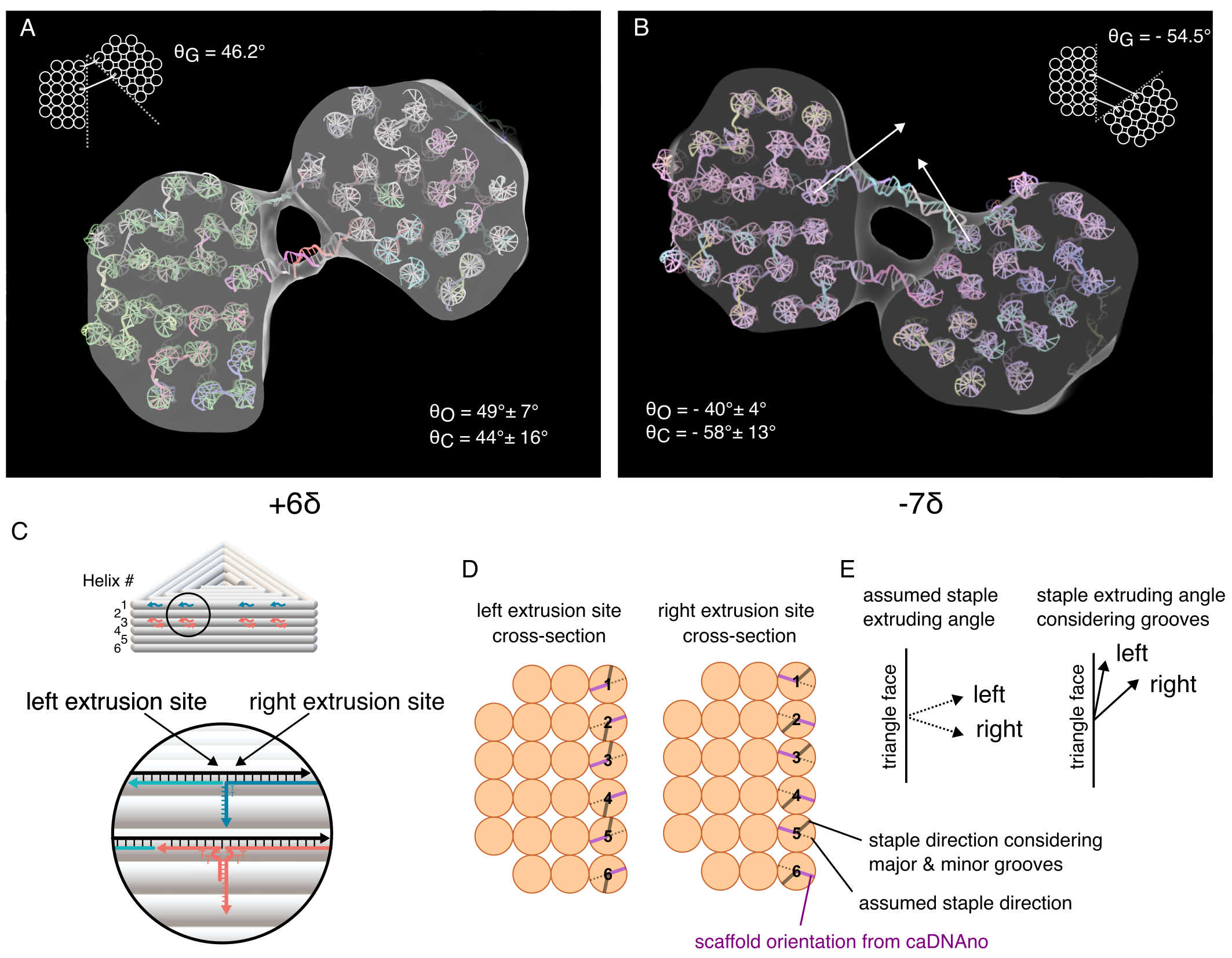}
 \caption{\textbf{Molecular view for short angle-domain lengths.}
Cross-section of two triangles binding with (A) $+6\delta$ and (B) $-7\delta$ configurations, seen in oxDNA, overlaid with cryo-EM reconstructions. Angles from geometric predictions, oxDNA simulations, and cryo-EM reconstructions are shown for each configuration. The white arrows in (B) show the initial direction of extrusion of the overhangs. (C) Schematics of the face of the triangle. The staples extruding from the right extrusion site terminate at 3'-end and have bond domains, while the staple extruding from the left extrusion site terminate at 5'-end and can hybridize to form the angle domain. (D) Cross-section of the side at the extrusion site. Helices 1 through 6 form the double-stranded DNA on the exterior of the triangle. The overhangs extrude from helices 1, 3, and 5. Scaffold positions along the helix for the left and right extrusion sites are labeled in purple, which is taken from the caDNAno design. The staple direction, assuming that staples are 180 degrees out of phase with the scaffold, is given as a dotted line, while the direction considering the major and the minor grooves is provided as a solid line. Note that the helix numbers on the exterior are 0 to 5 instead of 1 to 6 in the actual caDNAno design (Fig.~\ref{Sfig:CaDNAno}), but is shifted to maintain consistency with the main text (Fig.~2A). (E) The summary of extruding staple angles with different assumptions.
}
\label{sFIG:oxDNA_short}
\end{figure}

Observation of the SVD-derived mean configurations reveal that the overhang extrusion angles can bias the binding angle towards a smaller value.
In Fig.~\ref{sFIG:oxDNA_short}A and B, we show the cross-sectional view of the dimer interface for $+6\delta$ and $-7\delta$. Compared to the geometric prediction and cryo-EM reconstruction, the binding angle measurement by oxDNA matches reasonably well for $+6\delta$, but the same deviates substantially for $-7\delta$. 
Interestingly, through inspection of the SVD-derived mean structure for $-7\delta$, we unexpectedly observe that the overhangs on helix 3 extrude upward initially rather than connecting straight to the opposing helix, as indicated by the white arrows in Fig.~\ref{sFIG:oxDNA_short}B. Since the overhangs do not take the shortest possible path, the dimer ends up with smaller absolute opening angles in this case. This behavior of the overhangs is observed not just for this particular case but for all overhangs across all configurations.  This effect is more prominent for larger angle-domain lengths for positive angle configurations and across all negative angle configurations.

The angles of extruding overhangs can be tuned through interactive design combining oxDNA and cryo-EM reconstruction in the future. The origin of the biased extruding angles of the overhangs can be understood by going back to the caDNAno blueprint of the triangles \cite{douglas_rapid_2009}. At the site of staple extrusion (Fig.~\ref{sFIG:oxDNA_short}C, the scaffold orientation in caDNAno shows that helices 1, 3, and 5 have scaffolds oriented perpendicularly inwards (Fig.~\ref{sFIG:oxDNA_short}D). Assuming that staples are 180 degrees out-of-phase with the scaffold, they should extrude almost perpendicularly out from the face of the triangle. This abstraction is usually valid for multilayer DNA-origami designs~\cite{Castro2011Mar}. However, if we consider the major and minor grooves of the double-stranded DNA, the extruding angles are shifted 60 degrees counter-clockwise, justifying the oxDNA simulation results (Fig.~\ref{sFIG:oxDNA_short}E). Moreover, initially we have designed two bases long poly-T spacers to provide additional flexibility such that the binding angle does not become biased due to the extrusion angle of the overhangs. However, clearly we see that, two thymine base are not enough. Going forward, the binding conformation can be fixed in two ways. First, we can insert longer poly-T spacers, incorporating flexibility at the junctions to compensate for the biased extruding angles. Second, the location of the overhangs can be shifted to adjust for the extruding angles. Since the B-form DNA is right-handed and twists about 34.6 degrees per base-pair along the helix \cite{wang1979helical}, shifting the position of the overhangs along the helix leads to rotated overhangs. However, we note that these considerations are derived from analyzing oxDNA simulations rather than experimental results. With a subnanometer resolution of the interface, we may be able to characterize and fine-tune the dimer interface through experiments as well~\cite{kube2020revealing}.

\begin{figure*}[h]
 \centering
 \includegraphics[width=\textwidth]{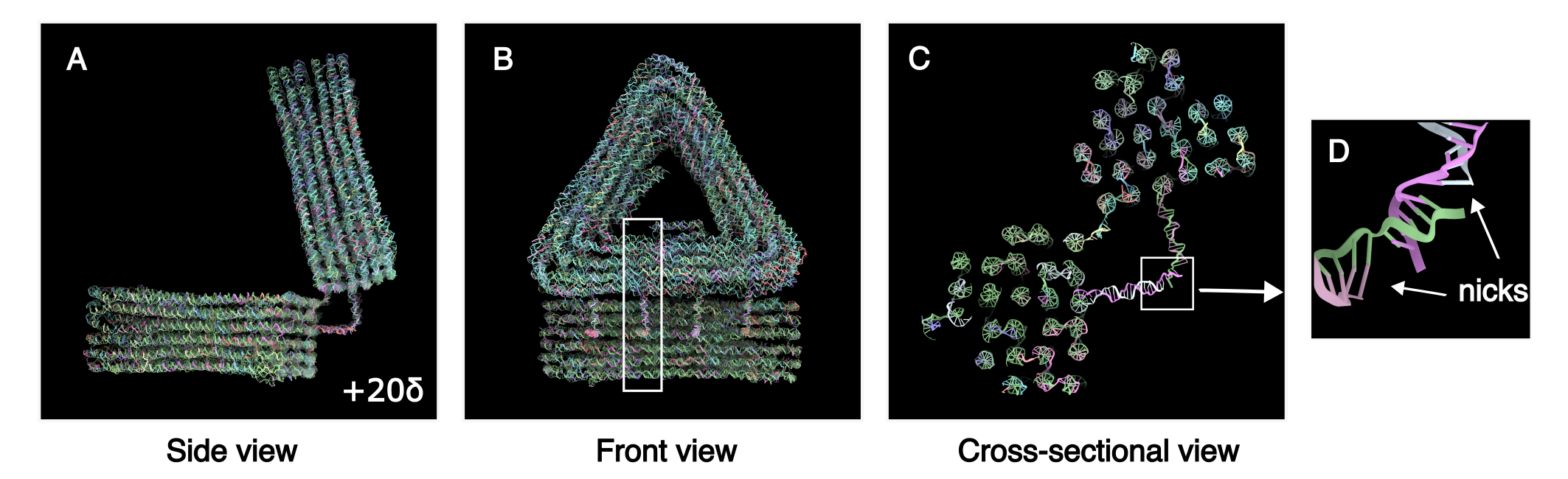}
 \caption{\textbf{Overhang configuration of $+15\delta$ $+20\delta$ dimer, simulated in oxDNA.}
 Side view (A), front view (B), and cross-sectional view (C) of a $+20\delta$ dimer. (D) Zoomed-in-view of the nicks at the buckled overhangs.
}
\label{sFIG:oxDNA_long}
\end{figure*}

We also show that larger deviations from the geometric prediction for large angle-domain lengths are due to the bending of the overhangs. As the angle-domain lengths become longer, we observe more significant deviations from the geometric prediction (Fig.~\ref{sFIG:oxDNA_angle}). In the main text, we indicate that this is due to the finite rigidity of the overhangs. MD simulations reveal that the double-helix overhangs bend for long arm-domain lengths, though they are still much shorter than the double-stranded DNA persistence length~\cite{Bednar1995Dec}. 
In Fig.~\ref{sFIG:oxDNA_long}, we show one example configuration with long angle domains. At $+20\delta$, we observe buckling of the double-stranded DNA in the form of nicks in between the bond and angle domain, contributing to even significant bending (Fig.~\ref{sFIG:oxDNA_long}). Interestingly, the binding angles between the two triangles in this case is 106.5 degrees, which exceeds the 180 degree limit imposed by the geometric model for angle-domain length longer than 15~base pairs. We hypothesize that when the overhangs bridging the helix 3 of two monomers are very long, they become strained as the monomers are held in close proximity by the overhangs of helix 1. Hence, they prefer to adopt a buckled conformation to alleviate the strain.

\clearpage

\section{Analyzing yield from gel intensities}
\label{sec:yield_analysis}

\begin{figure*}[h]
 \centering
 \includegraphics[width=0.8\textwidth]{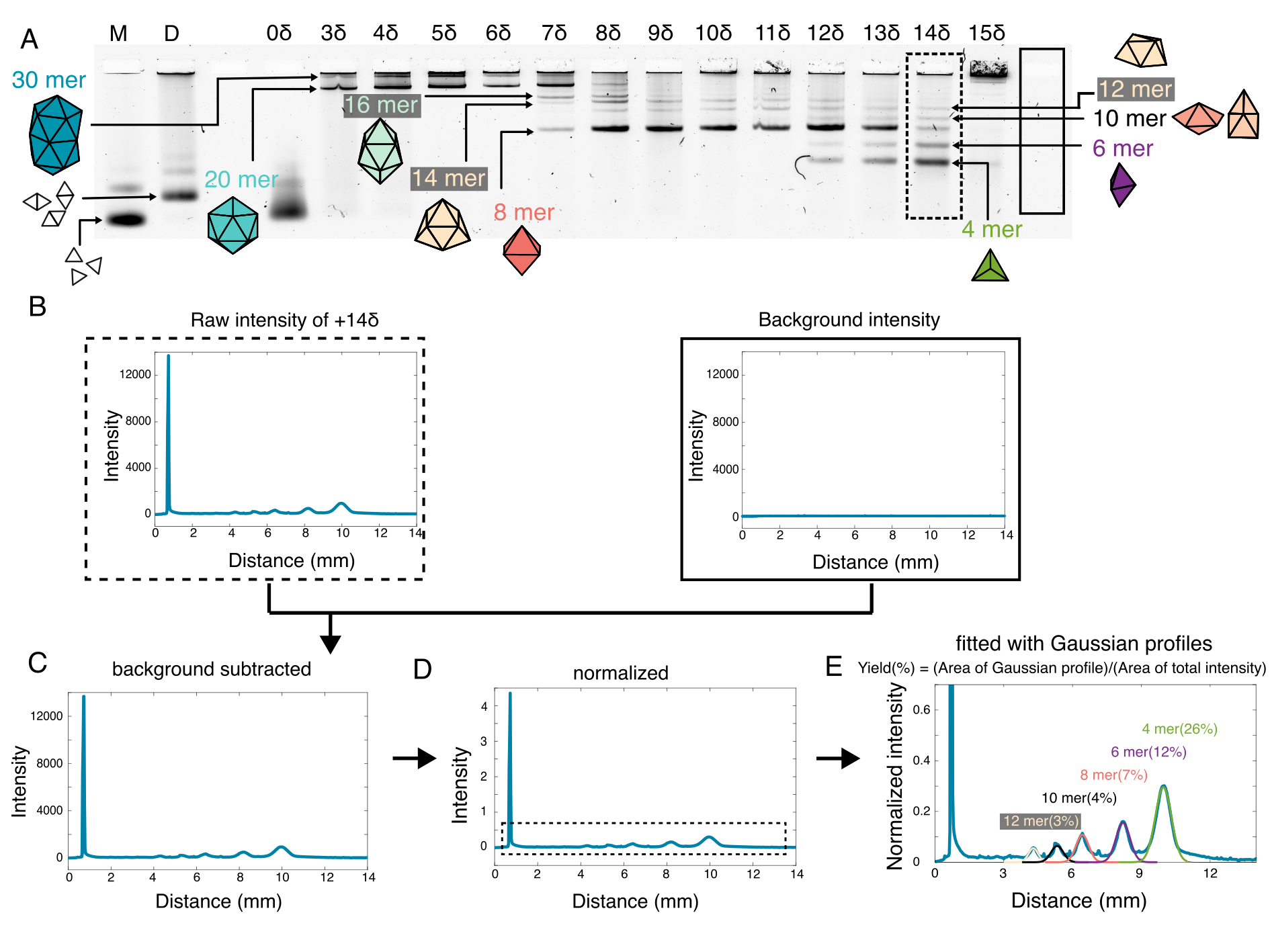}
 \caption{\textbf{Step-by-step analysis of the yield of deltahedral shells from gel electrophoresis. }
(A) Laser-scanned image of 1.5\% agarose gel ran at 20~mM MgCl$_{2}$ concentration. 'M' and 'D' represent control samples of pure monomers and pure dimers. (B) Raw intensity profile of +14$\delta$ (left) and empty lane (right), which is also known as Background. Intensity profile after background subtraction (C) and normalization (D). (E) The significant peaks of the normalized intensity profile are fit with Gaussian profile. 
}
\label{sFIG:yield_analysis}
\end{figure*}

Here, we provide the steps to determine the yield of assembly products using gel electrophoresis results, using deltahedral shell assembly results as an example (Fig.~\ref{sFIG:yield_analysis}A). To measure the yield of shell assembly products, the vertically integrated and horizontally averaged intensity profile is extracted from the laser-scanned gel image. Similarly, the intensity profile of an empty gel lane is extracted which is considered as background (Fig.~\ref{sFIG:yield_analysis}B). The raw intensity values of the background are subtracted from the intensity values of the assembly lanes (Fig.~\ref{sFIG:yield_analysis}C). After that, the intensity profile is normalized to make the area under the curve unity (Fig.~\ref{sFIG:yield_analysis}D). Then, the significant peaks are fitted with Gaussian profiles (Fig.~\ref{sFIG:yield_analysis}E). The area under the Gaussian profiles are calculated as `yield', indicating the fraction of monomers that assemble into each specific capsid. 

\begin{figure*}[th!]
 \centering
 \includegraphics[width=0.8\textwidth]{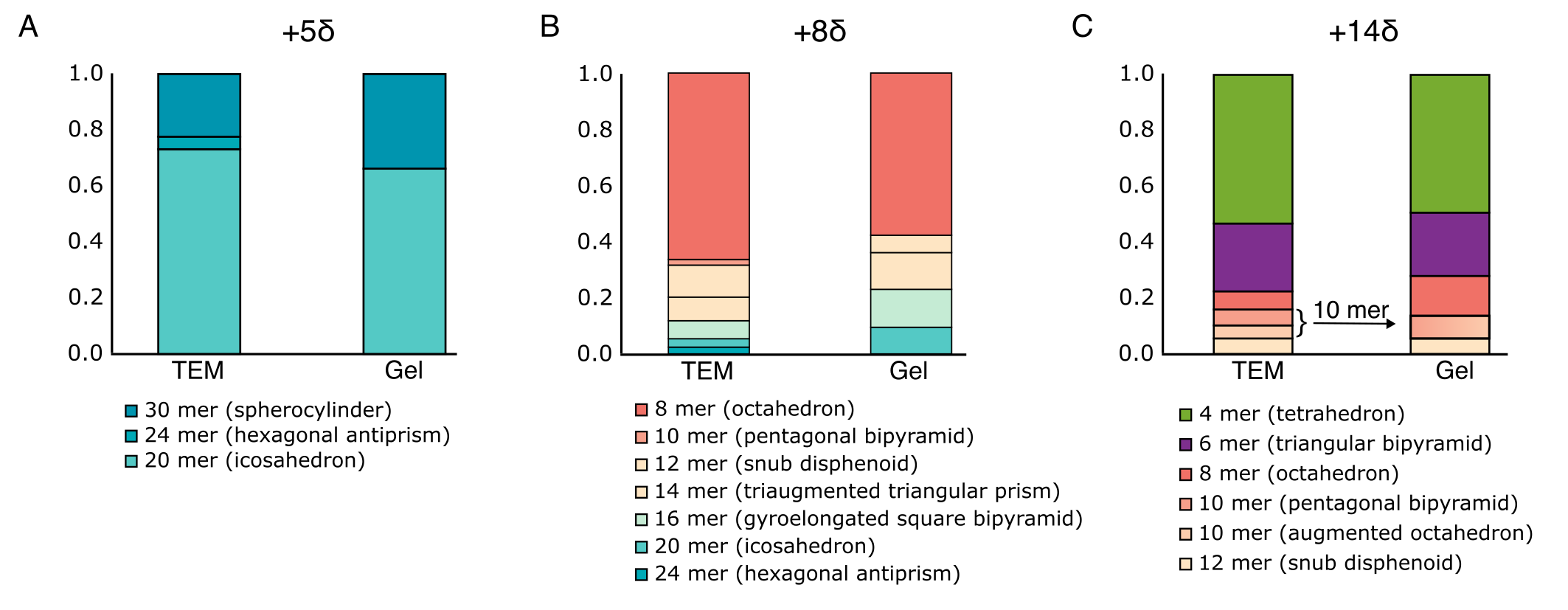}
  \caption{\textbf{Deltahedral shell assembly products are identified by comparing TEM and gel electrophoresis results. } +5$\delta$ (A), +8$\delta$ and +14$\delta$ (C) samples analyzed by particle counting from TEM micrographs (left) and yield analysis from laser-scanned gel image (right).
  }
\label{Sfig:capsid_identify}
\end{figure*}

\begin{figure*}[th!]
 \centering
 \includegraphics[width=0.8\textwidth]{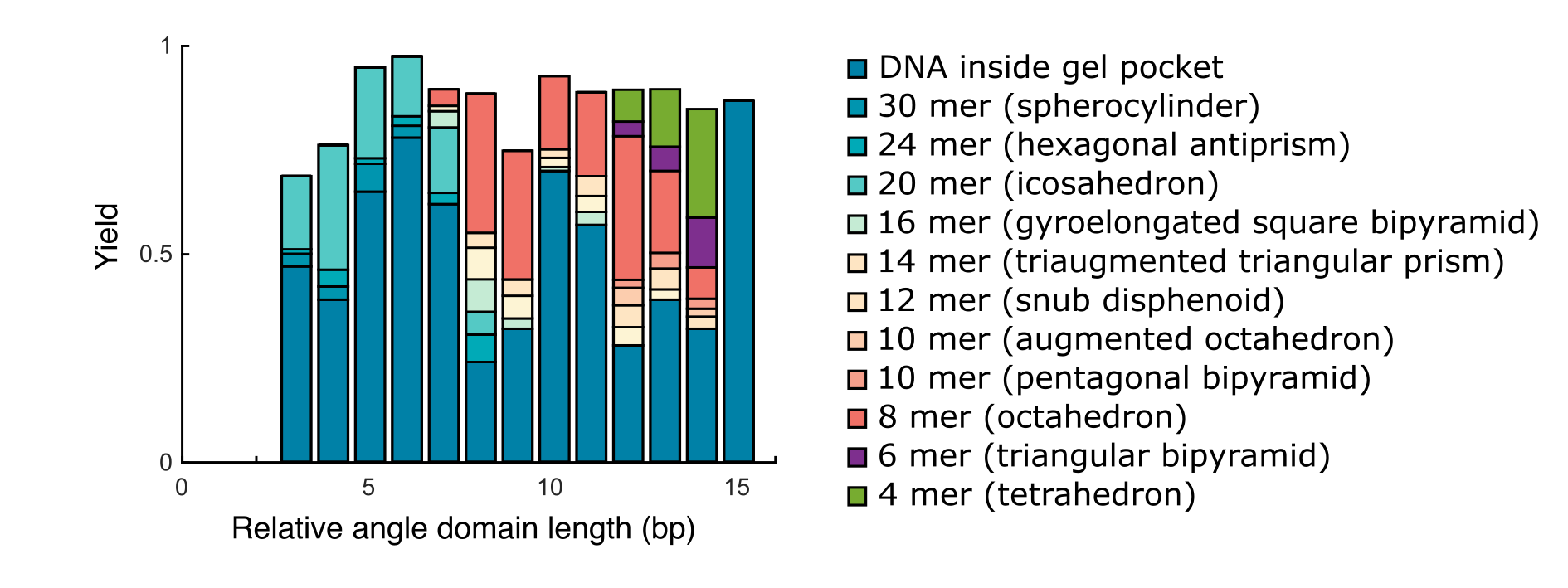}
  \caption{\textbf{Yield of deltahedral shells. } Plot of the yield various deltahedral shells analyzed using the gel-intensity plot and particle-counting from TEM micrographs.
  }
\label{Sfig:capsid_yield}
\end{figure*}

We also characterize the same samples using negative stain TEM, where we identify the deltahedral shells that may represent a significant intensity peak in the laser-scanned gel electrophoresis image. Here, we choose three samples such as +5$\delta$, +8$\delta$ and +14$\delta$ to show the comparison between the TEM result and gel electrophoresis result (Fig.~\ref{Sfig:capsid_identify}A-C). We find some shells such as spherocylinder (30-mer), hexagonal antiprism (24-mer), augmented octahedron (10-mer) and pentagonal bipyramid (10-mer) which are present in considerable quantities on the TEM grid, but analyzing their yield from the gel-intensity plot would be ambiguous or incorrect. For example, the spherocylinder band is found to be very close to the gel pocket that makes the Gaussian fitting ambiguous. We do not observe any significant peak for the hexagonal antiprism. We also observe two types of 10-mers such as the augmented octahedron and the pentagonal bipyramid in the TEM micrographs but they form a single peak in the gel. For these cases, we manually count the particles from the TEM micrographs and find their statistics in the unit of monomers. Finally, comparing with a known particle, such as an icosahedron that manifests as a peak in the gel, we determine the yield of the spherocylinder and the hexagonal antiprism. Similarly, as we know the yield of 10-mer from the gel, particle-counting statistics of 10-mer tell us the fraction of the augmented octahedron and the pentagonal bipyramid in the gel. Combining the gel electrophoresis and TEM, we plot the yield of various deltahedral shells in (Fig.~\ref{Sfig:capsid_yield}).


\begin{figure*}[th!]
 \centering
 \includegraphics[width=0.8\textwidth]{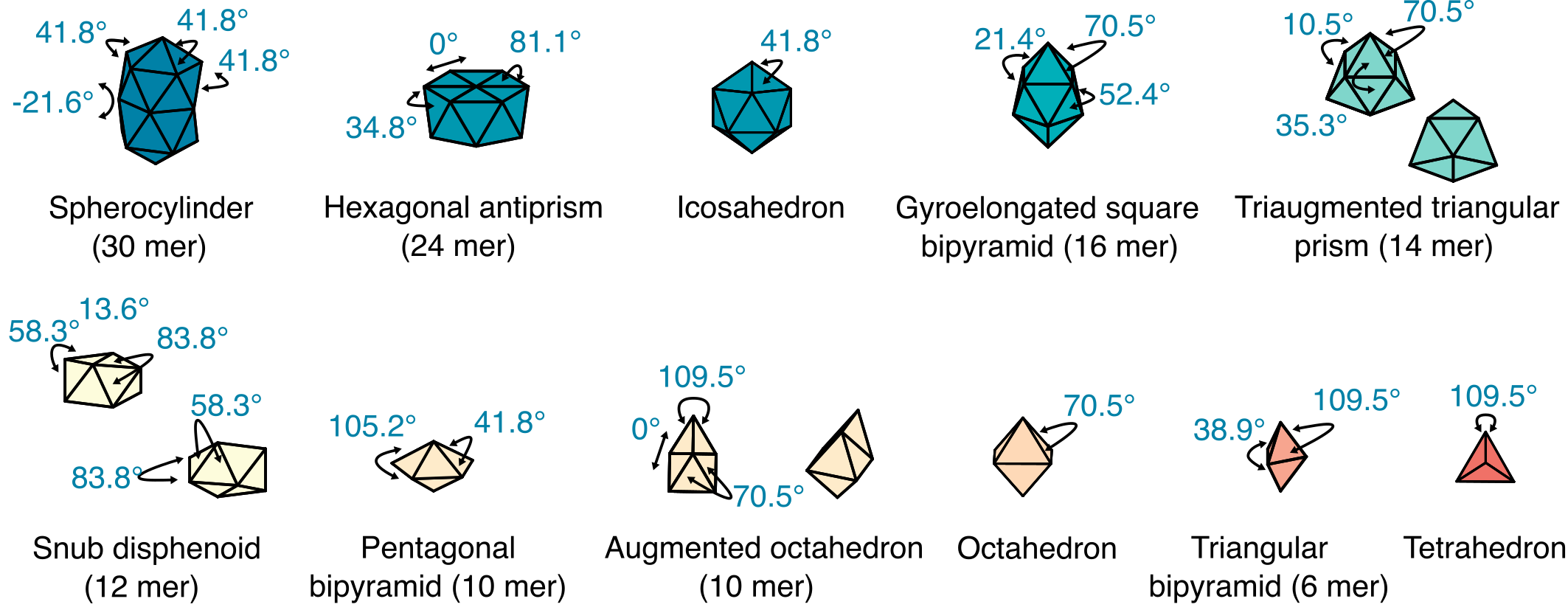}
  \caption{\textbf{Geometries and inter-monomer binding angles for all deltahedral shells.} Angles redundant due to the symmetry of the geometry are often omitted. 
  }
\label{Sfig:capsid_geometry}
\end{figure*}
\clearpage
\section{Bending analysis from cryo-EM data}
\label{sec:bending_analysis_cryo}

To get a more direct measurement of the opening angle and bending fluctuations for different angle-domain lengths, we reconstruct dimers of particles using single-particle cryo-EM. For each angle-domain length, we separately prepare two monomers, A and B, in which side 1 of monomer A binds to side 2 of monomer B \textit{via} complementary interactions. The average reconstructions for the designs with angle-domain lengths of +6$\delta$, +10$\delta$ and -7$\delta$ are shown in Fig.~\ref{Sfig:cryo-dimer}. From these final reconstructions, we can fit the separate triangular bodies and find the opening angle between them. However, to get a sense of the fluctuations in the system, we also perform a multibody analysis~\cite{nakane2018characterisation} on the dimer, assuming that each triangular monomer is a rigid body. For each design, we find that the first principle component analysis (PCA) mode of motion is nearly a pure bending mode, where the sides of the two monomers remain parallel and the binding strands act as a hinge.

\begin{figure*}[th!]
 \centering
 \includegraphics[width=\textwidth]{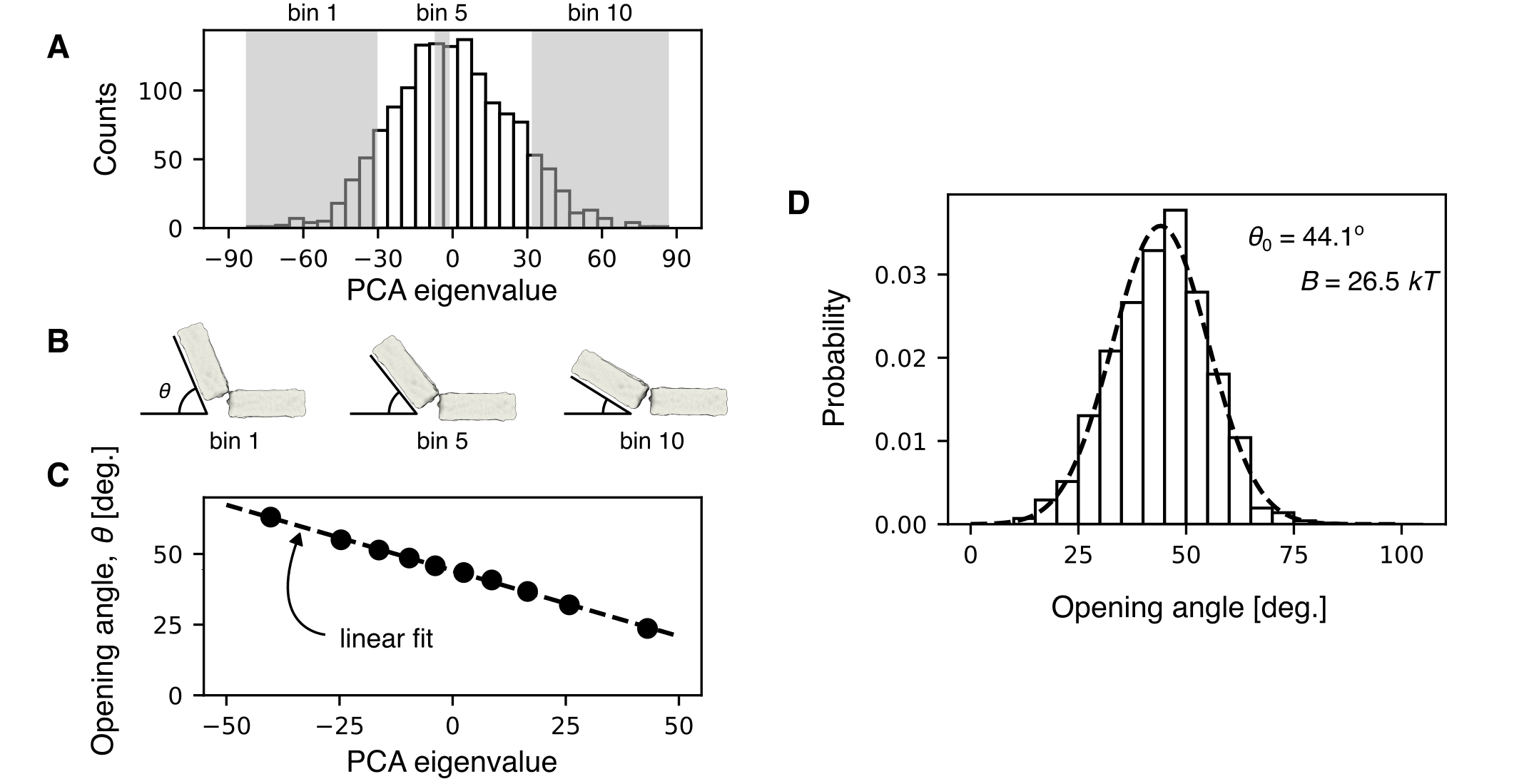}
 \caption{\textbf{Measuring bending from cryo-EM multibody analysis.} (A) Distribution of eigenvalues from a principle component analysis (PCA) mode that is close to a pure bending mode. (B) Views from the side of bins 1, 5, and 10 from the Multibody job in RELION-4. These bins show the average eigenvalue in equi-populated bins from the distribution, the shaded regions in A. From these, we can measure the opening angle, $\theta$. (C) From each bin of the PCA mode, we can measure an opening angle, giving us a fit that goes between the PCA mode's eigenvalues and an opening angle. (D) Using the fit found in C, we get the distribution of opening angles. A Gaussian is fit to this distribution, dotted line, from which we measure the average opening angle, $\theta_0$, and effective bending rigidity, $B$.}
\label{Sfig:bending-cryo}
\end{figure*}

\begin{figure*}[th!]
 \centering
 \includegraphics[width=\textwidth]{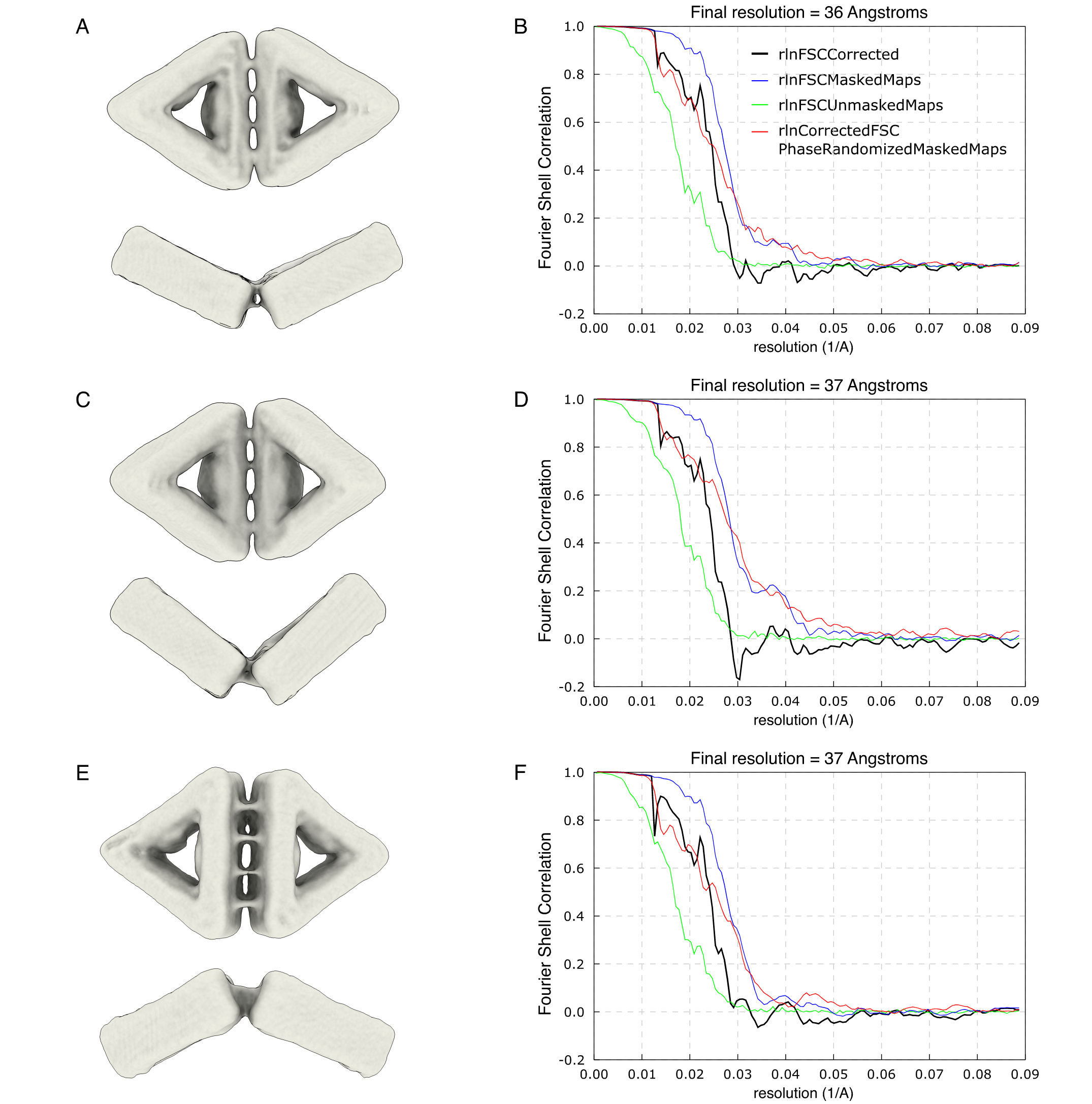}
 \caption{\textbf{Cryo-EM reconstruction of dimers with different binding angles.} (A,C,E) Images from top- and side-views of the dimer reconstructions for angle-domain lengths of +6$\delta$, $+10\delta$ and -7$\delta$ respectively. (B,D,F) Plots of the FSC curves used to estimate the resolution of the dimer reconstructions. 
}
\label{Sfig:cryo-dimer}
\end{figure*}

By looking at the range of motion in the bending mode we can extract both the average opening angle as well as an effective bending rigidity. Figure~\ref{Sfig:bending-cryo}A shows the distribution of eigenvalues for the PCA mode corresponding to bending. To get the opening angle we will need to find a way to go between the eigenvalues of the multibody PCA and a physically meaningful angle. Along with the distribution of eigenvalues, the RELION-4 multibody job also outputs configurations of the bodies for a range of eigenvalues, Fig.~\ref{Sfig:bending-cryo}B. We output 10 configurations corresponding to the average eigenvalues of 10 equi-populated bins of the PCA mode's eigenvalues, the extent of bins 1, 5, and 10 are shown as shaded regions in Fig.~\ref{Sfig:bending-cryo}A. From each of these configurations, we can extract coordinates from the triangular bodies and measure an opening angle for the dimer. Specifically, we measure the angles between the tip of the triangles and the center of the binding edge using ChimeraX \cite{goddard2018ucsf}. Since we know both the measured opening angle and eigenvalue for the PCA mode corresponding to each configuration, we obtain a fit function that converts the PCA mode's eigenvalue to the opening angle of the dimer, Fig.~\ref{Sfig:bending-cryo}C. Using this fit function, we can change our eigenvalue distribution into a distribution of the opening angle, Fig.~\ref{Sfig:bending-cryo}D. The average of this distribution gives us the average opening angle, $\theta_0$. To measure the effective bending rigidity at room temperature, we make the assumption that the distribution is a Boltzmann distribution that is related to the bending energy, $P(\theta)=\exp(-\frac{B}{2kT}(\theta - \theta_0)^2))/Z$, where $B$ is the bending rigidity and $Z$ is the partition function. From this, we see that the inverse of the variance of our opening angle distribution is directly the bending rigidity.

\begin{figure*}[th!]
 \centering
 \includegraphics[width=\textwidth]{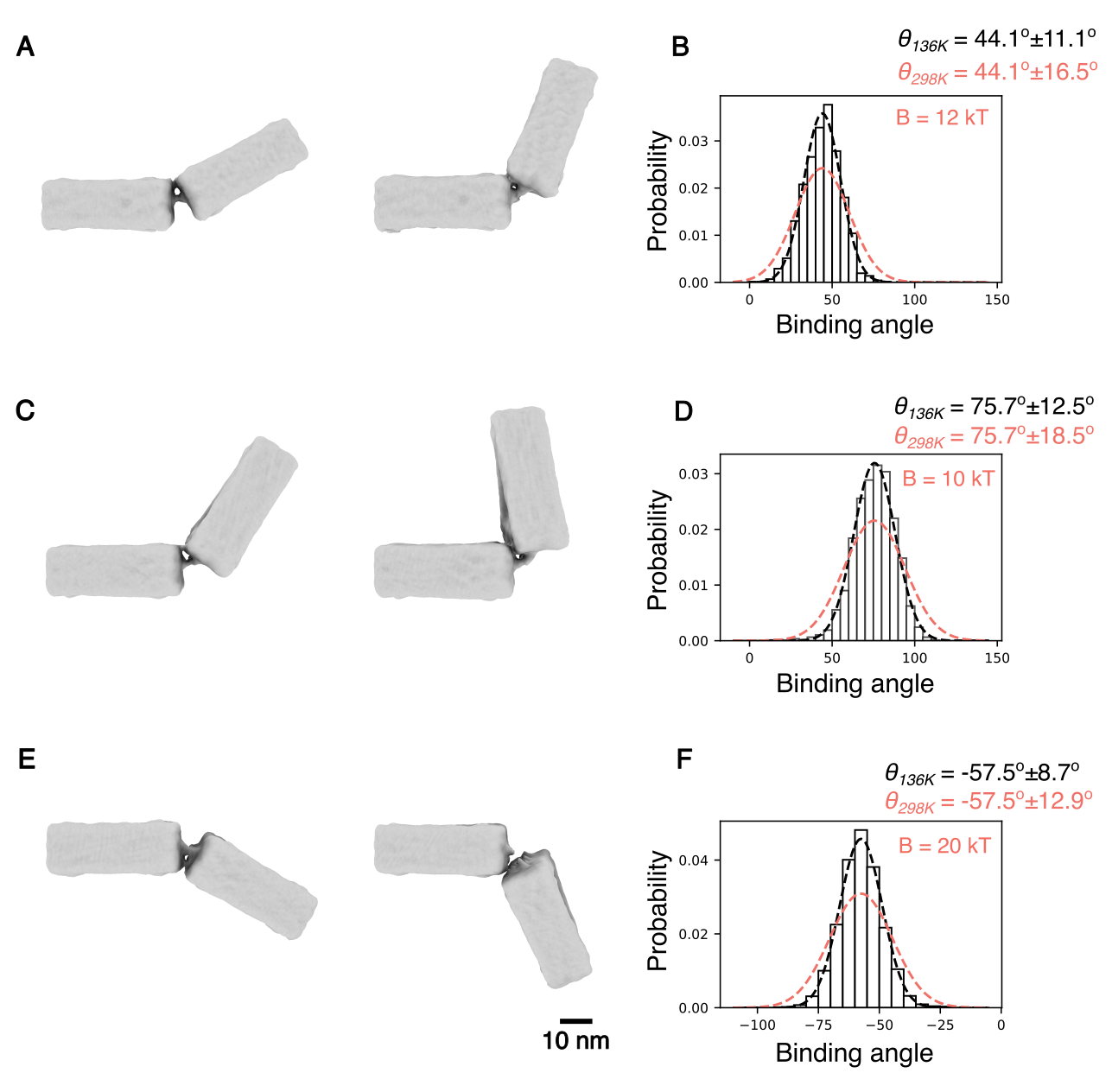}
 \caption{\textbf{Binding angle distribution of dimers, analyzed using cryo-EM multibody refinement.} (A,C,E) Images from side-views of the dimer reconstructions for relative angle-domain lengths of +6$\delta$, $+10\delta$ and -7$\delta$ respectively. Two extreme dimer conformations which are reconstructed from first 10\% and last 10\% of the binding angle distribution. (B,D,F) Plots of the binding angle distribution calculated from multibody analysis. The mean binding angle along with the standard deviation is reported for two temperatures: 136 K and 298 K.
}
\label{Sfig:cryo-dimer-binding-angle}
\end{figure*}

To accurately represent room-temperature behavior, we account for the rapid cooling during cryo-EM vitrification. The sample cools to 136 K in less than 200 microseconds, allowing dimer conformations to equilibrate due to their rapid rotational diffusion (a few nanoseconds)~\cite{dubochet1988cryo, bu2011proteins, bockhighres}.  To best represent the room-temperature ensemble we rescale the Boltzmann factor to 298 K. The bending modulus is calculated from the variance of these distributions, as shown in Fig.~\ref{Sfig:cryo-dimer-binding-angle} for all three dimers.

\clearpage
\section{DISCUSSION OF DISCREPANCY IN WIDTH OF DIMER ANGLE DISTRIBUTION BETWEEN \lowercase{ox}DNA AND CRYO-EM} \label{sec:discrepancy}

The geometric model (Fig.~2B) assumes that the poly-T sequences allow the bonds to rotate freely so that dsDNA bonds connecting a dimer are in a low-energy straightened configuration. In contrast, the 298~K oxDNA simulations predict that the dsDNA bonds are bent (Fig.~2D), which biases the binding angle to a smaller value compared to the geometric model. This effect increases with $n\delta$. The bottom row of Fig.~2D shows the case of a $+15\delta$ dimer, which forms a binding angle of 99.2 degrees, significantly less than predicted by the geometric model. High magnification inspection of the oxDNA configurations 
suggests that 
the two-thymine spacer located where the staples exit the core does not produce sufficient flexibility in the dsDNA connecting the two dimers to attain the designed, straightened configuration.  See SI section~\ref{sec:oxDNA} for a detailed analysis of MD simulations.

\begin{figure*}[h]
 \centering
 \includegraphics[width=\textwidth]{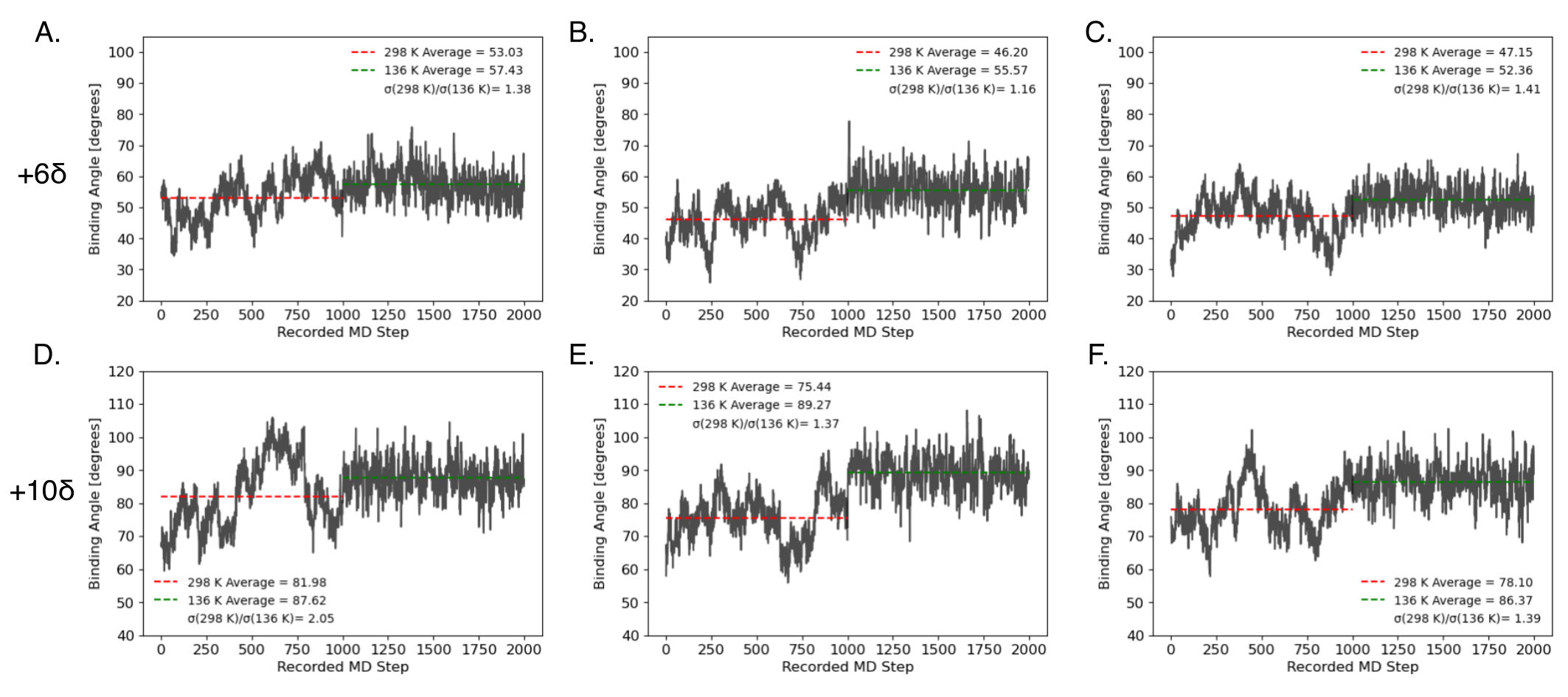}
 \caption{\textbf{OxDNA simulations were equilibrated at 298~K and subsequently quenched to 136~K.}
Panels (A–C) show three repetitions of +6$\delta$, while (D–F) correspond to +10$\delta$. The average binding angles at both temperatures and the ratio of their standard deviations ($\sigma$) are reported.
}
\label{sFIG:oxDNA_quench}
\end{figure*}
\begin{figure*}[h]
 \centering
 \includegraphics[width=\textwidth]{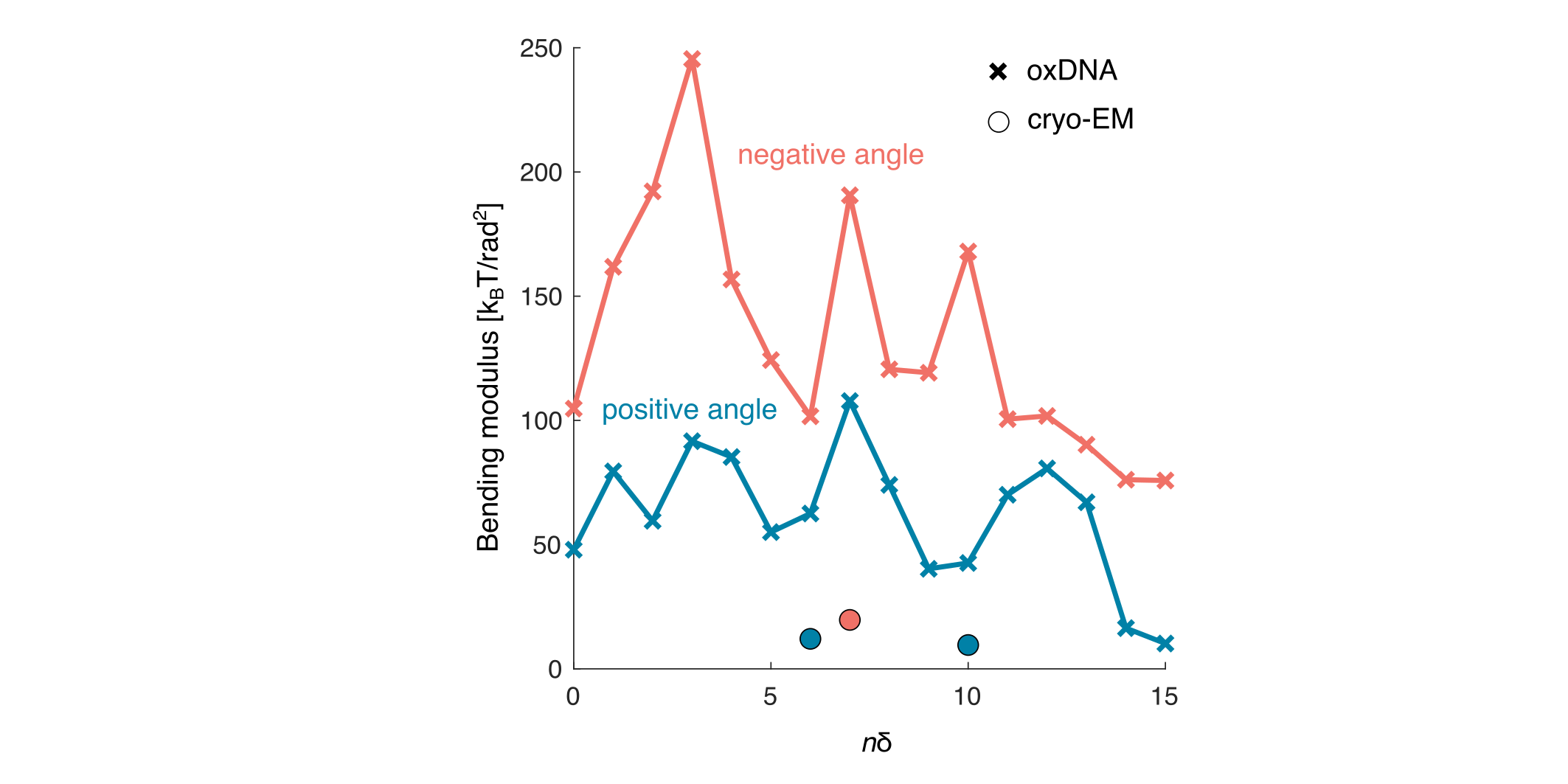}
 \caption{\textbf{Bending modulus of dimers}
Bending modulus vs. $n\delta$ for positive (blue) and negative (red) binding angles, as determined by oxDNA simulation (cross) and cryo-EM (circle) experiments.
}
\label{sFIG:oxDNA_bending_mod}
\end{figure*}

To compare cryo-EM data with oxDNA, we first equilibrate the dimer at 298 K, then quench it to 136 K (Fig. \ref{sFIG:oxDNA_quench}). 
When oxDNA results are compared with cryo-EM data, for both 298~K and 136~K temperatures, we find that oxDNA generates a narrower distribution of the binding angle. The standard deviations in the binding angle distributions obtained from cryo-EM data are more than twice larger than those of the oxDNA simulations (Fig.~2C, D). As a result, the bending modulus calculated from oxDNA simulations is approximately four times greater than that derived from cryo-EM results(Fig.~\ref{sFIG:oxDNA_bending_mod}). We hypothesize two categories of causes for the discrepancy between the angular distributions obtained from cryo-EM and oxDNA. 

The first category focuses on the cryo-EM process.   In cryo-EM, thousands of different dimers are sampled at the same time. There could be static heterogeneity in the cryo-EM samples due to the following two sources. First, the staples have error in the synthesis. The supplier, Integrated DNA Technologies, states 80\% of staples designed to have 60 bases are full length~\cite{Pazdernik_Speicher}.  Thus, some of the dimers in cryo-EM could have errors at the site of the bond and have more flexibility than the average. Second, in experiment the monomers self-assemble and heterogeneity in the assembly process provides the possibility of missing staples, or incompletely assembled staples. In contrast, these two sources of heterogeneity are absent in oxDNA simulations.

A second category focuses on oxDNA. OxDNA being a coarse-grained model, may not capture the nuances of the molecular details necessary to model fluctuations at the level of a single bond.   The discrepancies between the cryo-EM and oxDNA results, therefore, call for a more thorough oxDNA computational study that questions the assumptions of our cryo-EM analysis and the hypotheses discussed here. 


\section{Binding angle estimation of vertex and ring distributions} \label{sec:energetics}

To rationalize the distribution of vertices and rings in experiments, we consider a thermodynamic model accounting for binding and bending energies \cite{wagenbauer_gigadalton-scale_2017}. The predictable nature of the polymorphs in vertex and ring assemblies allows us to compute the free energies of all possible states. For both vertex and ring assemblies, monomers grow to larger clusters through the addition of other monomers, while a large enough cluster can close by binding the two exposed sides. Monomer addition to a cluster with open edges provides some free energy gain from bond energy.  Closure also benefits from additional bond formation, but if the preferred angle of a bond does not match  the closure angle, then there will be an additional cost of bending the bond. Given the monomer concentration, the bond energy, and the preferred binding angle, we compute the yields of allowed polymorphic structures at equilibrium using a thermodynamic model. To match the experiment, we set monomer concentration to 10~nM, bond free energy to $-17 kT$ \cite{Hayakawa2022Oct}, and bending modulus to $25 kT$/rad$^2$ as obtained from the cryo-EM multibody analysis of dimer assembly.

\subsection{Equilibrium estimation of vertex distributions}
By considering the binding energy of the bonds and bending modulus of a bound pair of triangles, we derive a simple energetics model to compute the yields and selectivity of self-closing structures, such as vertices and rings \cite{wagenbauer_gigadalton-scale_2017}. We first predict the yield of the vertices, by computing relative energies associated with all polymorphic structures. 

\begin{figure*}[th!]
 \centering
 \includegraphics[width=0.95\textwidth]{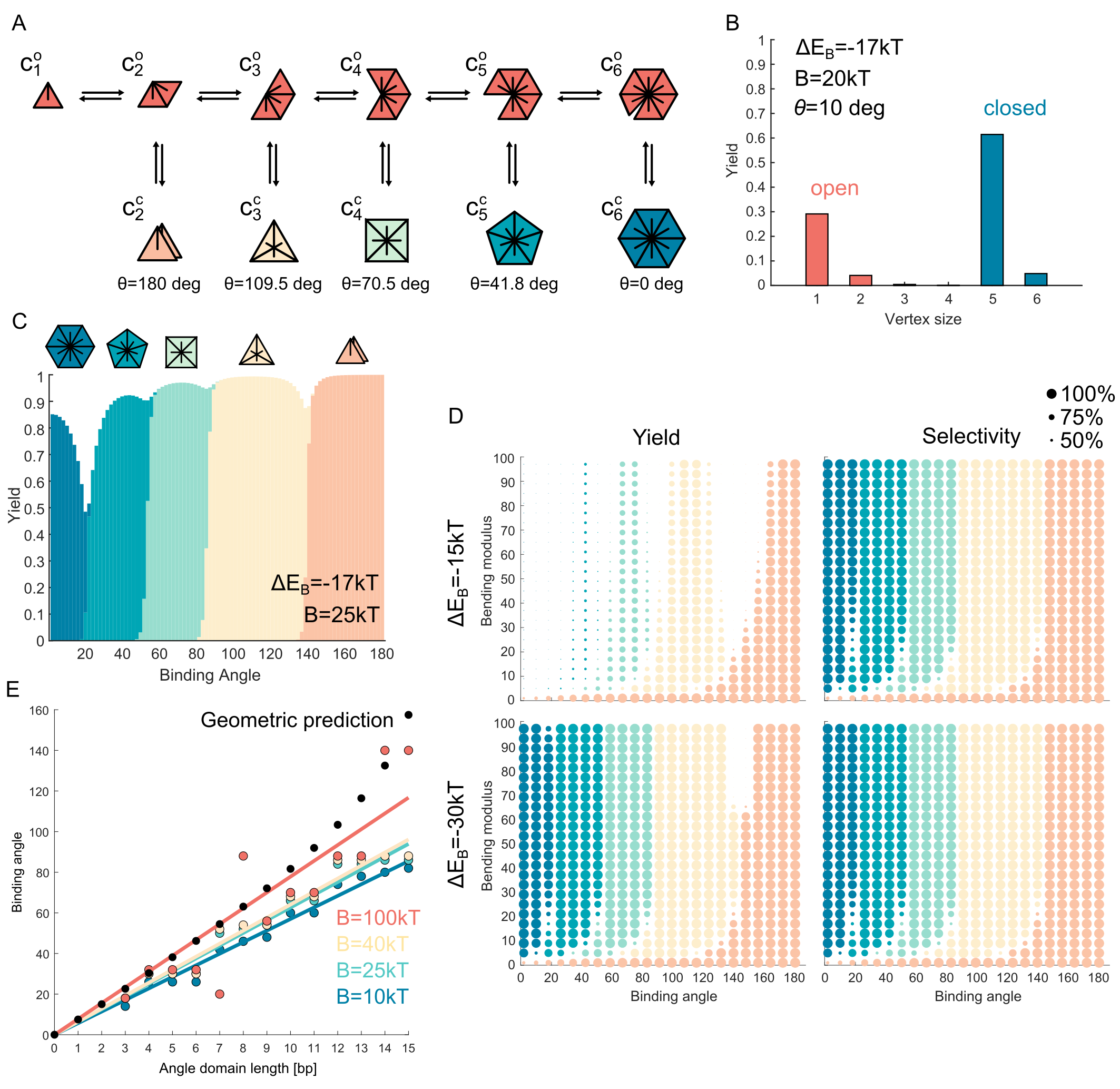}
 \caption{\textbf{Energetics and yield prediction for vertex assemblies.} 
(A) Simple pathways for assembly of closed vertices. Lateral arrows indicate monomer additions and removal, while vertical arrows indicate closure and opening of the vertices. The concentrations of each species are indicated by $c_N^s$, where $N$ is the cluster size and $s$ is the cluster state, which is $c$ for closed and $o$ for open structures. (B) An example calculation of yield of various open and closed states at equilibrium. (C) Yield prediction for closed vertices at $\Delta E_B=-17~kT$ and $B=25~kT$. (D) Prediction of yield and selectivity for various binding angles and bending modulus at $\Delta E_B=-15~kT$ and $\Delta E_B=-30~kT$. The color of the dots indicates the most popular vertex in the system, whereas the size of the circles represents the yield or selectivity of that most popular vertex. To emphasize the differences, the yield and the selectivity are correlated with the square of the disk size. The blank regions in the yield plot represent parameter space where closed vertices are less likely to assemble. (E) The binding angles derived from the distribution of vertices measured in the experiment, assuming different bending moduli. The data is only shown for positive angle configurations. Geometrically predicted angles assuming rigid struts are plotted as a reference. The slopes are 5.70, 6.26, 6.41, and 7.78 for $B=10~kT$, $25~kT$, $40~kT$, and $100~kT$, respectively. All simulations are run assuming the total monomer concentration of 10~nM.
}
\label{Sfig:vertexEnerg}
\end{figure*}

In Fig.~\ref{Sfig:vertexEnerg}A, we show all possible polymorphs we consider in the system. Starting from a monomer, all polymorphs are generated by applying either of the following two moves to the existing clusters: monomer addition or the closure of the vertex. Since a uniformly prescribed binding angle can only encode for positive Gaussian curvature, we disregard clusters above size $N=7$. As a result, we obtain 11 structures, as in Fig.~\ref{Sfig:vertexEnerg}A.

The equilibrium concentrations of the clusters are obtained by considering a simple set of chemical reaction equations along the reaction coordinate shown as arrows in Fig.~\ref{Sfig:vertexEnerg}A. For monomer addition onto an open vertex structure of size $N$, we consider reactions represented as 
\begin{equation}
    c_N^o + c_1^o \rightleftharpoons c_{N+1}^o,
\end{equation}
where superscript $o$ indicates open structures. The equilibrium constant can be associated with the bond free energy, $\Delta E_{B}$, as 
\begin{equation}
    K = \frac{c_{N+1}^o}{c_N^o c_1^o} = \exp\left({-\frac{\Delta E_{B}}{kT}}\right).
\end{equation}
Using this equation, we obtain the concentration of open clusters as
\begin{equation}
    c_{N}^o = c_{N-1}^o c_1^o\exp\left({-\frac{\Delta E_{B}}{kT}}\right)=c_{N-2}^o {c_1^o}^2\exp\left({-\frac{2\Delta E_{B}}{kT}}\right)=\cdots={c_1^o}^N\exp\left({-\frac{(N-1)\Delta E_{B}}{kT}}\right).
\end{equation}
For closing the vertices, we consider an unimolecular reaction
\begin{equation}
    c_N^o \rightleftharpoons c_{N}^c,
\end{equation}
where superscript $c$ indicates closed structures. The equilibrium constant can be associated with the free energy of closure, $\Delta G$, as 
\begin{equation}
    K = \frac{c_{N}^c}{c_N^o} = \exp\left({-\frac{\Delta G}{kT}}\right).
\end{equation}
In addition to forming the bond, closed vertices are penalized since their angles are now constrained. Specifically, we assume that the binding angles of all the sides are 180, 109, 70, 42, and 0 degrees for 2-, 3-, 4-, 5-, and 6-fold vertices, respectively. Any deviation of the target angle $\theta$ from closure angles for vertex size $N$, $\theta_N$, can be considered as a free energy penalty $B(\theta-\theta_N)^2/2$, where $B$ is the bending modulus in units of $kT$. Combined, the free energy associated with the closure of the vertex can be represented as 
\begin{equation}
    \Delta G = \Delta E_{B} + \frac{NB(\theta-\theta_N)^2}{2}.
\end{equation}
Therefore, the concentration of closed vertices with respect to open clusters is obtained as
\begin{equation}
    c_{N}^c = c_N^o\exp\left({-\frac{\Delta E_{B}+ \frac{NB(\theta-\theta_N)^2}{2}}{kT}}\right).
\end{equation}
Finally, since the sum of all monomers must be conserved to $c_\textrm{Tot}$, which we typically set to 10~nM, we solve for monomer concentration $c_1^o$ using the constraint,
\begin{equation}
    c_\textrm{Tot} = c_1^o + \sum_{i=2}^{6}i(c_i^o + c_i^c).
    \label{eq:monoConstraint}
\end{equation}

Given the bond free energy, the bending modulus, and the binding angle, we use Eq.~\ref{eq:monoConstraint} to generate the yield of the clusters and vertices. In Fig.~\ref{Sfig:vertexEnerg}B, we show an example of the calculated yield of clusters and vertices for $\Delta E_B=-17~kT$, $B=20~kT$, and $\theta=20$ degrees. Surprisingly, though the binding angle of the system is almost in the middle between the closure angle for a pentamer and a hexamer, we observe significant bias towards the closed pentamers. In general, the system seems to prefer smaller structures rather than larger structures to avoid the loss of entropic penalty that arises from having to incorporate an additional monomer in the cluster. This trend is more prominent for lower bending modulus, when the geometry is not as distinctly specified. By scanning through all binding angles, we obtain yield for all vertices as in Fig.~\ref{Sfig:vertexEnerg}C. We observe that the larger clusters assemble at lower yields due to the aforementioned entropic loss. Additionally, in between the vertex closure angles, the yield of the closed vertices decreases due to large bending energy penalties.

Finally, we show that the bond free energy does not significantly impact the selectivity of the vertices. In Fig.~\ref{Sfig:vertexEnerg}D, we show the yield and the selectivity of the closed vertices for two significantly different bond free energies, $-15~kT$ and $-30~kT$. The color of the dots for a given binding angle and bending modulus represents the most probable closed structure, whereas the size of the dots represents the probability of monomers in that most frequent structure. For bond free energy of $-15~kT$, we see that the yield of closed structure is lower than that of $-30~kT$ throughout the entire parameter space, as indicated by the smaller circles and blank domains. However, the selectivity of the two systems is very similar. Here, we note the selectivity is defined as the relative yield of a specific closed vertex among other closed vertices, or 
\begin{equation}
    s_N = N\frac{c_N^c}{\sum_{i=2}^{6}ic_i^c}.
\end{equation}
Therefore, as long as we characterize the selectivity of the closed vertices, we expect to obtain quantitatively similar results regardless of the bond free energies.

The selectivity data obtained from gel electrophoresis can be combined with the simulation results to fit for the binding angle for each angle-domain length. The assembly solution characterized in the gel is subjected to an annealing ramp, after which we assume that all structures turn into a closed vertex. A thin gel band, rather than a blurry band, observed for each closed vertex also supports that they are closed structures, rather than open floppy structures or transient assemblies. Therefore, we take the relative band intensity ratios as the selectivity for each angle-domain length solution. By comparing the selectivity from the experiment with that of simulation, obtained using bond free energy of $-17~kT$, we obtain the binding angles for each angle-domain length, given a bending modulus. Specifically, we characterize the root mean squared displacement of selectivity from the experiment, $s_N^\mathrm{Exp}$, with that from the simulation $s_N^\mathrm{Sim} (\Delta E_B,B,\theta)$, 
\begin{equation}
    \sigma(\Delta E_B,B,\theta) = \sqrt{\sum_{i=2}^{6}\left(s_i^\mathrm{Exp} - s_i^\mathrm{Sim} (\Delta E_B,B,\theta)\right)^2},
    \label{eq:RMSDDistribution}
\end{equation}
to obtain the binding angle that minimizes this value. We note that we simulate selectivity data with an interval of two degrees for this comparison. In Fig.~\ref{Sfig:vertexEnerg}E, we show the binding angles obtained by comparing the experimentally obtained selectivity (Fig.~\ref{Sfig:gel}) to that of the computed at different values of bending modulus. Other than the data compared against $B=100~kT$ distribution, the binding angles increase monotonically for increasing angle-domain length. This discontinuous behavior occurs since the selectivity at $B=100~kT$ is highly monodisperse; the computed distribution rarely exhibits coexistence, hindering good comparison when we actually observe coexistence in the experiment. We fit a linear line going through the origin to analyze the impact of changing angle-domain length on the binding angles. We find that the slope of the line increases as we increase the bending modulus.

\subsection{Ring distribution estimation at equilibrium}

\begin{figure*}[th!]
 \centering
 \includegraphics[width=0.95\textwidth]{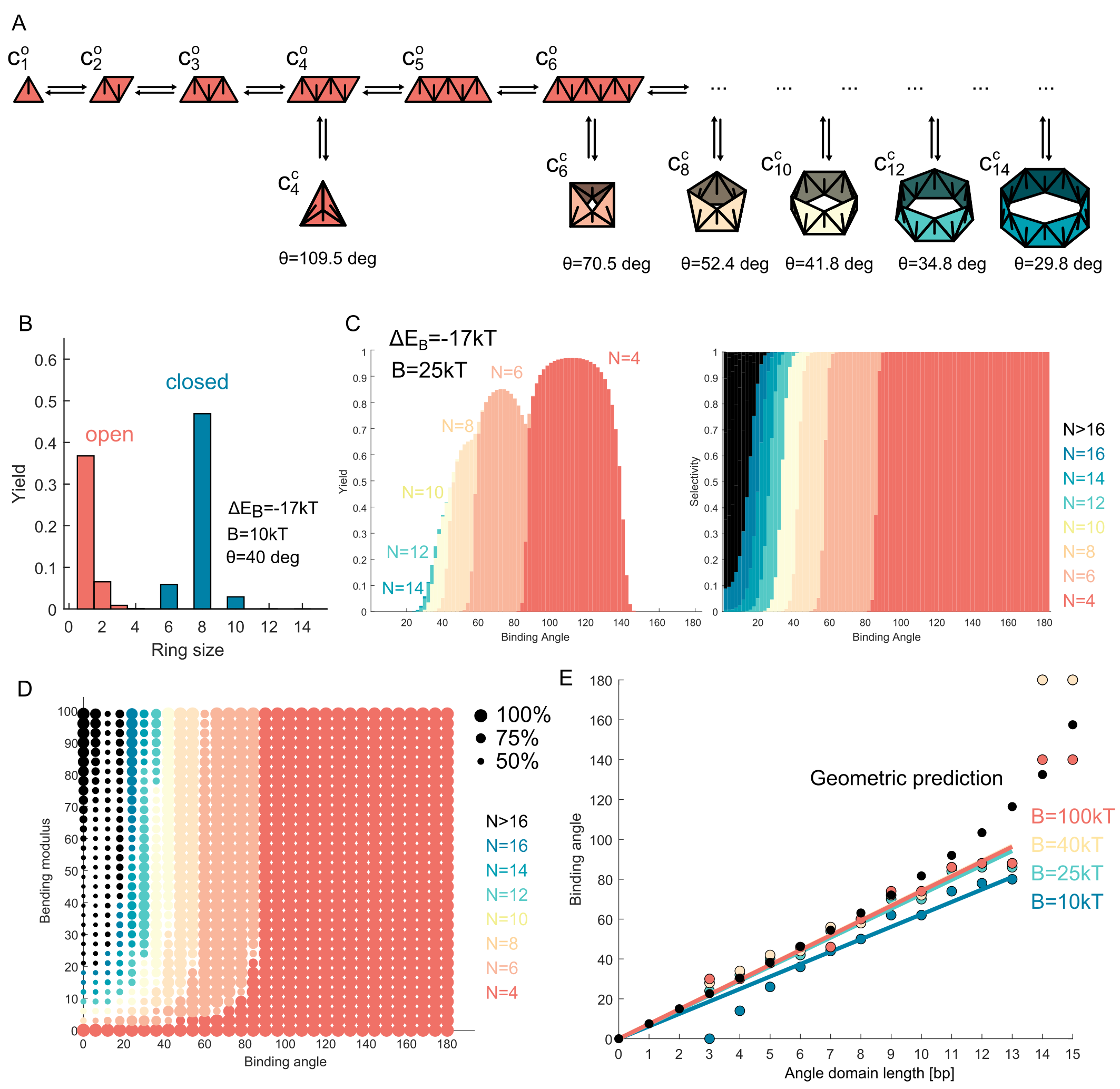}
 \caption{\textbf{Energetics and yield prediction for ring assemblies.} 
(A) Simple pathways for assembly of closed rings. Lateral arrows indicate monomer additions and removal, while vertical arrows indicate closure and opening of the rings. The concentrations of each species are indicated by $c_N^s$, where $N$ is the cluster size and $s$ is the cluster state, which is $c$ for closed and $o$ for open structures. The mean binding angles are also indicated for all closed rings. (B) An example calculation of yield of various open and closed states at equilibrium. (C) Yield and selectivity prediction for closed vertices at $\Delta E_B=-17~kT$ and $B=25~kT$. (D) Prediction of selectivity for various binding angles and bending modulus at $\Delta E_B=-17~kT$. The color of the dots indicates the most popular vertex in the system, whereas the size of the circles represents the yield or selectivity of that most popular vertex. The blank regions in the plot represent parameter space where rings above 16-mers are most probable. (E) The binding angles derived from the distribution of rings measured in the experiment, assuming different bending moduli. Geometrically predicted angles assuming rigid struts are plotted as a reference. The slopes are 6.23, 7.25, 7.45, and 7.41 for $B=10~kT$, $25~kT$, $40~kT$, and $100~kT$, respectively. All simulations are run assuming the total monomer concentration of 10~nM.}
\label{Sfig:ringEnerg}
\end{figure*}

Using a similar model, we compute the equilibrium distribution of the ring assemblies. To tailor the model for ring assemblies, we make two adjustments from the vertex assemblies. First, the rings can only assemble through closure of even-number sized chains, above $N=2$ (Fig.~\ref{Sfig:ringEnerg}A). This is simply due to the interaction encoded on the exposed edges of the chain. Therefore, odd-number sized chains and $N=2$ chain only has reaction pathways to add or remove a monomer. The binding angle for each ring structure can be obtained from the location of the vertices, or from the zigzag tubule geometry described in~\cite{Hayakawa2022Oct}. Also, unlike vertices, rings can close at an arbitrarily large size. This makes the computation intractable especially when the binding is strong and the binding angles are small. To avoid this issue, we set the largest possible assembly size to be $N=30$ in the calculation, which is much larger than the rings that are observed in the experiments. Additionally, we run the simulation with the bond free energy of $-17~kT$, where we still see the coexistence of monomers and closed rings in most of the parameter space. Finally, to reduce the computation error while solving Eq.~\ref{eq:monoConstraint}, we use a gradient descent method to obtain accurate monomer concentrations. 

Ring assembly simulation predicts notable assembly of rings up to size 14-mers, similar to what is observed in gel electrophoresis. For given input pararmeters, the distribution of open and closed structures can be estimated by solving the equilibrium model (Fig.~\ref{Sfig:ringEnerg}B). Especially at smaller binding angles, we observe the coexistence of more than two closed ring structures. By screening a wide range of binding angles, we find that the yield of the closed rings are not high throughout all binding angles, unlike the vertices (Fig.~\ref{Sfig:ringEnerg}C). The yield is low at small binding angle, since the large ring structures has to incorporate many monomers, accumulating an entropic penalty for assembly. On the other hand, there are no closed ring structures that correspond to binding angles above 109.5 degrees. Therefore, the bending energy cost wins over cohesion at large binding angles, contributing a catastrophe of the yield above 140 degrees. We note that the selectivity of the closed rings shows a monotonic change with the binding angle (Fig.~\ref{Sfig:ringEnerg}C and D).

\begin{figure*}[th!]
 \centering
 \includegraphics[width=\textwidth]{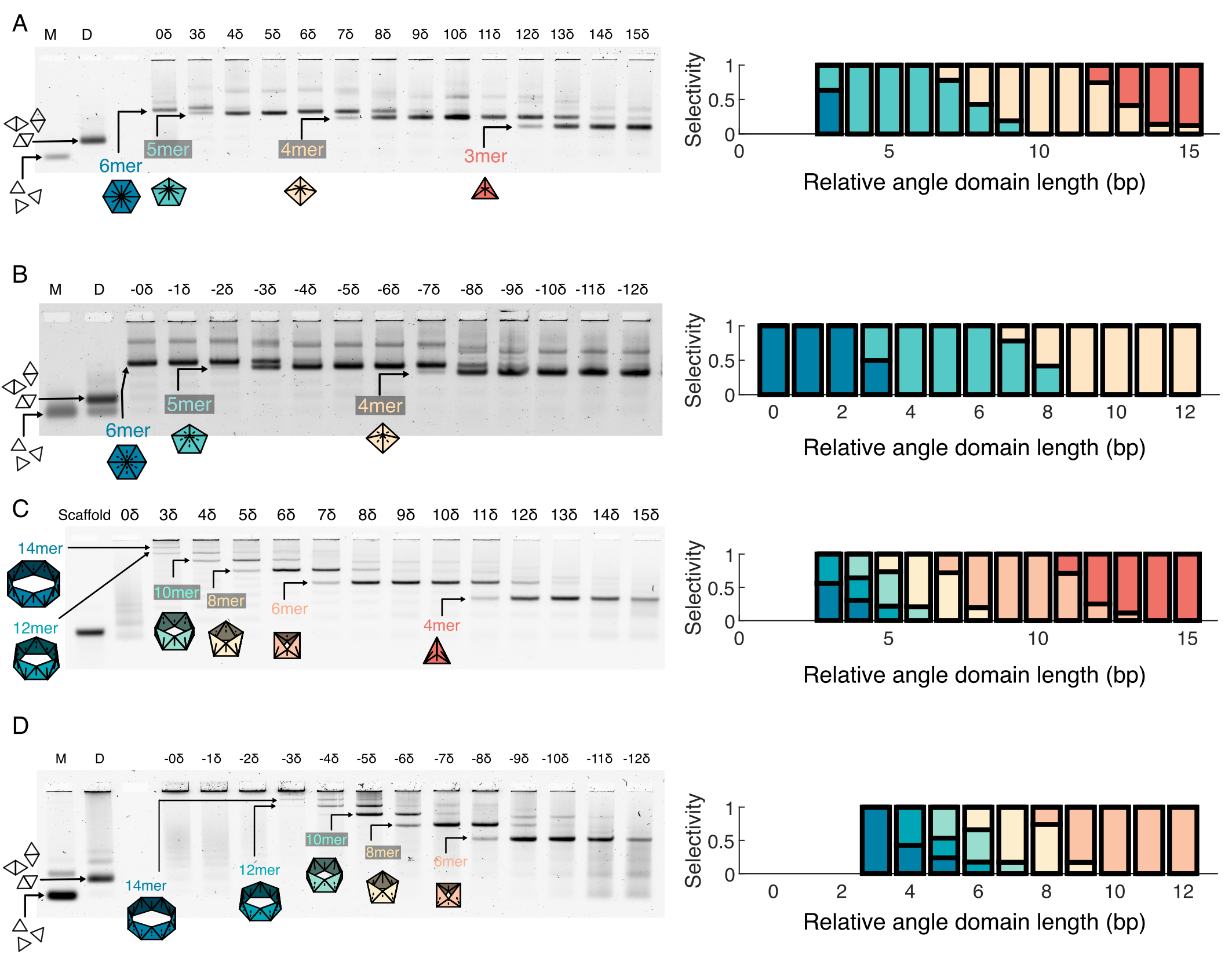}
 \caption{\textbf{Summary of the gel electrophoresis for vertices and rings.}
Summary of gel electrophoresis for (A) positive angle vertex, (B) negative angle vertex, (C) positive angle ring, (D) negative angle ring, and the selectivity obtained from respective band intensities. For all experiments, sides 1 and 2 are activated. The dashed and solid lines on the triangles indicate the triangles seen from the top and the bottom, respectively. For negative angle assemblies, the selectivity is plotted against the relative angle-domain length of strands on helices 3 compared to helices 5. The gels are all 1.5\% agarose, ran at 20~mM MgCl$_{2}$, except for positive vertices which is at 30~mM MgCl$_{2}$. The contrast of the images is enhanced. `M' and `D' represent control samples of pure monomers and pure dimers.
}
\label{Sfig:gel}
\end{figure*}

We compare the selectivity from the simulation to that from gel electrophoresis (Fig.~\ref{Sfig:gel}) to obtain the binding angle for each angle-domain length. By extending Eq.~\ref{eq:RMSDDistribution} for ring structures, we characterize the deviations between simulated ring distributions and experimental distributions. Using distributions simulated at $\Delta E_B=-17~kT$ and bending modulus ranging from $10~kT$ to $100~kT$, we obtain the binding angle that yields the least deviation in the distributions from the experiment (Fig.~\ref{Sfig:ringEnerg}E). For data points from each bending modulus, we fit a linear line going through the origin and obtain the slopes, which are 6.23, 7.25, 7.45, and 7.41 for $B=10~kT$, $25~kT$, $40~kT$, and $100~kT$, respectively. However, to fit this line, we ignore data points from 14 and 15~bp angle-domain lengths, since they are not accurate predictor of the actual angle. As seen from Fig.~4I, these two samples assemble into $N=4$ rings at 100\% selectivity. Unfortunately, the ring assembly system has a very low sensitivity in selectivity above the binding angle of 100 degrees, predicting 100\% selectivity from 100 degrees to 180 degrees (Fig.~\ref{Sfig:ringEnerg}D). In fact, the analysis shows the binding angle of 180 degrees for 14 and 15~bp angle-domain length samples when compared to distributions generated at $B=$10, 20, and $40~kT$ (Fig.~\ref{Sfig:ringEnerg}E). 

\clearpage

\section{Assembly of toroids} \label{sec:toroidExp}

\begin{figure*}[th!]
 \centering
 \includegraphics[width=0.9\textwidth]{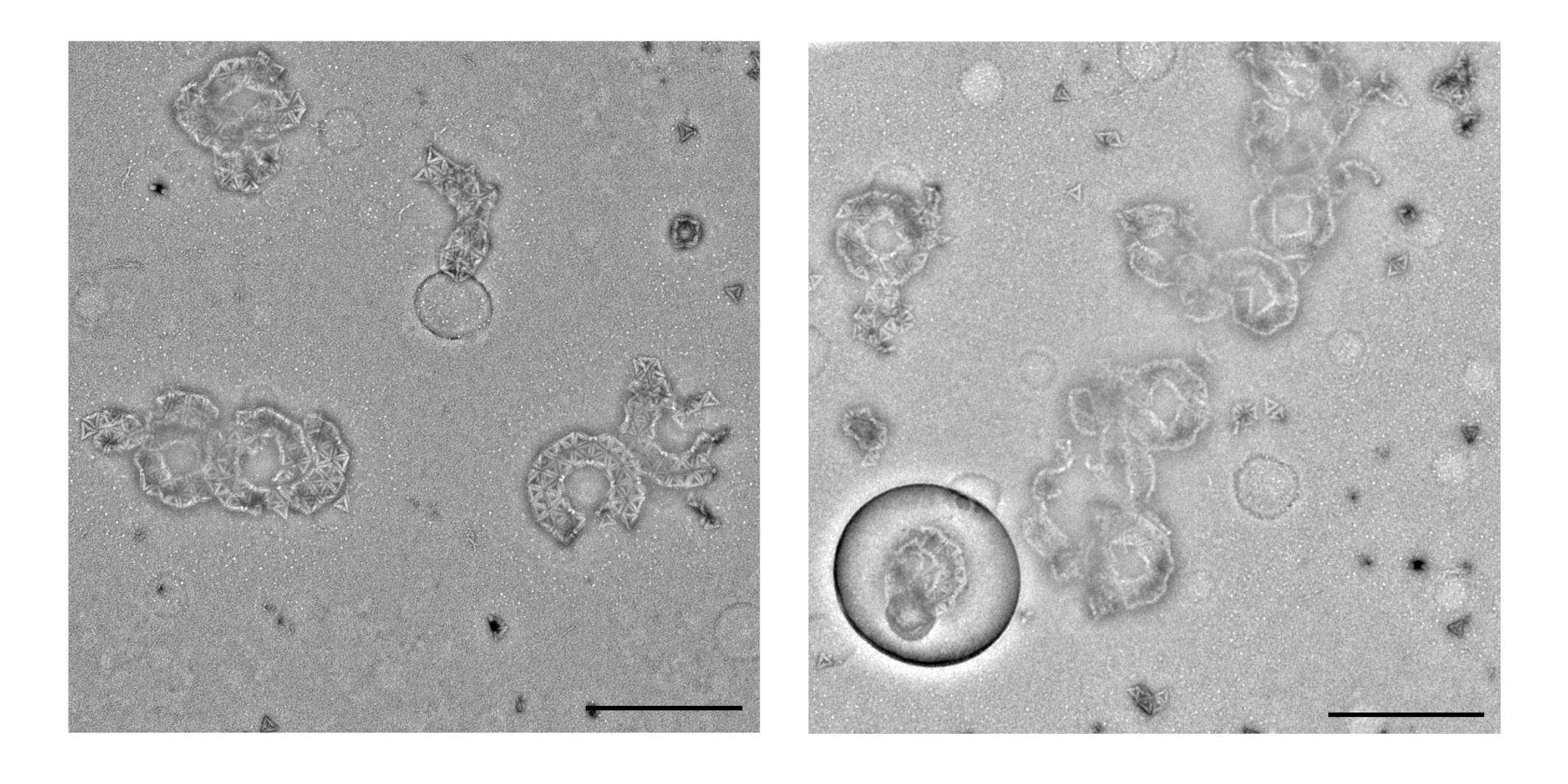}
 \caption{\textbf{Representative, wide-field images of toroids under TEM.} The scale bars are 500~nm.
}
\label{Sfig:wideViewToroids}
\end{figure*}

Although we succeed in assembling toroids, we observe two types of failure modes which give us insight into designing complex structures for the future. We assemble toroids through one-pot annealing of the purified monomers. As a result, we observe the assembly of complete toroids under TEM (Fig.~4B). However, we observe only a small fraction of fully-assembled toroids under TEM (Fig.~\ref{Sfig:wideViewToroids}). Note that quantitative analysis of the yield using gel electrophoresis is difficult due to the size of the structure. We attribute two factors that reduce the yield of the toroids: the kinetic arrest of many partially grown toroids~\cite{Deeds2012Feb} and the closure of 5-fold toroids as byproducts. While a slower ramp rate or hierarchical assembly of subunits may avert the kinetic deadlock, the polydispersity of toroids needs special attention in the structural design. 

\begin{figure*}[th!]
 \centering
 \includegraphics[width=0.9\textwidth]{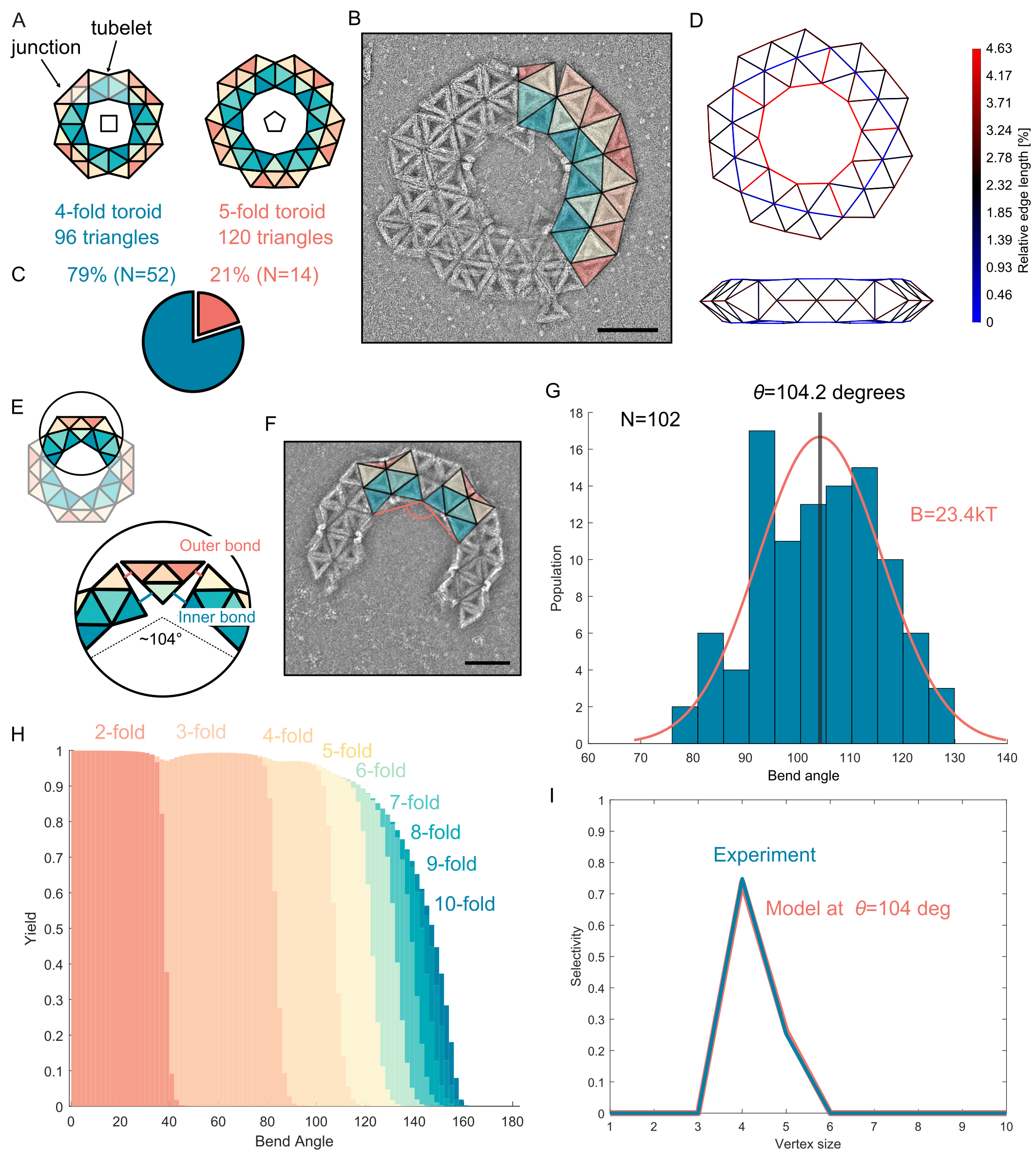}
 \caption{\textbf{Selectivity of the target toroid.} 
 (A) Top view of the 4-fold and 5-fold toroid. (B) Example TEM micrographs for 5-fold toroid. (C) The population of fully-assembled 4-fold and 5-fold toroids observed under TEM. Note that, to obtain the selectivity, the number of triangles in each structure must also be taken into count. (D) The amount of stretching per edge required to form a 5-fold toroid. (E) The inner and outer bonds between the tubelet and the junction have different lengths, which may lead to opening of the bend angle. (F) Bend angle is measured from TEM micrographs. The bend angle is defined as the relative orientation of two neighboring tubelets. (G) Distribution of bend angles measured from open toroids. The distribution is fitted with a Gaussian, which yields a mean of 104 degrees. (H) The theoretical yield of various-sized toroids depending on the programmed bend angle, as predicted from the thermodynamic model. Assembly of toroids larger than 10-fold is omitted from the simulation. The bending modulus used for this prediction is 23$kT$/rad$^2$. (I) The selectivity of closed toroid sizes measured from the TEM micrograph, compared to that from the thermodynamic model using a bend angle of 104 degrees and a bending modulus of 23$kT$/rad$^2$. All scale bars are 100~nm. 
}
\label{Sfig:toroidExp}
\end{figure*}

The toroid with 5-fold symmetry contains 120 triangles instead of 96 (Fig.~\ref{Sfig:toroidExp}A, B). By counting all fully closed structures in TEM micrographs, we observe that 21\% of the toroids are misassembled into 5-fold toroids (Fig.~\ref{Sfig:toroidExp}C). This comes as a surprise since 5-fold toroids, in theory, cannot be assembled using only equilateral triangles. Assuming that each edge has the same stretching modulus, we show that at maximum, the triangle edges have to stretch by about 4.7\% to close into a 5-fold toroid (Fig.~\ref{Sfig:toroidExp}D). Though the stretching of the triangles is one plausible mechanism to explain the presence of the 5-fold toroid, we think that the edges of the triangles are rigid, preventing much deformation. 

Another possible mechanism is the skewed bend angles between the tubelets due to the difference in the angle-domain length we assign to the triangles. Here, we refer to the two-layer, straight region of the toroid as a tubelet, whereas the eight-triangle-domain connecting the two tubelets is referred to as a junction (Fig.~\ref{Sfig:toroidExp}A). We hypothesize that the inner bond and the outer bond (Fig.~\ref{Sfig:toroidExp}E) connecting the tubelet and the junction region have different angle-domain lengths, leading to more opened toroids: the outer bond encodes a positive angle with $+3\delta$, and the inner encodes a negative angle with $-10\delta$. On average, this amounts to 6.5 bp for inner bonds and 1.5 bp for outer bonds. Since the inner bond is longer, the two tubelets are pushed outward, contributing to the opened angles. Simply comparing the average length of the angle domains for two edges, we expect an opening angle of 109.7 degrees. 

To test this hypothesis, we first measure the bend angle of the toroid and their bending modulus from opened toroids. Here, the bend angle is defined as the angle between two neighboring tubelets (Fig.~\ref{Sfig:toroidExp}F). Ideally, in a 4-fold toroid, the bend angle should be 90 degrees, and for a 5-fold toroid, 108 degrees. Using TEM micrographs of opened toroids, we measure the distribution of bend angles (Fig.~\ref{Sfig:toroidExp}G). By fitting the distribution with a Gaussian, we obtain the mean bend angle to be 104.2 degrees, which is not far from the predicted 109.7 degrees. Assuming the bend angle fluctuation follows a Boltzmann distribution with the energetic cost for angle deviation $B\Delta\theta^2/2$, where $B$ is the bending modulus of the bend angle, the standard deviation of the Gaussian, $\sigma$, can be associated with the bending modulus by $\sigma^2=1/B$. As a result, we obtain the bending modulus of 23.4~$kT/$rad$^2$. 

Using a thermodynamic model similar to that developed for vertex assembly, we confirm the bend angle from the distribution of 4- and 5-fold toroids. We develop a thermodynamic model to predict the yield and selectivity of the toroid of different sizes. We assume that the tubelet and junction constitute a rigid particle and that the preferred angles between the two of them are the bend angle. Inputting the bending modulus of 23~$kT/$rad$^2$, binding energy of -17~$kT$, we obtain the yield as shown in Fig.~\ref{Sfig:toroidExp}H. We note that we limit the maximum toroid size to 10-fold, but this does not impact the analysis of the experiment. In Fig.~\ref{Sfig:toroidExp}I, we show the comparison of the selectivity of the closed toroid from an experiment with that of the model at a bend angle of 104 degrees. The two selectivities agree well, supporting the hypothesis that relative inter-triangle bond lengths can contribute to the global assembly outcome. Going forward, preventing polymorphic structures while targeting the assembly of large complex structures may require the tuning of relative lengths of the bonds between multiple triangles.

\clearpage

\section{Symmetry-guided inverse design for finite 3D geometries} \label{sec:invdesign}

Given an arbitrary geometry, how do we design a set of unique subunits that can assemble the geometry? Moreover, how do we design such a set with a minimal number of unique subunits or interactions? This inverse-design method for self-assembly is usually a nontrivial task; the computation required diverges as the target structure becomes larger or more complex. Previously, we developed an algorithmic, inverse design strategy for constructing the minimal set of subunits that assemble into periodic 2D tilings, assuming specific, short-ranged, rigid bonds between the subunits~\cite{Hayakawa2024Mar}. The core idea is to utilize the translational and rotational symmetries that appear in the tilings to assign specific interactions to each bond deterministically. We note that reflection and glide reflection symmetries are omitted from the discussion here and onward since they require subunits to match their enantiomorphs. This symmetry-based design method can be extended to design the self-assembly of various geometries upon proper choice of the symmetries. To this extent, we show that the 3D structures we target in this paper, including toroids and non-Platonic shells, can be designed algorithmically by exploiting the spherical symmetry groups. 

\begin{figure*}[th!]
 \centering
 \includegraphics[width=0.6\textwidth]{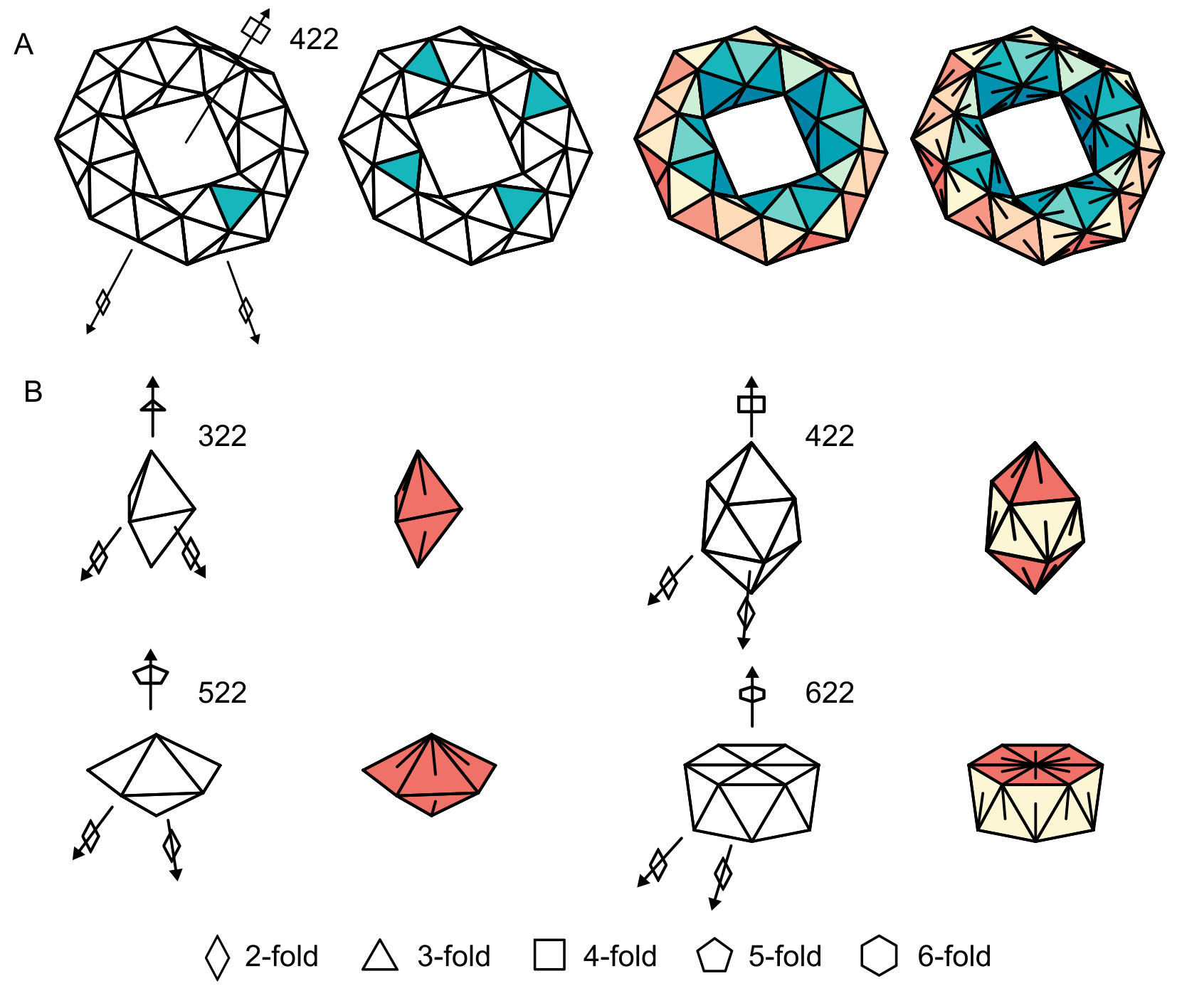}
 \caption{\textbf{Symmetry-guided inverse design method.}
(A) Step-by-step instructions for determining the species and orientations of the triangles within a toroid. Unique rotational symmetries are shown using an arrow for the rotating axis and a polygon to represent different orders of rotational symmetry. (B) The species and the orientations of triangles in triangular bipyramid, pentagonal bipyramid, gyroelongated square bipyramid, and hexagonal antiprism determined using the inverse-design method. The numbers indicate the symmetry groups used for designing each structure.}
\label{Sfig:invdesign}
\end{figure*}

As an example, the toroid with 96 equilateral triangles can be assembled from 12 unique triangles upon imposing the 422 symmetry. To carry out this inverse design, we first identify the symmetries that can be imposed on the target geometry. Remembering that reflection symmetries cannot exist, we find 4-fold rotational symmetry and two unique 2-fold rotational symmetries, shown in Fig.~\ref{Sfig:invdesign}A. Using an orbifold notation~\cite{conway_symmetries_2008}, this toroid thus belongs to one of the spherical symmetry groups, 422. We note that though this toroid also satisfies any subsymmetries of 422, the reduction in the order of the symmetry causes an increase in the number of unique interactions required~\cite{Hayakawa2024Mar}. Next, we choose a triangle and assign a color. Using the 422 symmetry, we identify the equivalent triangles upon symmetry operations and assign them the same color. We then choose another uncolored triangle, assign a color, and color all equivalent triangles. This operation continues until all triangles are colored. Finally, we determine the orientation of the triangles with a similar procedure; we find an orientation-free triangle, assign an orientation, and apply this to all equivalent triangles. As a result, the assembly components for this toroid are 12 unique triangles with 36 unique interactions. Applying this strategy, we also design non-Platonic shells, as shown in Fig.~\ref{Sfig:invdesign}B.

\clearpage
\section{Overhang design for the interface strands} \label{sec:overhang}

\begin{figure*}[th!]
 \centering
 \includegraphics[width=\textwidth]{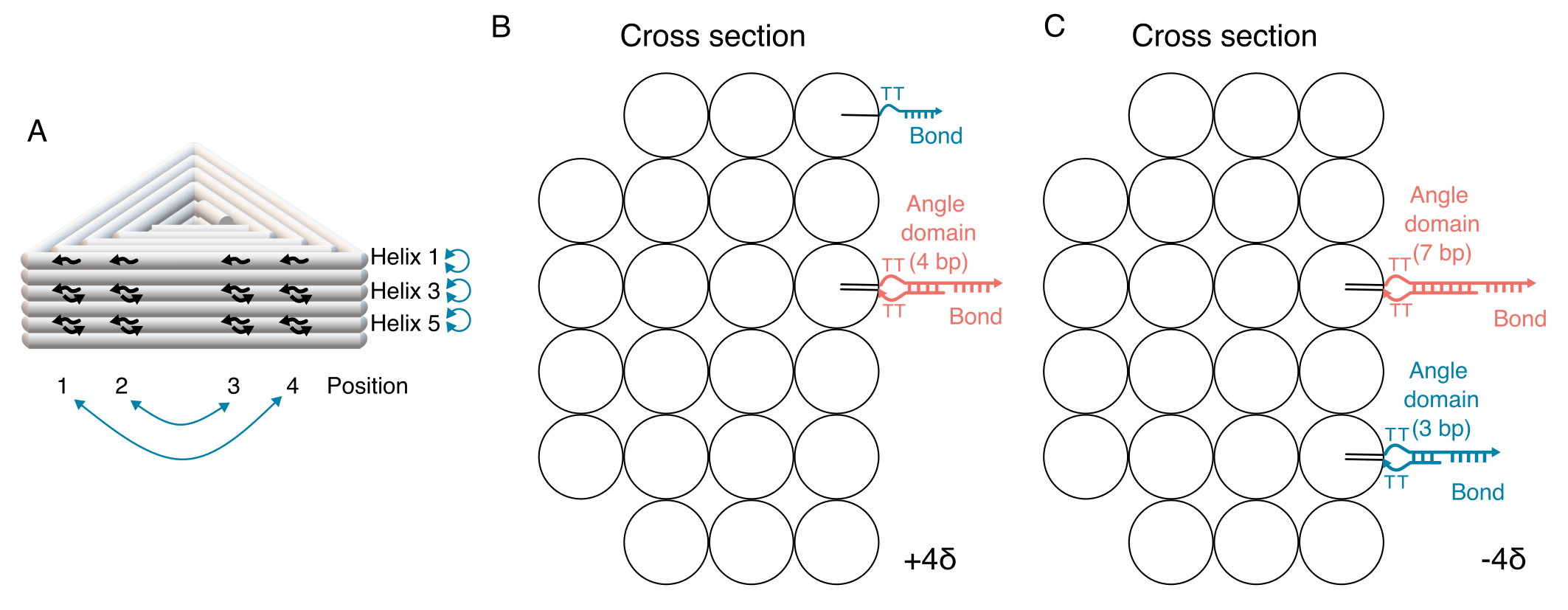}
 \caption{\textbf{Schematic of the interface design.}
(A) The overhang configuration on the side of the triangle. Black arrows indicate the locations where overhangs can extrude. The 3'-end is indicated by the arrow head. The blue arrows indicate the binding pairs of the strands. (B) Cross-sectional view of the overhangs for $+4\delta$ triangle. (C) Cross-sectional view of the overhangs for $-4\delta$ triangle.}
\label{sFIG:overhang}
\end{figure*}

Each triangle side can extrude up to 20 interface strands, distributed across 12 sites arranged in a 3-by-4 grid (Fig.~\ref{sFIG:overhang}A). These strands extrude from helices 1, 3, and 5, at the same four vertical positions on each helix referred to as positions 1 through 4. On helix 1, a single strand with a 3'-end extrudes at each site. On helices 3 and 5, two adjacent strands extrude at each site: one with a 3'-end and the other with a 5'-end. These adjacent strands hybridize to form the double-stranded DNA angle domain. Each triangle side can accommodate a total of 20 extruded strands, but only two helices are activated at a time to encode the desired binding angle. For positive angles, strands on helices 1 and 3 are extruded, while those on helix 5 are terminated at the scaffold. Conversely, for negative angles, strands on helices 3 and 5 are extruded, and strands on helix 1 are terminated.

\begin{figure*}[th!]
 \centering
 \includegraphics[width=0.9\textwidth]{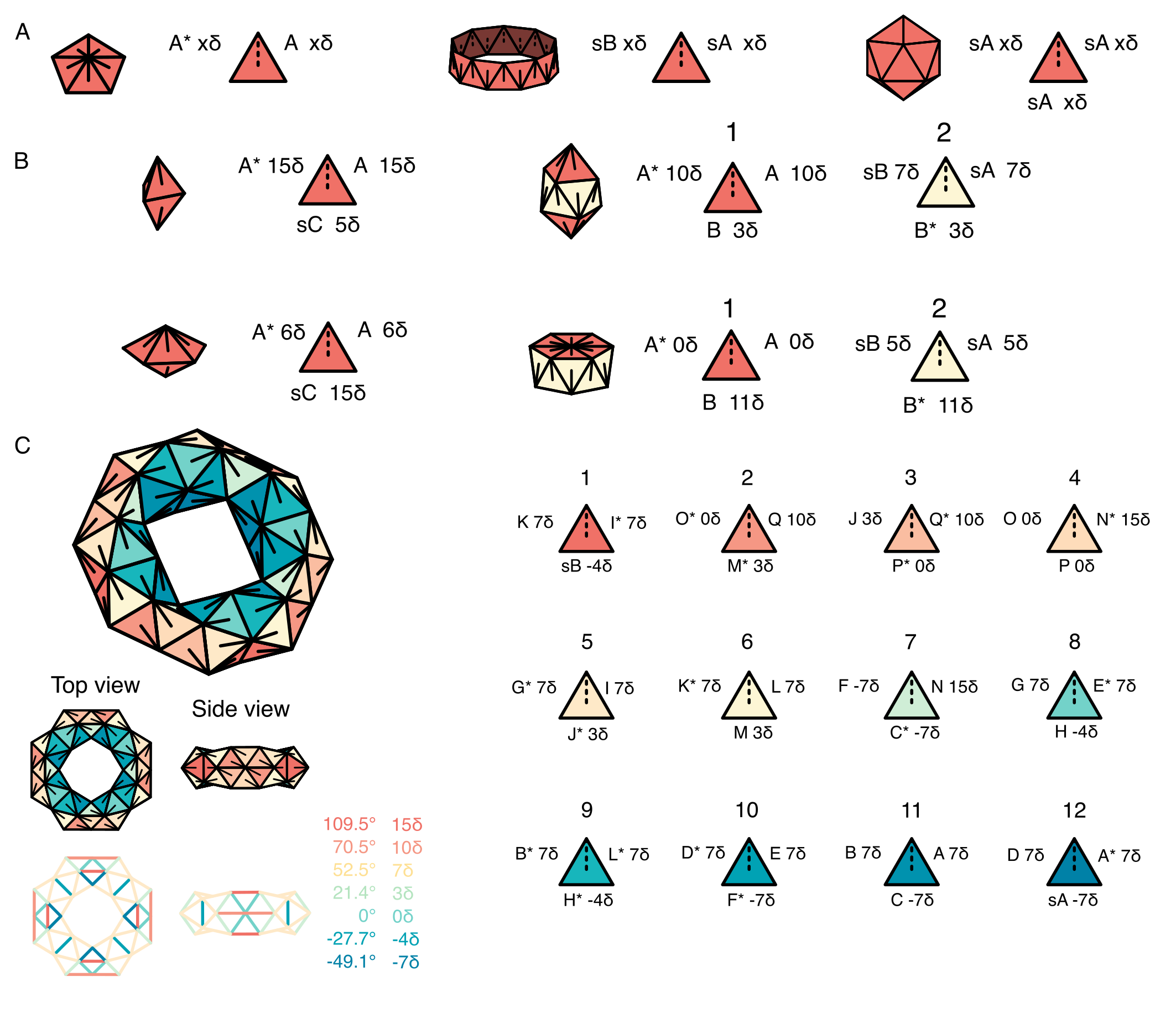}
  \caption{\textbf{Bond-domain sequences and angle-domain lengths used for various assemblies.} 
The bond-domain sequences and angle-domain lengths used to assemble (A) vertices, rings, shells, (B) non-Platonic shells, and (C) toroids. The line in the triangle helps identify the orientation of the triangle (Fig.~2F). The number above triangles indicates the species identification for multispecies assemblies. The angle-domain lengths for vertices, rings, and shells in (A) are varied, while the bond-domain sequences remain consistent, as indicated in the main text. All the angle-domain lengths for the non-Platonic shells and toroids are derived from a linear fit shown in Fig.~2C, except we use a slope of 7.18 degrees per base-pair for the positive binding angles. The geometries and binding angles of non-Platonic shells are explained in detail in Fig.~\ref{Sfig:capsid_geometry}.} 
\label{Sfig:design1}
\end{figure*}

The bond-domain sequences on the interface strands that extrude on the 3'-end encode the specific interactions necessary to assemble a target structure. To ensure the binding of triangles in the correct orientation, every 3'-extruder strand on a helix is assigned a unique sequence. The sequences are carefully designed to prevent unintended interactions, following a protocol previously developed~\cite{videbaek2023economical}. For two sides to bind, the bond-domain sequences of the two sides must be complementary in the configuration shown by the arrow in Fig~\ref{sFIG:overhang}A; strands on position 1 bind to position 4 and position 2 bind to position 3 on the same helices. Unique sequences following this same pattern of interaction are applied to helices 1, 3 and 5. See Fig.~\ref{Sfig:design1} for interactions encoded in each triangle for assembly, as well as Table~\ref{tab:hand_seq} for the corresponding bond-domain sequences.

Binding angles are programmed by adjusting the length of the double-stranded DNA angle domain. Positive angles are encoded by elongating the overhangs on helix 3, where two strands (3'-end and 5'-end) hybridize to form the angle domain (Fig.~\ref{sFIG:overhang}B). In these cases, the overhangs on helix 1 remain short, consisting of two thymines and a five-nucleotide bond-domain sequence. For negative angles, short overhangs of the form of helix 1 were initially extruded from helix 5, but steric hindrance from helix 6 at large negative angles limited the range of angles that could be realized. To address this limitation,  overhangs on helix 5 are extended by adding a 3-bp angle domain (Fig.~\ref{sFIG:overhang}C).  Positive and negative angles are denoted as $+n\delta$ and $-(n-3)\delta$, respectively, where $n$ is the angle-domain length on helix 3 in base pairs. Detailed bond-domain sequences and angle-domain lengths for specific geometries are provided in Fig.~\ref{Sfig:design1}. See Table~\ref{tab:arm_seq_row3} and Table~\ref{tab:arm_seq_row5} for angle-domain sequences.

The astute reader will realize that alternative configurations using helices 4 and 6 could mitigate these steric issues. However, by the time we realized this solution, implementing such changes would have required rerouting of the scaffold, which we chose not to do as we wanted all our triangular building blocks to be based on the same core.

\begin{longtable}{|c||c|c|c|c||c|c|c|c|}
\hline
& Helix3 & Helix3 & Helix3 & Helix3 & Helix1 or 5 & Helix1 or 5 & Helix1 or 5 & Helix1 or 5 \\ 
Name & Pos1 & Pos2 & Pos3 & Pos4 & Pos1 & Pos2 & Pos3 & Pos4 \\ 
\hline
\endfirsthead
\hline
     & Helix3 & Helix3 & Helix3 & Helix3 & Helix1 or 5 & Helix1 or 5 & Helix1 or 5 & Helix1 or 5 \\ 
Name & Pos1 & Pos2 & Pos3 & Pos4 & Pos1 & Pos2 & Pos3 & Pos4 \\ 
\hline 
\endhead

\endfoot

\endlastfoot

sA & ACTAG & AGTTA & TAACT & CTAGT & TTCAA & CCATT & AATGG & TTGAA \\ 
sB & TTAAC & TCGAC & GTCGA & GTTAA & TCAGA & GTCTA & TAGAC & TCTGA \\ 
sC & ATAAG & CTCGA & TCGAG & CTTAT & GATCT & CTGAT & ATCAG & AGATC \\ 
sD & GACAG & AGATT & AATCT & CTGTC & ATTGT & CAGGT & ACCTG & ACAAT \\ 
sE & ATGCA & ATTCT & AGAAT & TGCAT & GCCAA & TACGA & TCGTA & TTGGC \\ 
A & CAATA & TGATT & CTAGG & CACAT & ATGAC & TACAG & AACCT & GAGAC \\ 
A* & ATGTG & CCTAG & AATCA & TATTG & GTCTC & AGGTT & CTGTA & GTCAT \\ 
B & GCATC & TATTC & AGATT & TTCTC & AGATA & TTCCT & TTCCA & GATAT \\ 
B* & GAGAA & AATCT & GAATA & GATGC & ATATC & TGGAA & AGGAA & TATCT \\ 
C & AGTTC & CGATT & ATTCT & ATTCA & GGATA & TCATC & GGTAT & GGTAA \\ 
C* & TGAAT & AGAAT & AATCG & GAACT & TTACC & ATACC & GATGA & TATCC \\ 
D & CACAT & CAATA & ACGAA & TGATT & GAGAC & ATGAC & GACAG & TACAG \\ 
D* & AATCA & TTCGT & TATTG & ATGTG & CTGTA & CTGTC & GTCAT & GTCTC \\ 
E & TTCTC & GCATC & CTGTG & TATTC & GATAT & AGATA & ATGCA & TTCCT \\ 
E* & GAATA & CACAG & GATGC & GAGAA & AGGAA & TGCAT & TATCT & ATATC \\ 
F & ATTCA & AGTTC & CTTGA & CGATT & GGTAA & GGATA & ACTGA & TCATC \\ 
F* & AATCG & TCAAG & GAACT & TGAAT & GATGA & TCAGT & TATCC & TTACC \\ 
G & ACGAA & CTAGG & ACCTG & CAATA & GACAG & AACCT & ACTAA & ATGAC \\ 
G* & TATTG & CAGGT & CCTAG & TTCGT & GTCAT & TTAGT & AGGTT & CTGTC \\ 
H & CTGTG & AGATT & TCGTA & GCATC & ATGCA & TTCCA & AACAT & AGATA \\ 
H* & GATGC & TACGA & AATCT & CACAG & TATCT & ATGTT & TGGAA & TGCAT \\ 
I & CTTGA & ATTCT & GTAGA & AGTTC & ACTGA & GGTAT & AGAGA & GGATA \\ 
I* & GAACT & TCTAC & AGAAT & TCAAG & TATCC & TCTCT & ATACC & TCAGT \\ 
J & CAATA & GACCT & GGTAT & ATTCA & ATGAC & TGATT & CCTAT & GGTAA \\ 
J* & TGAAT & ATACC & AGGTC & TATTG & TTACC & ATAGG & AATCA & GTCAT \\ 
K & GCATC & AGTCA & AACCT & CACAT & AGATA & TATTC & CGATG & GAGAC \\ 
K* & ATGTG & AGGTT & TGACT & GATGC & GTCTC & CATCG & GAATA & TATCT \\ 
L & AGTTC & CGTCC & TTCCA & TTCTC & GGATA & CGATT & CTTGT & GATAT \\ 
L* & GAGAA & TGGAA & GGACG & GAACT & ATATC & ACAAG & AATCG & TATCC \\ 
M & ATTCA & CAATA & GACAG & GACCT & GGTAA & ATGAC & CTTGA & TGATT \\ 
M* & AGGTC & CTGTC & TATTG & TGAAT & AATCA & TCAAG & GTCAT & TTACC \\ 
N & CACAT & GCATC & TAACA & AGTCA & GAGAC & AGATA & ACGAA & TATTC \\ 
N* & TGACT & TGTTA & GATGC & ATGTG & GAATA & TTCGT & TATCT & GTCTC \\ 
O & TTCTC & AGTTC & AGTAT & CGTCC & GATAT & GGATA & CTGTG & CGATT \\ 
O* & GGACG & ATACT & GAACT & GAGAA & AATCG & CACAG & TATCC & ATATC \\ 
P & GACAG & GGTAT & AACAT & CAATA & CTTGA & CCTAT & AACTA & ATGAC \\ 
P* & TATTG & ATGTT & ATACC & CTGTC & GTCAT & TAGTT & ATAGG & TCAAG \\ 
Q & TAACA & AACCT & AGAGA & GCATC & ACGAA & CGATG & TCTTC & AGATA \\ 
Q* & GATGC & TCTCT & AGGTT & TGTTA & TATCT & GAAGA & CATCG & TTCGT \\ 
R & AGTAT & TTCCA & ACTAA & AGTTC & CTGTG & CTTGT & GTATG & GGATA \\ 
R* & GAACT & TTAGT & TGGAA & ATACT & TATCC & CATAC & ACAAG & CACAG \\ 
S & GTACA & AGTCA & CGATG & CTTAC & ACAAT & CGTCC & CTTGT & CTACA \\ 
S* & GTAAG & CATCG & TGACT & TGTAC & TGTAG & ACAAG & GGACG & ATTGT \\ 
T & TTGGC & GACCT & CCTAT & CTTAG & CTTAC & GTACA & AGTAT & AGTCA \\ 
T* & CTAAG & ATAGG & AGGTC & GCCAA & TGACT & ATACT & TGTAC & GTAAG \\ 
U & CTACA & ACAAT & GACAG & CGTCC & CTTAG & TTGGC & TAACA & GACCT \\ 
U* & GGACG & CTGTC & ATTGT & TGTAG & AGGTC & TGTTA & GCCAA & CTAAG \\ 
V & AGTAT & CGATG & GTATG & GTACA & GACAG & CTTGT & AACTA & ACAAT \\ 
V* & TGTAC & CATAC & CATCG & ATACT & ATTGT & TAGTT & ACAAG & CTGTC \\ 
W & TAACA & CCTAT & TCTTC & TTGGC & GTACA & TACAG & CTAGG & CTTAG \\ 
W* & GCCAA & GAAGA & ATAGG & TGTTA & CTAAG & CCTAG & CTGTA & TGTAC \\ 
X & ACAAT & TTCCT & AGATT & CTTAC & TTGGC & TCATC & ATTCT & CTACA \\ 
X* & GTAAG & AATCT & AGGAA & ATTGT & TGTAG & AGAAT & GATGA & GCCAA \\ 
Y & CTTAG & GTACA & ACTGA & TACAG & CTTAC & ACAAT & GACAG & TTCCT \\ 
Y* & CTGTA & TCAGT & TGTAC & CTAAG & AGGAA & CTGTC & ATTGT & GTAAG \\ 
Z & CTACA & TTGGC & ATGCA & TCATC & ACTGA & CTAGG & GTAGA & GTACA \\ 
Z* & GATGA & TGCAT & GCCAA & TGTAG & TGTAC & TCTAC & CCTAG & TCAGT  
\\
\hline
\caption{\textbf{Bond-domain sequences for multispecies assemblies.} A lowercase ‘s’ in the bond-domain sequence name indicates that the sequence is self-complementary. An asterisk denotes a sequence that is complementary to the original.}
\label{tab:hand_seq}
\end{longtable}

\begin{longtable}{|c|c|c|c|c|c|}
\hline
Side & arm length (bp) & Pos1 & Pos2 & Pos3 & Pos4 \\ 
\hline
\endfirsthead
Side & arm length (bp) & Pos1 & Pos2 & Pos3 & Pos4 \\ 
\hline 
\endhead

\endfoot

\endlastfoot
1 & 15 & AACTATCATACATAT & TCTACCTAATCTCTT & CTCATCACCTCCTAC & ATCCATACAACACCA \\ 
1 & 14 & AACTATCATACATA & TCTACCTAATCTCT & CTCATCACCTCCTA & ATCCATACAACACC \\ 
1 & 13 & AACTATCATACAT & TCTACCTAATCTC & CTCATCACCTCCT & ATCCATACAACAC \\ 
1 & 12 & AACTATCATACA & TCTACCTAATCT & CTCATCACCTCC & ATCCATACAACA \\ 
1 & 11 & AACTATCATAC & TCTACCTAATC & CTCATCACCTC & ATCCATACAAC \\ 
1 & 10 & AACTATCATA & TCTACCTAAT & CTCATCACCT & ATCCATACAA \\ 
1 & 9 & AACTATCAT & TCTACCTAA & CTCATCACC & ATCCATACA \\ 
1 & 8 & AACTATCA & TCTACCTA & CTCATCAC & ATCCATAC \\ 
1 & 7 & AACTATC & TCTACCT & CTCATCA & ATCCATA \\ 
1 & 6 & AACTAT & TCTACC & CTCATC & ATCCAT \\ 
1 & 5 & AACTA & TCTAC & CTCAT & ATCCA \\ 
1 & 4 & AACT & TCTA & CTCA & ATCC \\ 
1 & 3 & AAC & TCT & CTC & ATC \\ 
2 & 15 & ATCCACTATTACAGT & CTACCATATCCTTAC & TAACTTATTACTATA & TCTTCTAACAACTCA \\ 
2 & 14 & ATCCACTATTACAG & CTACCATATCCTTA & TAACTTATTACTAT & TCTTCTAACAACTC \\ 
2 & 13 & ATCCACTATTACA & CTACCATATCCTT & TAACTTATTACTA & TCTTCTAACAACT \\ 
2 & 12 & ATCCACTATTAC & CTACCATATCCT & TAACTTATTACT & TCTTCTAACAAC \\ 
2 & 11 & ATCCACTATTA & CTACCATATCC & TAACTTATTAC & TCTTCTAACAA \\ 
2 & 10 & ATCCACTATT & CTACCATATC & TAACTTATTA & TCTTCTAACA \\ 
2 & 9 & ATCCACTAT & CTACCATAT & TAACTTATT & TCTTCTAAC \\ 
2 & 8 & ATCCACTA & CTACCATA & TAACTTAT & TCTTCTAA \\ 
2 & 7 & ATCCACT & CTACCAT & TAACTTA & TCTTCTA \\ 
2 & 6 & ATCCAC & CTACCA & TAACTT & TCTTCT \\ 
2 & 5 & ATCCA & CTACC & TAACT & TCTTC \\ 
2 & 4 & ATCC & CTAC & TAAC & TCTT \\ 
2 & 3 & ATC & CTA & TAA & TCT \\ 
3 & 15 & ACCAACTTCTTATAA & TAATTCAACTCTTGC & CAACCTCTCCACTGT & TTCCACTCACCAAAT \\ 
3 & 14 & ACCAACTTCTTATA & TAATTCAACTCTTG & CAACCTCTCCACTG & TTCCACTCACCAAA \\ 
3 & 13 & ACCAACTTCTTAT & TAATTCAACTCTT & CAACCTCTCCACT & TTCCACTCACCAA \\ 
3 & 12 & ACCAACTTCTTA & TAATTCAACTCT & CAACCTCTCCAC & TTCCACTCACCA \\ 
3 & 11 & ACCAACTTCTT & TAATTCAACTC & CAACCTCTCCA & TTCCACTCACC \\ 
3 & 10 & ACCAACTTCT & TAATTCAACT & CAACCTCTCC & TTCCACTCAC \\ 
3 & 9 & ACCAACTTC & TAATTCAAC & CAACCTCTC & TTCCACTCA \\ 
3 & 8 & ACCAACTT & TAATTCAA & CAACCTCT & TTCCACTC \\ 
3 & 7 & ACCAACT & TAATTCA & CAACCTC & TTCCACT \\ 
3 & 6 & ACCAAC & TAATTC & CAACCT & TTCCAC \\ 
3 & 5 & ACCAA & TAATT & CAACC & TTCCA \\ 
3 & 4 & ACCA & TAAT & CAAC & TTCC \\ 
3 & 3 & ACC & TAA & CAA & TTC

\\ 
\hline
\caption{\textbf{Angle-domain sequence on 3'-end strand located on helix 3.} The angle-domain sequences on 5'-end strand are complementary to the sequences listed here.}
\label{tab:arm_seq_row3}
\end{longtable}

\begin{longtable}{|c|c|c|c|c|c|}
\hline
Side & arm length (bp) & Pos1 & Pos2 & Pos3 & Pos4 \\ 
\hline
\endfirsthead

Side & arm length (bp) & Pos1 & Pos2 & Pos3 & Pos4 \\ 
\hline 
\endhead

\endfoot

\endlastfoot
1 & 3 & TCC & TCG & ACC & ACG \\ 
2 & 3 & CTC & CTG & GTG & GCA \\ 
3 & 3 & CCT & GCT & CCA & GTC 

\\
\hline
\caption{\textbf{Angle-domain sequence of 3'-end strand located on helix 5.} The angle-domain sequences on 5'-end strand are complementary to the sequences listed here.}
\label{tab:arm_seq_row5}
\end{longtable}

\clearpage



\clearpage

\begin{figure*}[th!]
 \centering
 \includegraphics[width=\textwidth]{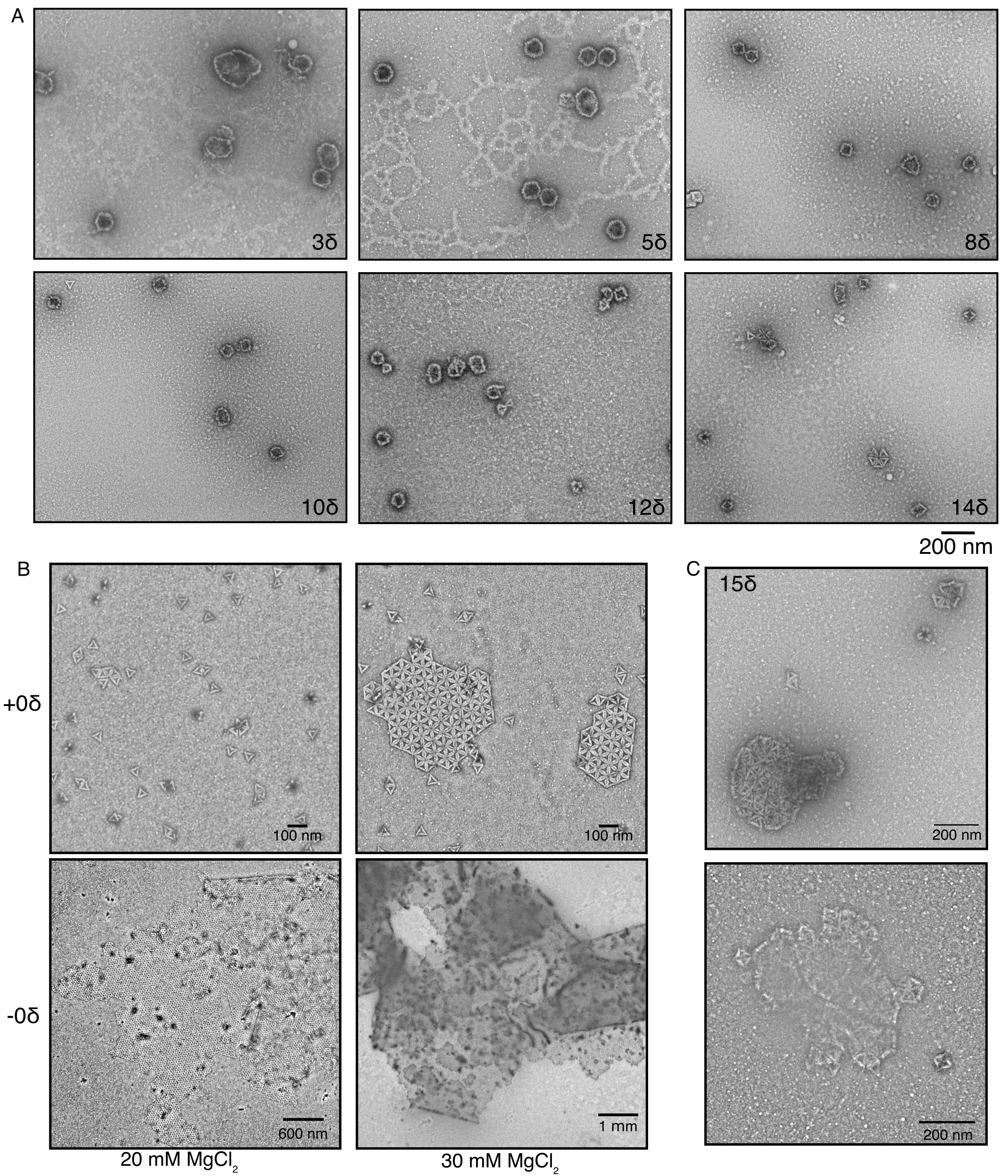}
 \caption{\textbf{Example TEM micrographs of shell assemblies. }
(A) Wide-field-of-view negative stain TEM images of the assembly of $+3\delta$, $+5\delta$, $+8\delta$, $+10\delta$, $+12\delta$, $+15\delta$ triangles at 10~nM concentration at 20~mM MgCl$_{2}$ concentrations. (B) Negative stain TEM images of the sheet assembly using $+0\delta$ and $-0\delta$ triangles at 10~nM concentration at 20~mM and 30mM MgCl$_{2}$ concentrations. Whereas $-0\delta$ assemble to sheets over a micrometer at 20~mM  MgCl$_{2}$, we do not observe any assemblies using $+0\delta$ triangles at 20~mM  MgCl$_{2}$. We think the lack of blunt-end stacking interaction between the bond and angle domain for $+0\delta$ assemblies leads to weaker interactions. (C) Examples of aggregates we observe for $+15\delta$ assemblies at 20~mM  MgCl$_{2}$. 
}
\label{sFIG:shellTEM}
\end{figure*}

\begin{figure*}[th!]
 \centering
 \includegraphics[width=\textwidth]{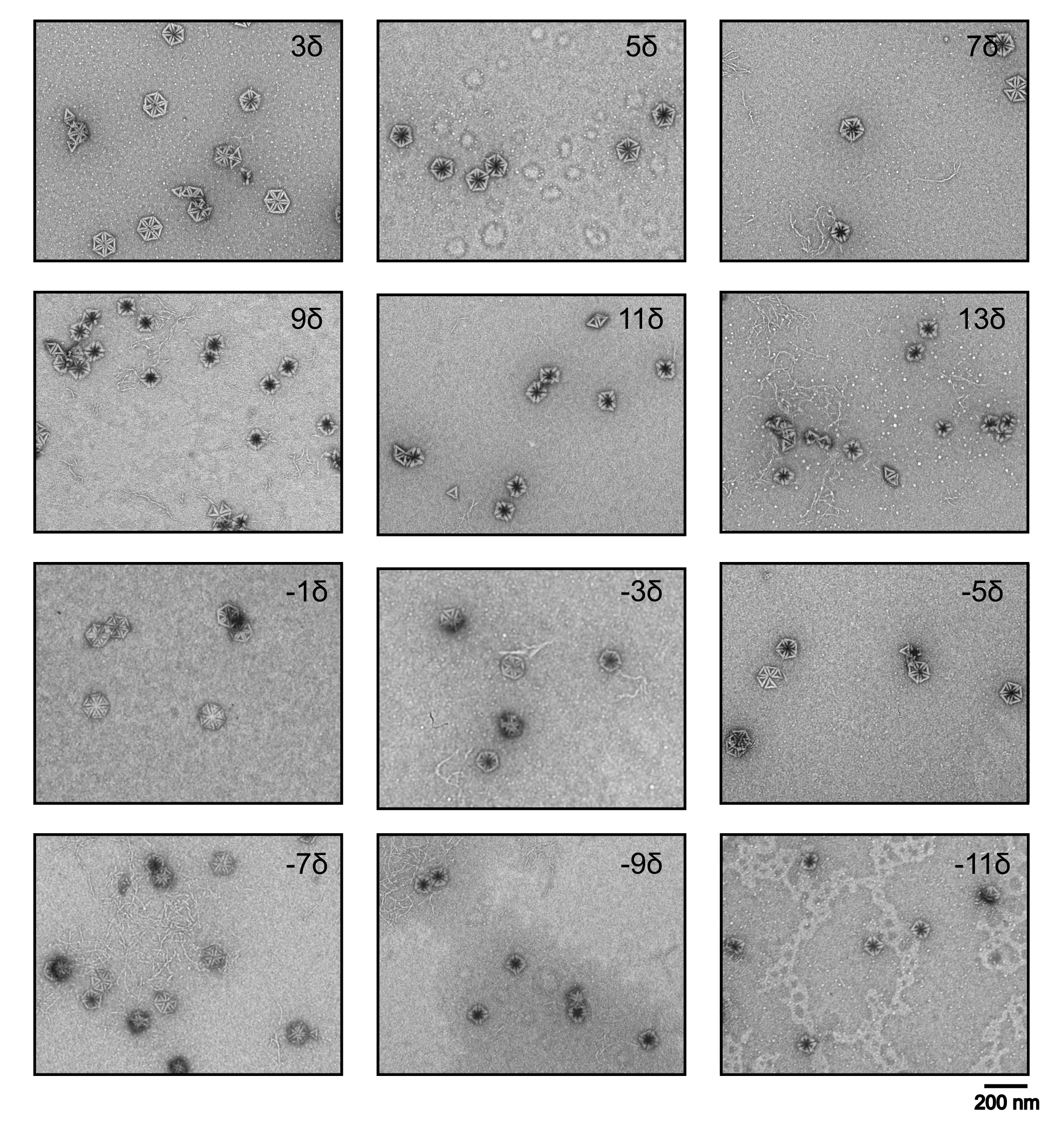}
 \caption{\textbf{TEM images of vertex assemblies.}
Wide-field-of-view TEM images of vertex assemblies with various angle-domain lengths. Positive angle vertices are assembled at 30~mM  MgCl$_{2}$ and negative angle vertices are assembled at 20~mM  MgCl$_{2}$.
}
\label{Sfig:vertexTEM}
\end{figure*}

\begin{figure*}[th!]
 \centering
 \includegraphics[width=0.8\textwidth]{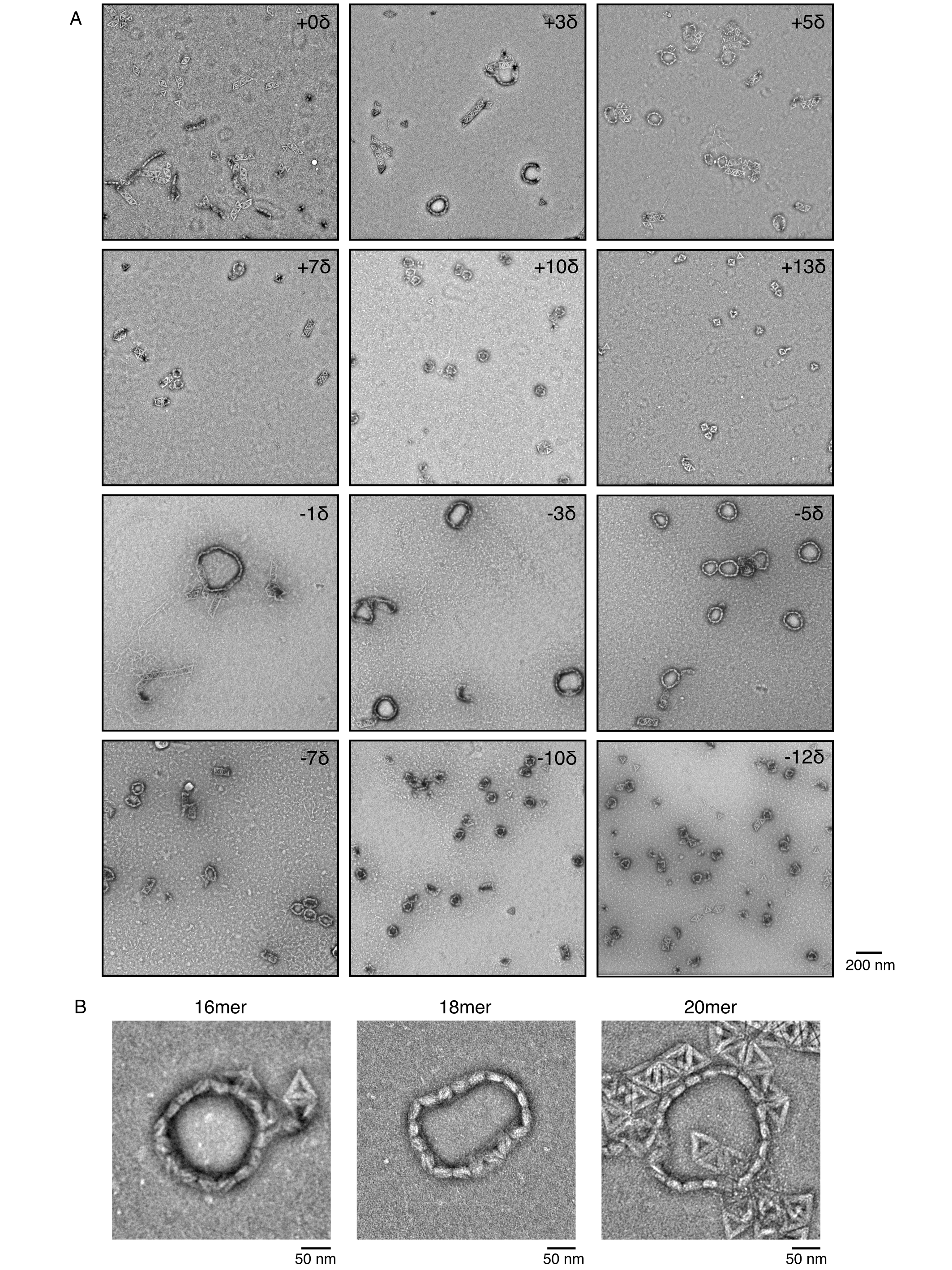}
 \caption{\textbf{TEM images of ring assemblies.}
(A) Wide-field-of-view TEM images of ring assemblies with various angle-domain lengths. All rings are assembled at 20~mM  MgCl$_{2}$. (B) Large rings observed in $+3\delta$ sample.
}
\label{Sfig:ringTEM}
\end{figure*}


\begin{figure*}[th!]
 \centering
 \includegraphics[width=\textwidth]{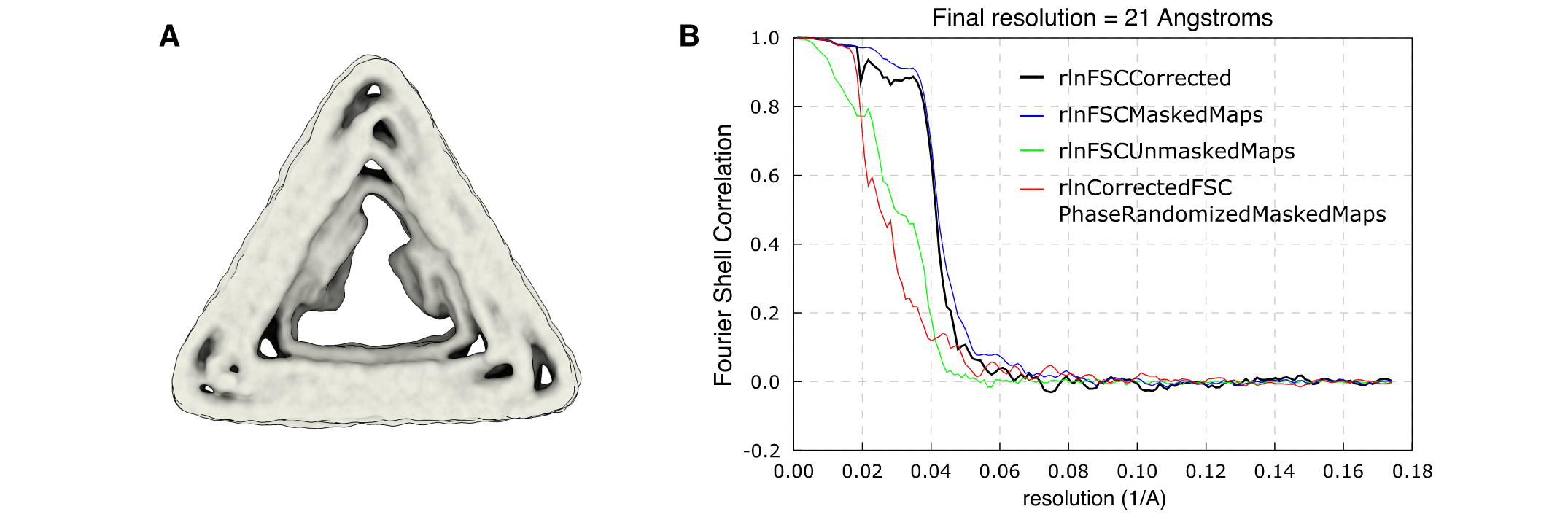}
 \caption{\textbf{Cryo-EM single particle reconstruction of the modular block.} (A) Top-down view of the reconstruction. (B) Plot of the FSC curves used to estimate the resolution of the reconstruction. 
}
\label{Sfig:cryo-monomer}
\end{figure*}

\begin{figure*}[th!]
 \centering
 \includegraphics[width=\textwidth]{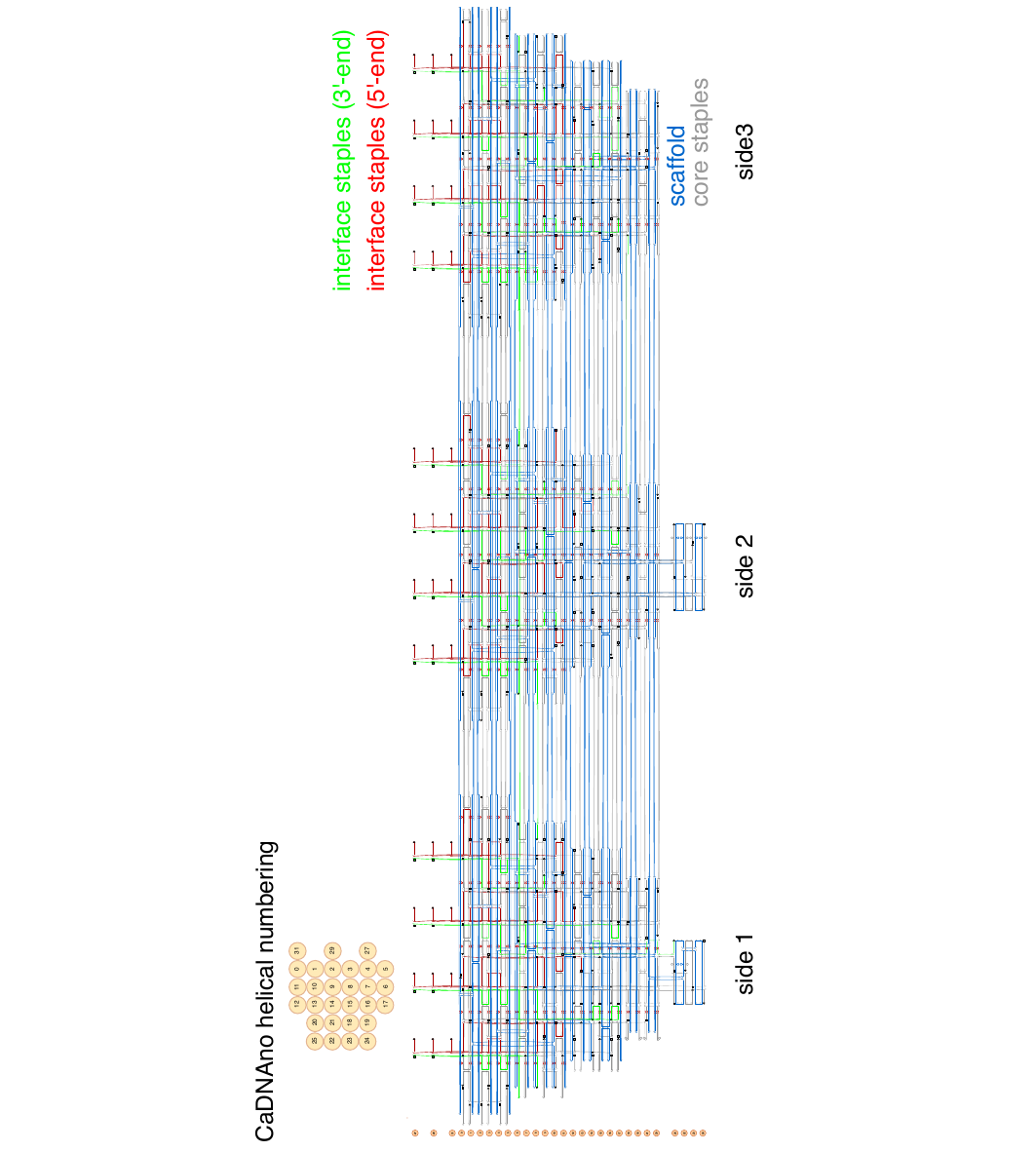}
 \caption{\textbf{caDNAno design of DNA origami triangle with helical numbering.} 
}
\label{Sfig:CaDNAno}
\end{figure*}


\clearpage

\bibliography{main.bib}